\documentclass[preprint2]{aastex}
\usepackage{color}
\usepackage{epsfig}
\usepackage[fleqn]{amsmath}
\usepackage{multirow}
\usepackage{multicol}
\usepackage{mathptmx}
\usepackage{graphicx}
\usepackage{bm}
\usepackage{mathrsfs}
\usepackage{lineno}

\shorttitle{Equivalence between two charged black holes}
\shortauthors{Zhang et al.}
\begin{document}
\title{\large Equivalence between two charged black holes in dynamics of orbits  outside the event horizons}
\author{Hongxing Zhang$^{1,2}$, Naying Zhou$^{1,2}$, Wenfang Liu$^{1}$, Xin Wu$^{1,2,3, \dag}$}
\affil{
1. School of Mathematics, Physics and Statistics, Shanghai University of Engineering Science, Shanghai 201620,
China \\
2.
Center of Application and Research of Computational Physics, Shanghai University of Engineering Science,
Shanghai 201620, China \\
3. Guangxi Key Laboratory for Relativistic Astrophysics, Guangxi University, Nanning 530004, China\\} \email{$\dag$ Corresponding
Author: wuxin$\_$1134@sina.com, xinwu@gxu.edu.cn} 
\begin{abstract}

Using the Fermi--Dirac distribution function, Balart and Vagenas
gave a charged spherically symmetric regular black hole, which is
a solution of Einstein field equations coupled to a nonlinear
electrodynamics. In fact, the regular black hole is a
Reissner-Nordstr\"{o}m (RN)  black hole when a metric function is
given a Taylor expansion to first order approximations. It does
not have a curvature singularity at the origin, but the RN black
hole does. Both black hole metrics have horizons and similar
asymptotic behaviors, and satisfy the weak energy conditions
everywhere. They are almost the same in photon effective
potentials, photon circular orbits and photon spheres outside the
event horizons. Due to the approximately same photon spheres, the
two black holes have no explicit differences in the black hole
shadows and constraints of the black hole charges based on the
First Image of Sagittarius A$^{*}$. There are relatively minor
differences between effective potentials, stable circular orbits
and innermost stable circular orbits of charged particles outside
the event horizons of the two black holes immersed in external
magnetic fields. Although the two magnetized black holes allow
different construction methods of explicit symplectic integrators,
they exhibit approximately consistent results in the regular and
chaotic dynamics of charged particles outside the event horizons.
Chaos gets strong as the magnetic field parameter or the magnitude
of negative Coulomb parameter increases, but becomes weak when the
black hole charge or the positive Coulomb parameter increases. A
variation of dynamical properties is not sensitive dependence on
an appropriate increase of the black hole charge. The basic
equivalence between the two black hole gravitational systems in
the dynamics of orbits outside the event horizons is due to the
two metric functions having an extremely small difference. This
implies that the RN black hole is reasonably replaced by the
regular black hole without curvature singularity in many
situations.

\end{abstract}

\section{Introduction}

The standard black holes in General Relativity, such as the
Schwarzschild black hole, the Reissner-Nordstr\"{o}m (RN) black
hole and the Kerr black hole, are classical solutions of the
Einstein field equations. Most of them possess curvature
singularities dressed by event horizons. The existence of
singularities causes the exterior region of a black hole to have
no causal connection with the interior of black hole. The singular
behavior of the known black hole solutions makes it impossible to
describe physical objects falling process in the interior of black
hole. The General Relativity Theory belongs to a classical field
theory with an inherent property of singularities at the center of
black holes and has the inconsistency of the theory with quantum
field gravity for singularity avoidance [1-5]. For the sake of
solving the problem on the breakdown of General Relativity and
well explaining some observations, modified or alternative
theories of gravity [6-8] are necessarily proposed.

Bardeen [9] first introduced a regular black hole that  does not
possess any curvature singularities but has horizons. The Bardeen
black hole is not an exact solution of the Einstein field
equations coupled to some known physical sources in the usual
sense. The regular  black hole solution of Ay\'{o}n-Beato and
Garc\'{i}a [10] is an exact solution of the Einstein equations
coupled to a nonlinear electrodynamic field as a physically
reasonable source. The obtainment of such a regular black hole
solution is based on the scalar-tensor-vector modified gravity
theory [11]. The black hole mass and the black hole electric
charge are two parameters in the solution. It is worth noting that
a regular center exists in the Ay\'{o}n-Beato--Garc\'{i}a solution
with the electric charge and without magnetic charge. This
property of the solution evidently seems to contradict the result
of Ref. [12] on the regular center requiring a zero electric
charge. In fact, the two results  have no conflict. Such a
globally regular metric in Ref. [12] is based on nonlinear
electrodynamics (NED) with a nonzero magnetic charge and a
particular Lagrangian $\mathcal{L}(F)$ tending to a finite limit
as $F\rightarrow\infty$, where $F=F_{\mu\nu}F^{\mu\nu}$ and
$F_{\mu\nu}$ denotes an electromagnetic field tensor. However, the
Ay\'{o}n-Beato--Garc\'{i}a solution is not a solution of the
Lagrangian $\mathcal{L}(F)$ and needs different Lagrangians in
different ranges of the radial coordinate. It uses an alternative
form of NED (called the $P$ framework) through a Legendre
transformation from the original $F$ framework.  The
Ay\'{o}n-Beato--Garc\'{i}a spacetime satisfies the weak energy
condition, and  asymptotically behaves like the RN black hole.
Another regular black hole solution satisfying these properties
was presented by Balart and Vagenas [13]. There are also other
regular black hole solutions [14-16]. For example, the solutions
with the asymptotical behavior of the RN solution and the weak
energy condition dissatisfied were given in Refs. [12,17,18]. The
Bardeen regular black hole [9] was also regarded to satisfy the
weak energy condition, but does not asymptotically behave like the
RN solution when it was considered as gravity coupled to a theory
of NED for a self-gravitating magnetic monopole in Ref. [19]. The
regular black hole solution of Hayward [20] is also similar to the
Bardeen solution. A family of regular black hole metrics for
satisfying the weak energy condition can be found in Refs.
[21-23]. Several charged regular black hole metrics were
constructed in terms of mass continuous probability distribution
functions [13]. A noncharged regular black hole solution
satisfying the weak energy condition was given by Dymnikova in
Ref. [24]. Regular charged rotating black holes in NED
non-minimally coupled to gravity were presented by Dymnikova in
Ref. [25]. The dynamics of  circular orbits of charged particles
around the regular black hole or other modified gravity black
holes in the presence of magnetic fields was investigated in the
literature (see e.g. [26-29]). Of course, the dynamics of charged
particles around standard general-relativity black holes in the
presence of external magnetic fields has naturally been focused on
some references. For example, the authors of [30] discussed the
escape velocity of charged particles in the innermost stable
circular orbits near a slowly rotating Kerr black hole immersed in
an external  magnetic field.

As is mentioned above, the charged regular black hole of Balart
and Vagenas [13] was obtained by its metric function represented
in terms of the Fermi--Dirac-type distribution function. It is
similar to the RN black hole that satisfies the weak energy
condition and has asymptotic behavior. A typical difference
between the two black holes is that the regular solution is not
singular in its curvature invariants at the origin but the RN
solution is. Now, a question is whether the dynamical behavior of
physical objects moving in the outer region of the regular black
hole is equivalent to that of physical objects outside the RN
black hole. In order to answer this question, we compare effective
potentials and circular orbits of photons at the equatorial plane
in the two black hole solutions. Photon spheres corresponding to
the two black hole shadows and constraints of the two black hole
charges based on the latest observation of the image of
Sagittarius A$^{*}$ [31] are also compared. On the other hand, the
effective potentials, stable circular orbits and innermost stable
circular orbits of charged particles at the equatorial plane
moving around the two black holes with the inclusion of external
magnetic fields are considered. Explicit symplectic integration
methods are directly available  for the off-equatorial motions of
charged particles in the RN black hole system [32]. However, they
are not for the off-equatorial motions of charged particles in the
regular black hole system but for an appropriate time-transformed
Hamiltonian [33] corresponding to the magnetized regular black
hole. The proposed explicit symplectic integrators are used to
explore the effects of dynamical parameters on the regular and
chaotic dynamics of charged particles in the two black hole
gravitational problems. In a word, an extensive comparison in the
dynamical behavior of physical objects moving outside the two
black holes is the main aim of the present paper.

The paper is organized as follows. In Section 2, we introduce  a
charged regular black hole [13], which is immersed in an external
magnetic field. In Section 3, we focus on effective potentials and
circular orbits of photons or charged particles at the equatorial
plane in the two black hole solutions. The two black hole shadow
sizes are discussed and the two black hole charges are constrained
by using the latest observation of the image of Sagittarius
A$^{*}$ [31]. In Section 4, we design several explicit symplectic
integrators for a time-transformed Hamiltonian of the magnetized
regular black hole. The effects of dynamical parameters on the
regular and chaotic dynamics of charged particles in the two black
hole systems are compared in Section 5. Finally, the main results
are summarized in Section 6.

\section{Regular charged black hole immersed in an external magnetic
field }

The RN spacetime has curvature singularity and multiple horizons.
It can be altered as a class of regular charged black  holes,
which still satisfy Einstein's field equations and have horizons
but avoid the curvature singularities. For example, a static
spherically symmetric regular charged black hole metric is written
in Ref. [13] (see also Ref. [26])  as
\begin{eqnarray}
ds^2=-f(r)dt^2+\frac{dr^2}{f(r)}+r^2(d\theta^2+\sin^2\theta
d\varphi^2),
\end{eqnarray}
where $f(r)$ is  a metric function of radial distance $r$ in the
following expression
\begin{eqnarray}
   f(r) = 1-\frac{2M}{r}\frac{2}{\exp(Q^2/Mr)+1}.
\end{eqnarray}
$M$ is the mass of the black hole, and $Q$ stands for the charge
of the black hole. In fact, the speed of light $c$ and the
gravitational constant $G$ are taken as geometric units: $c = G =
1$. For simplicity, dimensionless operations are implemented by
scale transformations: $r\rightarrow Mr$ and $Q\rightarrow MQ$.
Proper time $\tau$ and coordinate $t$ also have similar
transformations: $\tau\rightarrow M\tau$ and $t\rightarrow Mt$. In
this case, $M$ in Eq. (2) is replaced with 1.

If $r$ is large enough (i.e., $Q^2/r\rightarrow 0$), then
$\exp(Q^2/r)\approx 1+Q^2/r$. The Fermi-Dirac-type distribution
function $1/[\exp(Q^2/r)+1]\approx1/(2+Q^2/r)\approx
[1-Q^2/(2r)]/2$. Based on the Taylor expansion to the second
order, Eq. (2) is rewritten as
\begin{eqnarray}
    f(r)\approx 1-\frac{2}{r}+\frac{Q^{2}}{r^2}+\frac{Q^{4}}{4r^3}=f^{\star}(r)+\frac{Q^{4}}{4r^3}.
 \end{eqnarray}
Metric function $f^{\star}(r)$ just corresponds to the RN black
hole. Namely, Eq. (1) is  the RN metric when $ f(r)$ is replaced
with $f^{\star}(r)$ for the case of $r\gg 1$. If $f^{\star}(r)=0$,
$r_{\pm}=1\pm\sqrt{1-Q^{2}}$ with $|Q|\leq 1$ are the horizon(s)
of the RN black hole.  For $Q=0$, $r=2$ satisfies $f^{\star}(r)=0$
and $f(r)=0$. This means that $r=2$ is a horizon of the
Schwarzschild black hole. When $Q\neq0$ and $r>1$, the difference
between the regular black hole (1) and the RN black hole, the term
$Q^{4}/(4r^3)$, is relatively small. When $Q\neq0$ and $0<r<1$,
the difference is very large. The black hole spacetime (1) has the
horizons $r_{\pm}=4/[\exp(Q^2/r_{\pm})+1]<2$ for $Q\neq0$. $r=0$
is a curvature singularity of the Schwarzschild black hole, but is
not that of the regular black hole (1). In fact, there are no
curvature singularities in the regular black hole (1) for all
charges satisfying the range $0<|Q|\leq 1$. This result is easily
checked as follows. The Ricci scalar is
$R=2(2\sigma'(r)+r\sigma''(r))/(r^2\sigma_{\infty})$, and the
Ricci squared is
$R_{\mu\nu}R^{\mu\nu}=2(4\sigma'(r)^2+r^2\sigma''(r))^2/(r^4\sigma^2_{\infty})$
[13], where the function $\sigma(r)$ is determined by
$f(r)=1-2\sigma(r)/(r\sigma_{\infty})$ and
$\sigma_{\infty}=\lim_{r\rightarrow\infty}\sigma(r)$.
$\sigma(r)=1-Q^2/(2r)$ and $\sigma_{\infty}=1$ for the RN metric;
$\sigma(r)=1/(e^{Q^2/r}+1)$ and $\sigma_{\infty}=1/2$ for the
regular  black hole metric. As $r\rightarrow 0$ in the RN metric,
the Ricci scalar is $R=\frac{2}{r^2}\times 0=\infty\times
0$=\emph{a uncertain number}, and the Ricci squared is
$R_{\mu\nu}R^{\mu\nu}=\frac{4Q^4}{r^8}\rightarrow\infty$. When
$r\rightarrow 0$ in the regular  black hole metric, the Ricci
scalar is
$R=-\frac{8Q^4}{r^5}[(e^{Q^2/r}+1)^{-2}-(e^{Q^2/r}+1)^{-3}]=0$,
and the Ricci squared is
$R_{\mu\nu}R^{\mu\nu}=\frac{8Q^4}{r^8}e^{Q^2/r}(e^{Q^2/r}+1)^{-4}
\{4+[2+\frac{Q^2}{r}-2Q^2e^{Q^2/r}(e^{Q^2/r}+1)^{-1}/r ]^2\}=0$.
Thus, the Ricci scalar and its squared are regular everywhere
except the origin $r=0$ for the RN metric, and are also regular
everywhere including the origin $r=0$ for the regular black hole.
Notice that the metric (1) with Eq. (2) is one of the solutions of
Ay\'{o}n-Beato and Garc\'{i}a [17-19]. The existence of a regular
center in this metric allows the black hole having the electric
charge in the $P$ framework, as is mentioned in the introduction.
Such a regular electric solution corresponds to different
Lagrangians  in different parts of space. Although the regular
solution of the field equations does not allow the electric charge
for a given Lagrangian $\mathcal{L}(F)$ in the $F$ framework, it
allows a magnetic charge [12]. The metric (1) with Eq. (2) is
given by a Legendre transformation from the original $F$ framework
with the  magnetic charge to the $P$ framework with the electric
charge. Here, the Legendre transformation relates to a duality
between spherically symmetric solutions in the $F$ and $P$
frameworks, which connects solutions of the two different
theories. For the Lagrangian $\mathcal{L}(F)$, a Hamiltonian-like
quantity as a function of $P$ is
$H(P)=2F\mathcal{L}_F-\mathcal{L}(F)$, where
$\mathcal{L}_F=d\mathcal{L}(F)/dF$ and the tensor
$P_{\mu\nu}=\mathcal{L}_FF_{\mu\nu}$ with its invariant
$P=P_{\mu\nu}P^{\mu\nu}$. This implies that any regular magnetic
solution obtained from $\mathcal{L}(F)$ can always correspond to a
purely electric counterpart with a similar dependence $H(P)$. The
sufficient and necessary condition for the existence of a regular
center is that $H(P)$ tends to a finite limit as
$P\rightarrow\infty$. Three choices of $H(P)$ were listed in Eqs.
(20)-(22) of Ref. [12]. See Section 4 of Ref. [12] for more
details on the $FP$ duality and electric solutions. In addition,
the regular black hole (1) satisfies the weak energy condition
everywhere and asymptotically behaves as the RN black hole.

The regular black hole is immersed in an external electromagnetic
field with 4-vector potential having two nonvanishing components
[34]:
\begin{eqnarray}
   A_t &=&-\frac{Q}{r}, \\
   A_\varphi  &=&\frac{1}{2} Br^2\sin^2 \theta,
\end{eqnarray}
where $B$ represents a constant strength of the magnetic field.
The electromagnetic 4-potential is an exact solution of Maxwell's
equations in the background metric (1).  The electromagnetic field
is asymptotically uniform and does not affect the gravitational
field.

The motion of a particle with charge $q$ and mass $m$ around the
regular charged black hole immersed in the magnetic field can be
described by the super-Hamiltonian
\begin{eqnarray}
    H &=& \frac{1}{2m}g^{\mu \nu }(p_\mu-qA_\mu)(p_\nu-qA_\nu) \nonumber \\
    &=&-\frac{1}{2mf(r)}(E-\frac{Q^*}{r})^2+\frac{f(r)}{2m}p_r^2+\frac{p_\theta
    ^2}{2mr^2} \nonumber \\
&&   +\frac{1}{2mr^2\sin^2\theta }(L-\frac{\beta}{2
    }r^2\sin^2\theta)^2,
\end{eqnarray}
where $\beta =Bq$ and $Q^*=qQ$. $E$ is a constant of motion,
called as the particle's energy $E=-p_t$. $L$ is another constant
of motion, called as the particle's angular momentum
$L=p_{\varphi}$. In practice, the constant angular momentum is
because the system (6) is axially symmetric with respect to the
$z$-axis\footnote{The inclusion of asymptotically uniform external
electromagnetic field makes the system (6) be axially symmetric,
whereas it still causes the spacetime (1) to be spherically
symmetric because such an electromagnetic field does not exert any
influence on the spacetime geometry.} or does not explicitly
depend on the angle $\varphi$. The two constants are expressed as
\begin{eqnarray}
   E  &=& -mg_{tt}\dot{t}-qA_t= mf(r)\dot{t}+\frac{Q^*}{r}, \\
   L &=& mg_{\varphi\varphi}\dot{\varphi}+qA_\varphi \nonumber \\
   &=& mr^2\sin^2\theta \dot{\varphi } +\frac{1}{2}\beta r^2\sin^2
   \theta.
\end{eqnarray}
Dimensionless treatments are $B \to B/M$, $E \to mE$, $L \to mML$,
$p_r \to mp_r$, $p_\theta \rightarrow mMp_\theta$, $q \to mq$, $H
\to mH$. In this way, $m=1$ is taken in Eqs. (6)-(8).

Besides the two constants of motion (7) and (8), a third constant
of motion exists in the Hamiltonian system (6). The constant is
the conserved Hamiltonian because the system (6) does not
explicitly depend on the proper time $\tau$. Based on the
normalization condition of the particle's 4-velocities or the
particle's rest mass, the conserved Hamiltonian is always
identical to -1/2, i.e.,
\begin{eqnarray}
    H=-\frac{1}{2}.
\end{eqnarray}
The Hamilton-Jacobi equation of the Hamiltonian (6) does not allow
for the separation of variables. This means that the Hamiltonian
(6) has no fourth constant of motion. Thus, the motions of charged
particles around the regular black hole are nonintegrable.

When the regular black hole gives place to the RN black hole, the
three constants given in Eqs. (7)-(9) still exist in the system
(6) with $f(r)\rightarrow f^{\star}(r)$. The energy of particles
near the RN black hole also uses $f^{\star}(r)$ to substitute for
$f(r)$ in Eq. (7). The external magnetic field of Eqs. (4) and (5)
also causes the nonintegrable dynamics of charged particles moving
around the RN black hole.

In what follows, we make several comparisons between the two black
hole gravitational systems. These comparisons involve effective
potentials, circular orbits, constructions of explicit symplectic
integrators, and dynamical properties of ordered and chaotic
orbits.

\section{Effective potentials and circular orbits}

At first, we compare effective potentials and circular orbits of
photons moving outside the event horizons of the regular and RN
two black holes with the same parameters. In particular, the
ranges of the charges of the two black holes are constrained in
terms of the latest observations of Sagittarius A$^{\star}$ black
hole shadows [31]. Then, we check possible differences between the
two black holes in effective potentials and circular orbits of
charged particles.

\subsection{Circular orbits and spherical orbits of photons}

For a photon moving in the vicinity of the regular black hole at
the equatorial plane $\theta=\pi/2$, $A_t=A_\varphi=0$ correspond
to $\beta=Q^*=0$ in Eq. (6). Eq. (9) becomes $H=0$, and $\tau$ is
not the proper time but is an affine parameter. Taking
$p_r=p_\theta=0$, we use  Eq. (6) to obtain an effective potential
of  the photon  in the regular black hole problem
\begin{eqnarray}
   U_{eff}=E=\frac{L}{r}\sqrt{f(r)}.
\end{eqnarray}
When $f(r)\rightarrow f^{\star}(r)$, Eq. (10) is an effective
potential of  the photon  in the RN black hole system.

Figs. 1 (a) and (b) plot the photon effective potentials for the
two black holes  with the angular momentum $L=4.6$ and three
different values of the charge $|Q|$. The shape of the  photon
effective potential goes toward the regular black hole in Fig.
1(a) as the black hole charge increases. This means that such an
increase of the charge leads to increasing the potential energy.
The result on the energy $E$ increasing with $|Q|$ can be shown
directly through Eq. (7) or Eq. (10). There is the same rule for
the  photon effective potential varying with the RN black hole
charge in Fig. 1(b). In fact, the two photon effective potentials
have no typical differences in the two black hole problems  for a
given charge.

Each of the photon effective potentials has a maximum value, which
corresponds to a unstable  photon circular orbit satisfying the
condition
\begin{eqnarray}
   \frac{dU_{eff}}{dr}=0.
\end{eqnarray}
The charges $|Q|=$0.1, 0.3 and 0.5 correspond to the radii of the
unstable  photon circular orbits: $r_p=$ 2.82306070, 2.93875679,
2.99331846 in sequence for the regular black hole in Fig. 1(a),
and $r_p=$ 2.82287566, 2.93874945, 2.99331846 for the RN black
hole in Fig. 1(b). This indicates that the two black holes have
almost the same photon circular orbits when the parameters $L$ and
$|Q|$ are given.

Each photon circular orbit at the equatorial plane is associated
with a photon sphere in the three-dimensional space. The radius of
the photon sphere for the regular black hole is determined by an
impact parameter
\begin{eqnarray}
   b(r,Q)=\frac{L}{E}=\frac{r}{\sqrt{f(r)}},
 \end{eqnarray}
where $E$ is the energy $E_p$ of a photon circular orbit or $r$
takes the radius $r_p$ of  a photon circular orbit. In this case,
$b$ is a critical impact parameter labeled as $b_p$. If $Q$ is
given, then $b$ varies with $L$. When $b<b_p$, the photon is
absorbed and falls into the center of the black hole. When
$b>b_p$, the photon is deflected and scattered to the infinity.
Therefore, the photon sphere radius $b_p$ just corresponds to the
radius of black hole shadow. The regular black hole charges $|Q|=$
0.1, 0.3 and 0.5 respectively correspond to the shadow radii
$b_p=$ 4.96807193, 5.11680106, 5.18747526 in Fig. 1 (a). The RN
black hole charges $|Q|=$ 0.1, 0.3 and 0.5 respectively correspond
to the shadow radii $b_p=$ 4.96791433, 5.11679475, 5.18747526 in
Fig. 1 (b). The shadow radii for the two black holes have only
minor differences.

We consider the Event Horizon Telescope (EHT) data of Sagittarius
A$^{*}$ black hole [31] for reference. Sagittarius A$^{*}$ black
hole has mass $M=4\times 10^{6}$ $M_{\odot}$ and distance  $D=8$
kpc, where $M_{\odot}$ is the mass of the Sun. The angular
diameter of observing the black hole shadow is $\Omega=51.8\pm2.3$
$\mu$as. The observing black hole shadow radius is calculated by
\begin{eqnarray}
   b_o=\frac{1}{2}D\Omega.
 \end{eqnarray}
Suppose both black holes are Sagittarius A$^{*}$ black hole.
Letting $b_o=b_p$, we can estimate the regular black hole charge
constrained in the range $0\leq |Q|\leq 0.47442268$, as shown in
Fig. 1(c). The constrained range $0\leq |Q|\leq 0.47442228$ is
given to the RN black hole charge.

The main result concluded from the above demonstrations is given
here. When the two black holes have same parameters, their
corresponding effective potentials, photon circular orbits and
photon spheres are almost the same.

\subsection{Stable circular orbits of charged particles}

For a charged particle moving at the equatorial plane, Eq. (6)
with Eq. (9) is considered and $\tau$ is the proper time. The
radial effective potential for the regular black hole system reads
\begin{eqnarray}
    U_{eff}=E=\sqrt{f(r)[1+\frac{1}{r^2}(L-\frac{\beta}{2} r^2)^2]}+\frac{Q^*}{r}.
\end{eqnarray}
When $f(r)\rightarrow f^{\star}(r)$, Eq. (14) is the radial
effective potential of a charged particle moving in the vicinity
of the regular black hole.

Taking the parameters  $L=4.6$, $Q^*=10^{-4}$ and $\beta=10^{-3}$,
we draw the  effective potentials of charged particles in the two
black hole systems with different charges $Q=$ 0.1, 0.3 and 0.5 in
Fig. 2. For a given charge $Q$, the effective potentials in the
two cases have no explicit differences. For a given separation
$r$, the potential increases with the black hole charge
increasing.

The local extremums of the effective potentials correspond to
particle circular orbits. The local minimum indicates the presence
of a stable particle circular orbit, which satisfies the
conditions
 \begin{eqnarray}
       \frac{d U_{eff}}{d r} &=& 0, \\
  \frac{d^2U_{eff}}{d r^2}&\geq& 0.
     \end{eqnarray}
Eq. (16) takes the equality
\begin{eqnarray}
       \frac{d^2U_{eff}}{d r^2}= 0,
     \end{eqnarray}
which shows the  innermost stable circular orbit (ISCO). The radii
of stable circular orbits are $r_S=$17.9686, 17.7441, 17.6308, and
those of ISCOs are $r_I=$5.60654, 5.86239, 5.98454 for the regular
black hole charges $Q=$0.5, 0.3, 0.1 in Fig. 2(a). The radii of
stable circular orbits are $r_S=$17.9686, 17.7441, 17.6308, and
those of ISCOs are $r_I=$5.60630, 5.86238, 5.98454 for the RN
black hole charges $Q=$0.5, 0.3, 0.1 in Fig. 2(b). These facts
show the radii of stable circular orbits and ISCOs for the regular
black hole are approximately consistent with those for the RN
black hole when the two black hole systems have the same
parameters.

In brief, the regular and RN two black holes with the same
parameters are almost the same in the effective potentials and
circular orbits of photons, the sizes of black hole shadows, and
the stable circular orbits and the  innermost stable circular
orbits of charged particles.

If charged particles move off the equatorial plane, then the
system (6) as well as the magnetized RN black hole system is
nonintegrable. Numerical integration methods are convenient to
solve this nonintegrable problem.

\section{Explicit symplectic integrators}

The magnetized RN black hole system given in Eq. (6) with
$f(r)\rightarrow f^{\star}(r)$ allows for 5 explicitly integrable
splitting pieces constructing explicit symplectic integrators
[32]. However, the Hamiltonian (6) for the magnetized regular
black hole system does not allow for such a splitting form.
Introducing a time-transformed Hamiltonian to the Kerr spacetime,
Wu et al. [33] split the time-transformed Hamiltonian into several
explicitly integrable parts and proposed several explicit
symplectic integrators. Following this idea, we search for a
time-transformed Hamiltonian of Eq. (6).

\subsection{Splitting and composition methods}

The proper time $\tau$ is regarded as a new coordinate $q_0=\tau$,
and its corresponding momentum is  $p_0=-H=1/2$. In the extended
phase space made of $(r,\theta,q_0,p_r,p_{\theta},p_0)$, the
Hamiltonian (6) becomes
\begin{eqnarray}
   \mathbb{H} =H+p_0.
\end{eqnarray}
This Hamiltonian is identical to zero, $\mathbb{H} =0$. Setting a
time transformation
\begin{eqnarray}
    d\tau =g(r)dw,
\end{eqnarray}
where $g(r)$ is a time transformation function
\begin{eqnarray}
    g(r)=\frac{1}{f(r)},
\end{eqnarray}
we obtain  a time transformation Hamiltonian with respect to the
new time $w$ as follows:
\begin{eqnarray}
    \mathcal{H} &=& g\mathbb{H} \nonumber \\
    &=&-\frac{1}{2f^2(r)}(E-\frac{Q^*}{r})^2  \nonumber \\ && +\frac{1}{2r^2f(r)\sin^2\theta}
    (L-\frac{1}{2}\beta r^2 \sin^2\theta ) ^2 \nonumber \\ && +\frac{1}{2}p_r^2+\frac{1}{2r^2f(r)}p_\theta
    ^2+\frac{p_0}{f(r)}.
\end{eqnarray}

The time-transformed Hamiltonian is divided into three parts
\begin{eqnarray}
    \mathcal{H} =\mathcal{H}_1 +\mathcal{H}_2 +\mathcal{H}_3,
\end{eqnarray}
where the sub-Hamiltonian parts are expressed as
\begin{eqnarray} \nonumber
    \mathcal{H}_1&=&-\frac{1}{2f^2(r)}(E-\frac{Q^*}{r})^2
    +\frac{p_0}{f(r)} \nonumber \\
   && +\frac{(L-\frac{\beta}{2} r^2\sin^2\theta)^2}{2r^2f(r)\sin^2\theta},\\
       \mathcal{H}_2&=&\frac{1}{2}p_r^2,\\
    \mathcal{H}_3&=&\frac{1}{2r^2f(r)}p_\theta ^2.
\end{eqnarray}
Obviously, each of the three parts has an analytical solution as
an explicit function of the new time $w$. Operators for
analytically solving the sub-Hamiltonians  $\mathcal{H}_1$,
$\mathcal{H}_2$ and $\mathcal{H}_3$ are labeled as $\xi_1$,
$\xi_2$ and $\xi_3$, respectively.

Setting $h$ as a time step, we have a solution of the
time-transformed Hamiltonian (21) or (22), which is approximately
provided by a second-order explicit symplectic integrator
\begin{eqnarray}
S_2 (h)= \chi^*(\frac{h}{2})\times \chi(\frac{h}{2}),
\end{eqnarray}
where two first-order operators are
\begin{eqnarray}
    \chi(h) &=& \xi_3(h)\times \xi_2(h)\times \xi_1(h), \\
    \chi^*(h) &=& \xi_1(h)\times \xi_2(h)\times \xi_3(h).
\end{eqnarray}
The second-order method can be raised to a fourth-order explicit
symplectic algorithm of Yoshida [35]
\begin{eqnarray}
    S_4=S_2(\gamma h)\times S_2(\delta  h)\times S_2(\gamma h),
\end{eqnarray}
where $\gamma =1/(1-\sqrt[3]{2}) $ and $\delta =1-2\gamma$. More
first-order operators  $\chi$ and $\chi^*$ can be used to
symmetrically compose optimized higher-order  partitioned
Runge-Kutta (PRK) and Runge-Kutta-Nystr\"{o}m (RKN) symplectic
algorithms [36]. For instance, an optimized fourth-order PRK
algorithm in Ref. [37] is
\begin{eqnarray}
PRK_64 &=& \chi^* (\alpha _{12} h)\times \chi(\alpha _{11}
h)\times \cdots \nonumber \\ && \times \chi^* (\alpha _2 h) \times
\chi(\alpha _1 h),
\end{eqnarray}
where time coefficients are
\begin{eqnarray}
    \nonumber
    &&\alpha _1=\alpha _{12}= 0.0792036964311597,   \\ \nonumber
    &&\alpha _2=\alpha _{11}= 0.1303114101821663,   \\ \nonumber
    &&\alpha _3=\alpha _{10}= 0.2228614958676077,    \\ \nonumber
    &&\alpha _4=\alpha _9=-0.3667132690474257,    \\ \nonumber
    &&\alpha _5=\alpha _8= 0.3246484886897602,   \\ \nonumber
    &&\alpha _6=\alpha _7= 0.1096884778767498.   \nonumber
\end{eqnarray}
An optimized fourth-order RKN algorithm in [37] reads
\begin{eqnarray}
    RKN_64 &=&\chi^* (\eta _{12} h)\times \chi(\eta _{11} h)\times
    \cdots \nonumber \\ && \times\chi^*(\eta _2 h)\times\chi(\eta _1 h),
\end{eqnarray}
where time coefficients are
\begin{eqnarray}
    \nonumber
    &&\eta _1=\eta _{12}= 0.082984402775764,   \\\nonumber
    &&\eta _2=\eta _{11}= 0.162314549088478,   \\\nonumber
    &&\eta _3=\eta _{10}= 0.233995243906975,    \\\nonumber
    &&\eta _4=\eta _9= 0.370877400040627,    \\\nonumber
    &&\eta _5=\eta _8= -0.409933704882860,   \\\nonumber
   && \eta _6=\eta _7= 0.059762109071016.  \\\nonumber
\end{eqnarray}

A notable point is that the time transformation (19) has two
roles. Such a time transformation plays an important role in
splitting the time-transformed Hamiltonian fit for explicit
symplectic integrators. It also makes these symplectic integrators
be adaptive time-steps when the particle moves in the vicinity of
the black hole's horizons. However, there is no dramatic
difference between the proper time $\tau$ and the new time $w$
when the particle runs away from the horizons. In this case, the
adaptive proper time step control becomes useless.

Another notable point is that no time transformation is necessary
to construct the explicit symplectic integrators for the
magnetized RN black hole, as is mentioned above. The Hamiltonian
(6) with $f(r)\rightarrow f^{\star}(r)$ is split into five
explicit integrable parts, whose analytical solutions explicitly
depend on the proper time  $\tau$ in Ref. [32]. The algorithms
$S_4$, $PRK_64$ and $RKN_64$ have no differences but only the two
first-order operators  $\chi$ and $\chi^*$ are slightly modified.
In addition, these methods work in the proper time $\tau$ for the
RN case.

\subsection{Numerical Evaluations}

The time step is $h=1$. The parameters are given by  $E=0.995$,
$L=4.6$, $Q=0.1$, $Q^*=0.06$ and $\beta=8\times 10^{-4}$. An orbit
has the initial conditions $\theta =\pi/2$, $p_{r}=0 $, $r=15$ and
$p_\theta>0$ obtained from Eq. (9). The Hamiltonian errors $\Delta
H=H+1/2$ for the three fourth-order algorithms in Fig. 3(a) have
no secular drifts. This accords with  one of the properties of
symplectic methods. In addition, the errors for $PRK_64$ are
slightly smaller than those for $RKN_64$. Both algorithms have two
orders of magnitude smaller errors than $S_4$. In other words,
$PRK_64$ performs the best accuracy, while $S_4$ shows the poorest
accuracy. Because of this, $PRK_64$ is employed  in the later
computations.

The relation between the proper time $\tau$ and the new time $w$
is described in Fig. 3(b) and Table 1. It is clear that $w$ is
approximately equal to $\tau$. The main motivation  of the  time
transformation is a successful  splitting of the  time
transformation Hamiltonian, as  is aforementioned. This result is
because $f(r)\approx 1$ when the particle runs away from the black
hole's horizons; in fact, the radial distances $r$ are always
larger than 15 in Fig. 3(c).

The numerical performance of the explicit symplectic integrator
$PRK_64$ for the regular black hole are similar to that for the RN
black hole (not plotted). In what follows, we provide some
insights into the orbital dynamics of charged particles.

\section{Regular and chaotic dynamics of charged particle orbits}

Note that the integration times are $w$ for the regular black hole
and $\tau$ for the RN black hole. For comparison, the obtained
numerical results in the new time $w$ should be changed into those
in the proper time $\tau$  for the regular black hole. The
velocity $\dot{r}_{\tau}$ for the regular black hole is calculated
by
 \begin{eqnarray}
     \dot{r}_{\tau}=\frac{d r}{d \tau }=\frac{d r}{d w} \frac{d w}{d \tau
     }=\frac{dr}{dw}g(r).
 \end{eqnarray}
Although $PRK_64$ uses a fixed time step $h=1$ in the new time
$w$, it adopts slightly variable proper time steps, as shown in
Table 1. Using the linearly inserting value method, we easily
obtain  a series of solutions
$(r_{\tau},\theta_{\tau},\dot{r}_{\tau})$ of the system (6) at
proper times $\tau=1,2,\cdots$. That is,
\begin{eqnarray}
r_{i} &=& r_{\tau_{i-1}}+\frac{r_{\tau_{i}}-r_{\tau_{i-1}}}{\tau_{i}-\tau_{i-1}}(i-\tau_{i-1}),\\
\dot{r}_{i} &=& \dot{r}_{\tau_{i-1}}+\frac{\dot{r}_{\tau_{i}}-\dot{r}_{\tau_{i-1}}}{\tau_{i}-\tau_{i-1}}(i-\tau_{i-1}),\\
\theta_{i} &=&
\theta_{\tau_{i-1}}+\frac{\theta_{\tau_{i}}-\theta_{\tau_{i-1}}}{\tau_{i}-\tau_{i-1}}(i-\tau_{i-1}),
 \end{eqnarray}
where $i=1,2,\cdots$, $\tau_{0}=0$, and $\tau_{i}$ corresponds to
$\tau$ in Table 1 (e.g., $\tau_{1}=0.96$, $\tau_{2}=1.95$). In
this way, the point $(R_{\tau},\dot{R}_{\tau})$ on the
Poincar\'{e} section $\theta=\pi/2$ with $d\theta/dw>0$ is
determined and is labeled as $(r_{\tau},\dot{r}_{\tau})$. Note
that $r_{\tau}$ on the section is $R_{\tau}$ but not $r_{i}$ in
Eq. (33), and $\dot{r}_{\tau}$ on the section is $\dot{R}_{\tau}$
but not $\dot{r}_{i}$ in Eq. (34).

Figs. 4 (a)-(c) plot Poincar\'{e} sections for  three different
choices of the magnetic field parameter $\beta$ in the regular
black hole system. The other parameters are $E=0.9975$, $L=4.6 $,
$ Q=0.3$ and $Q^*=0.001$. The orbits have the initial conditions
(except the initial separations $r$ and the initial angular
momenta $p_{\theta}$) same as those in Fig. 3. The orbits are
regular tori for the magnetic field parameter
$\beta=2\times10^{-4}$ in Fig. 4(a). Several orbits are chaotic
for $\beta=3\times10^{-4}$ in Fig. 4(b). Only one of the orbits is
nonchaotic for $\beta=4\times10^{-4}$ in Fig. 4(c). This shows
that a transition from order to chaos occurs and chaos gets
stronger as the magnetic field strength increases. When $f(r)$ for
the regular black hole is replaced by $f^{\star}(r)$ for the RN
black hole, the phase space structures are shown in Figs. 4
(d)-(f). The dynamical properties in the two black hole systems
are almost the same under the same parameters.

In addition to the method of Poincar\'{e} sections, the technique
of Lyapunov exponents is often used to distinguish between the
regular and chaotic dynamical properties. A Lyapunov exponent is
an indicator measuring an average exponential deviation between
two nearby orbits. If a bounded orbit has a positive Lyapunov
exponent, it is chaotic; it is regular if its largest Lyapunov
exponent tends to zero with a long enough integration time. The
variational method and the two-particle one are two methods
calculating the largest Lyapunov exponent [38]. The two-particle
method has a convenience of application. Considering this point,
the authors of Ref. [39] suggested using the proper time $\tau$
and a proper distance $d(\tau)$ between two nearby orbits to
define the largest Lyapunov exponent as follows:
\begin{eqnarray}
\lambda=\lim_{\tau\rightarrow\infty}\frac{1}{\tau}\ln\frac{d(\tau)}{d(0)},
 \end{eqnarray}
where $d(0)$ represents the initial separation between two nearby
orbits. Such a definition is independent of a choice of spacetime
coordinates. Fig. 5(a) displays that the Lyapunov exponents of
orbits with initial radius $r=30$ for $\beta=2\times10^{-4}$ and
$\beta=3\times10^{-4}$ in the regular black hole system tend to
zeros and show the regularity of the orbits. The Lyapunov exponent
for $\beta=4\times10^{-4}$ tending to the stable value
$10^{-0.375}$ indicates the onset of chaos. That is to say, the
orbital dynamical behaviors described by the Lyapunov exponents in
Fig. 5(a) are consistent with those given by the Poincar\'{e}
sections in Figs. 4 (a)-(c). The Lyapunov exponents in Fig. 5(c)
also show the dynamical properties of orbits in the RN system, as
the Poincar\'{e} sections in Figs. 4 (d)-(f) do.

Notice that a longer time is necessary for us to obtain  the
desired stable value of the Lyapunov exponent. Compared with the
Lyapunov exponent, a fast Lyapunov indicator (FLI) is a more
sensitive tool to detect chaos from order. The FLI uses completely
different rates on the lengths of deviation vectors increasing
with time as the distinguishment of the ordered and chaotic two
cases. An algebraical increase of the lengths with time
corresponds to the features of ordered orbits, while an
exponential increase of the lengths with time corresponds to the
features of chaotic bounded orbits. Based on tangential vectors,
several FLIs are defined in Ref. [40]. In terms of the
two-particle method, the FLI is modified in Ref. [39] as a
spacetime coordinate independent definition
\begin{eqnarray}
\textrm{FLI}=\log_{10}\frac{d(\tau)}{d(0)}.
 \end{eqnarray}
The FLI for the regular black hole with $\beta=4\times10^{-4}$ in
Fig. 5(b) describes the existence of chaos of a charged particle
in the magnetized regular black hole system because it grows
exponentially with time $\log_{10} \tau$. The FLIs for
$\beta=2\times10^{-4}$ and $\beta=3\times10^{-4}$ correspond to
the orbital chaoticity because they grow slowly with time. The
FLIs for the RN black hole in Fig. 5(d) are consistent with those
for the regular black hole in Fig. 5(b) when the considered
parameters and orbits are the same.

When the proper time $\tau=10^{6}$, the threshold of FLIs between
the ordered and chaotic cases is FLI$=5$. FLIs less than the
threshold correspond to order, but those more than the threshold
correspond to chaos. Following this idea, we employ FLIs to trace
the dependence of the transition from order to chaos on a
variation of the magnetic parameter $\beta$. The two black hole
gravitational systems have consistent results that chaos occurs
when $\beta>3.7\times10^{-4}$ in Fig. 6.

The dependence of the dynamical behavior on the regular black hole
charge $Q$ in Figs.  7 (a)-(c) and 8(a) is in agreement with that
on the RN charge in Figs.  7 (d)-(f) and 8(b). A variation of the
dynamical properties is not relatively sensitive dependence on an
appropriate increase of the black hole charge from a global phase
space structure. The strength of chaos seems to be slightly
weakened with the black hole charge increasing. The two black hole
gravitational systems also have the basically same effects of the
charge parameters $Q^{*}$ on the orbital dynamics, as shown in
Figs. 9-11. Chaos seems to get weaker as a positive value of
$Q^{*}$ increases, whereas it seems to become stronger with the
magnitude of negative charge parameter $Q^{*}$ increasing.

An explanation to the effect of each parameter on chaos can be
given here. When $Q^2/r\ll 1$, the sum of the first term and the
fourth term in Eq. (6) has the approximate expression
\begin{eqnarray}
\Gamma &=& -\frac{1}{2}(E^2+L\beta)+\frac{1}{8}\beta^2 r^2-
\frac{E^2}{r} \nonumber \\ && +\frac{EQ^*}{r}
+\frac{L^2}{2r^2}+\frac{Q^2E^2}{2r^2}+\cdots.
 \end{eqnarray}
The second term in Eq. (37) gives a gravitational force to the
charged particle from the magnetic field. The third term exerts a
gravitational effect from the black hole. The fourth term
corresponds to a Coulomb repulsive force for the Coulomb parameter
$Q^*>0$, but  a Coulomb gravitational force for $Q^*<0$. The fifth
term relates to the particle angular momentum yielding an inertial
centrifugal force. The sixth term acts as a gravitational
repulsive force caused by the black hole charge. The increase of
$\beta$ or $|Q^*|$ with $Q^*<0$ means enhancing the gravitational
effects and chaos is much easily induced. On the contrary,  the
increase of $Q^*$ with $Q^*>0$ or $Q$ means enhancing the
repulsive force effects, equivalently, weakening the gravitational
effects. Thus, the strength of chaos decreases. Because the sixth
term is smaller than one of the second, third, fourth and fifth
terms, an appropriate increase of the black hole charge does not
sensitively change  the dynamical properties.

The approximate equivalence between the two black hole black hole
gravitational systems in the dynamics of orbits outside the event
horizons is due to the two metric functions $f(r)$ and
$f^{\star}(r)$ having only minor differences. As Eq. (3) shows,
the difference between  the two metric functions $f(r)$ and
$f^{\star}(r)$ is $Q^4/(4r^3)<Q^4/(4\times 2^3)\leq 1/32$ as $r>2$
and $0<|Q|\leq 1$. In fact, Fig. 12(a) describes that the
difference $|f(x)-f^{\star}(x)|$ is about $10^{-8}\sim 10^{-6}$
when  $r$ belongs to the range $2.5\leq r\leq 6$ and $|Q|=0.3$.
Given $r=6$ and $0\leq |Q|\leq 1$ in Fig. 12(b), the difference is
about $10^{-10}\sim 10^{-5}$. Because of such small differences,
the RN black hole is not clearly distinguished from the regular RN
black hole via the motions of particles or photons outside the
event horizons. Thus, the two black holes are basically equivalent
in these photon circular orbits, photon spheres outside the event
horizons, and constraints of the black hole charges based on the
First Image of Sagittarius A$^{*}$. Their equivalences are also
suitable for the stable circular orbits, innermost stable circular
orbits, and regular and chaotic dynamics of charged particles
outside the event horizons. However, the two black hole systems
are completely different as particles or photons move in the
vicinity of  the origin $r=0$. In this case, the metric function
$f(r)$ is not expanded like Eq. (3). Although we cannot provide
the analytical solutions of particles or photons moving near the
origin, we easily estimate
\begin{eqnarray}
\lim_{r\rightarrow 0}f(r)=1, ~~~~ \lim_{r\rightarrow
0}f^{\star}(r)=\infty.
 \end{eqnarray}
Typically larger differences between the two metric functions for
smaller values of $r$ are also shown in Figs. 12 (c) and (d).
Therefore, the RN black hole can be distinguished clearly from the
regular RN black hole via such dramatically different changes of
the two metric functions in the motions of particles or photons
around the origin.

 \section{Summary}

The aforementioned regular black hole is a solution of the
Einstein field equations coupled to a suitable NED. It is unlike
the RN  black hole with curvature singularities, but resembles the
RN black hole with horizons. The two black holes satisfy the weak
energy condition everywhere and have similar asymptotic behaviors.

When the two  black hole systems have the same parameters, they
also have the same photon effective  potentials, photon circular
orbits and photon spheres outside the event horizons. Due to the
same photon spheres in the two  black hole systems, the black hole
shadows and constraints of the black hole charges based on the
First Image of Sagittarius A$^{*}$ are almost the same.

The radial motions of charged particles outside the horizons of
the two magnetized black holes, including the effective
potentials, stable circular orbits and innermost stable circular
orbits, have no explicit differences. The constructions of
explicit symplectic integrators for the off equatorial plane
motions of charged particles  in the two magnetized black hole
systems are different. The splitting and composition methods are
easily available for the RN  black hole. However, they are not for
the regular black hole unless an appropriate time transformation
is given to the regular black hole. In fact, such a
time-transformed Hamiltonian exists five explicitly integrable
splitting pieces. In spite of these differences, the two black
hole systems have the same regular and chaotic dynamical features
of charged particles. An increase of the magnetic field parameter
or the magnitude of negative Coulomb parameter equal to the
product of the black hole charge and the particle charge easily
induces chaos due to the gravitational effects increasing.
However, an increase of the black hole charge or the positive
Coulomb parameter weakens the strength of chaos due to the
repulsive force effects increasing. The variation of dynamical
properties is not very sensitive dependence on an appropriate
increase of the black hole charge.

In sum, the basic equivalence between the two black hole
spacetimes exists in the photon circular orbits, photon spheres
outside the event horizons and constraints of the black hole
charges. This equivalence is also suitable for the stable circular
orbits, innermost stable circular orbits, and regular and chaotic
dynamics of charged particles  outside the event horizons. These
equivalences in the dynamics of orbits outside the event horizons
of the two charged black holes are due to the extremely small
differences of the two metric functions. In view of this, the RN
black hole can be simulated by using the regular black hole
without curvature singularity in some situations.

\section*{Data Availability Statements}

All data generated or analysed during this study are included in
this published article.

\section*{Acknowledgments}

The authors are grateful to two referees for useful suggestions.
This research was supported by the National Natural Science
Foundation of China (Grant No. 11973020) and the National Natural
Science Foundation of Guangxi (No. 2019JJD110006).

\begin{table*}
    \begin{center}
\caption{Relationship between the new time $w$ and the proper time
$\tau$ for the magnetized regular black hole system in Fig. 3(b).
$\tau^\star$ is the proper time in the magnetized RN black hole.
The difference between the new time $w$ and the proper time $\tau$
in the regular black hole system is small.}
    \begin{tabular}{cccccccccccccccc}
        \hline
        $w $&1&2&3&4&5&6&7&8&9&10\\
        \hline
        $ \tau $&0.96&1.95&2.94&3.92&4.87&5.85&6.84&7.82&8.78&9.76
        \\
        \hline
        $\tau^\star$ &1&2&3&4&5&6&7&8&9&10\\
        \hline
    \end{tabular} 
 \end{center}
 \end{table*}

\begin{figure*}
    \centering{
        \includegraphics[width=12pc]{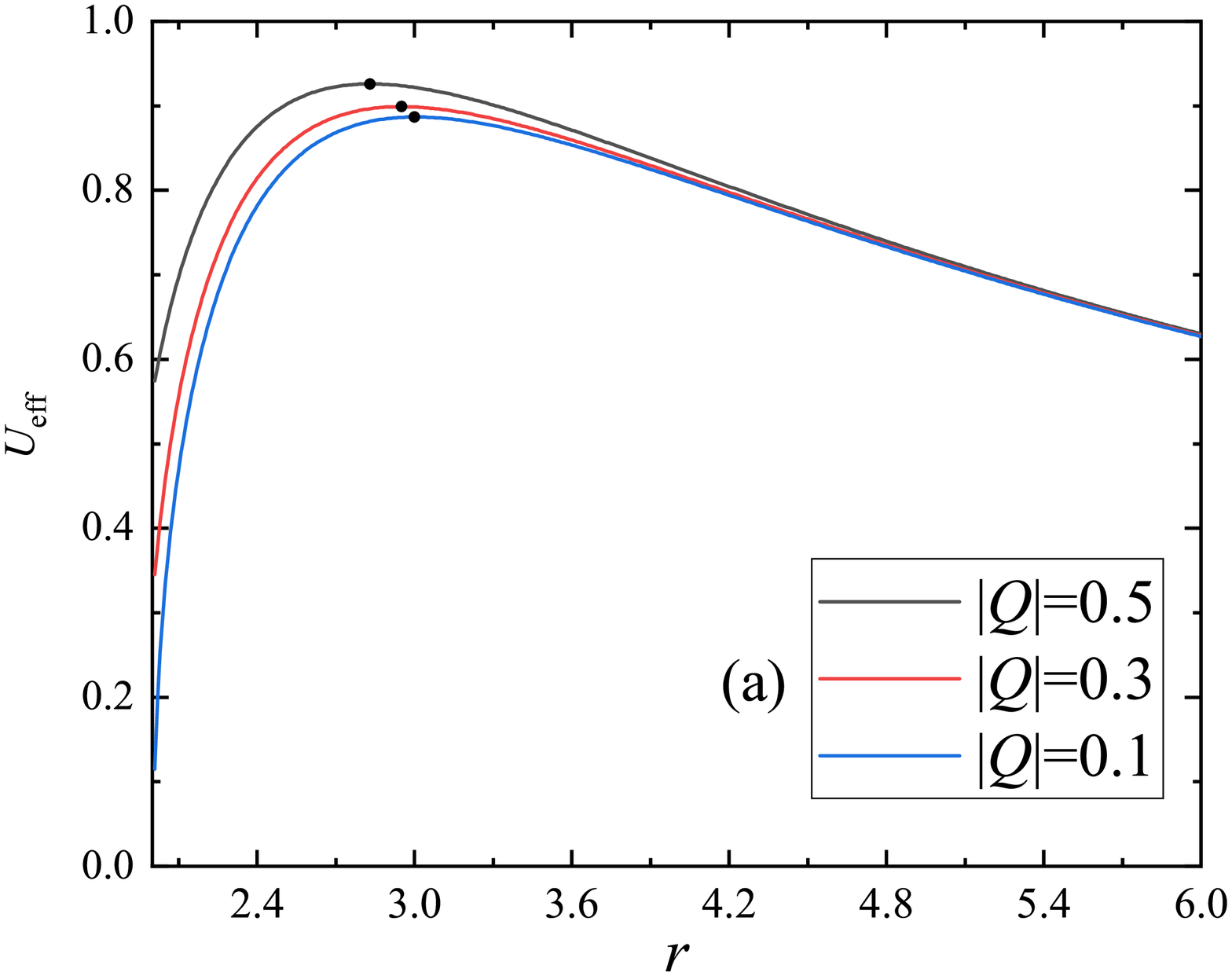}
        \includegraphics[width=12pc]{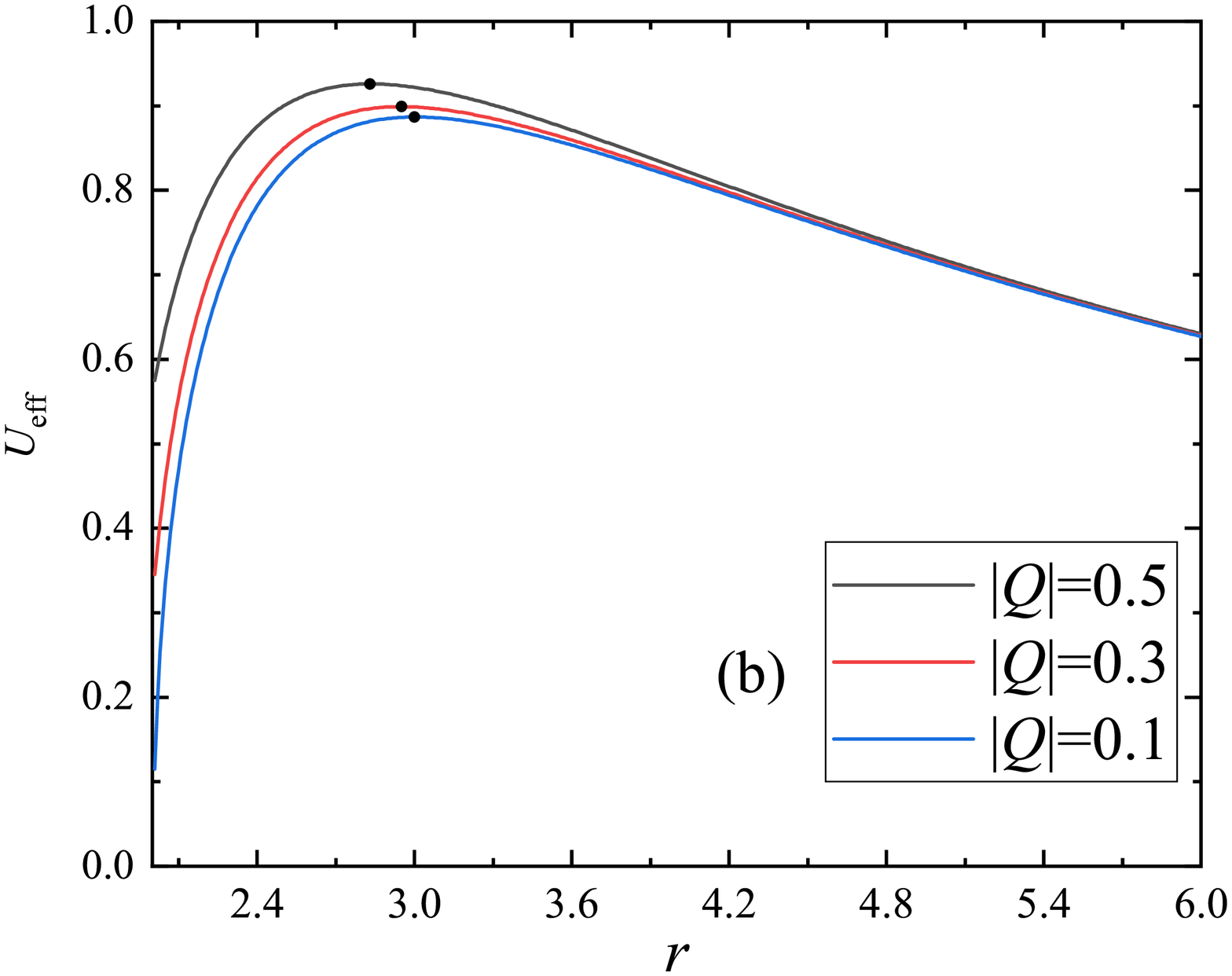}
        \includegraphics[width=12pc]{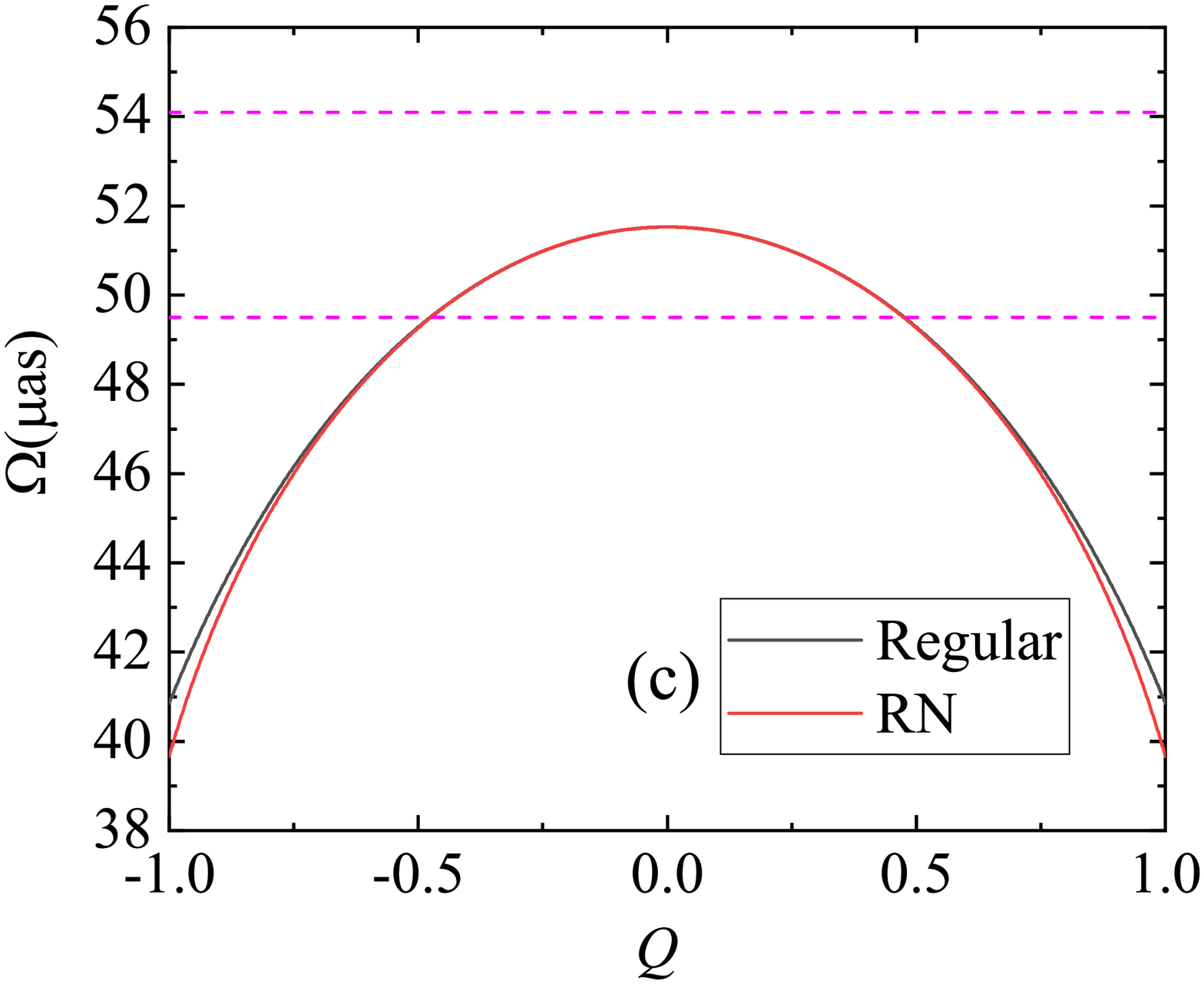}
\caption{(a): Effective potentials of photons moving around the
regular black hole at the equatorial plane $\theta=\pi/2$ for
several values of the black hole charges $|Q|$. (b): Effective
potentials of photons moving around the RN black hole. (c):
Relation between the observed angular diameter $\Omega$ of black
hole shadows and the black hole charges $|Q|$.
            }}
\end{figure*}

\begin{figure*}
    \centering{
        \includegraphics[width=18pc]{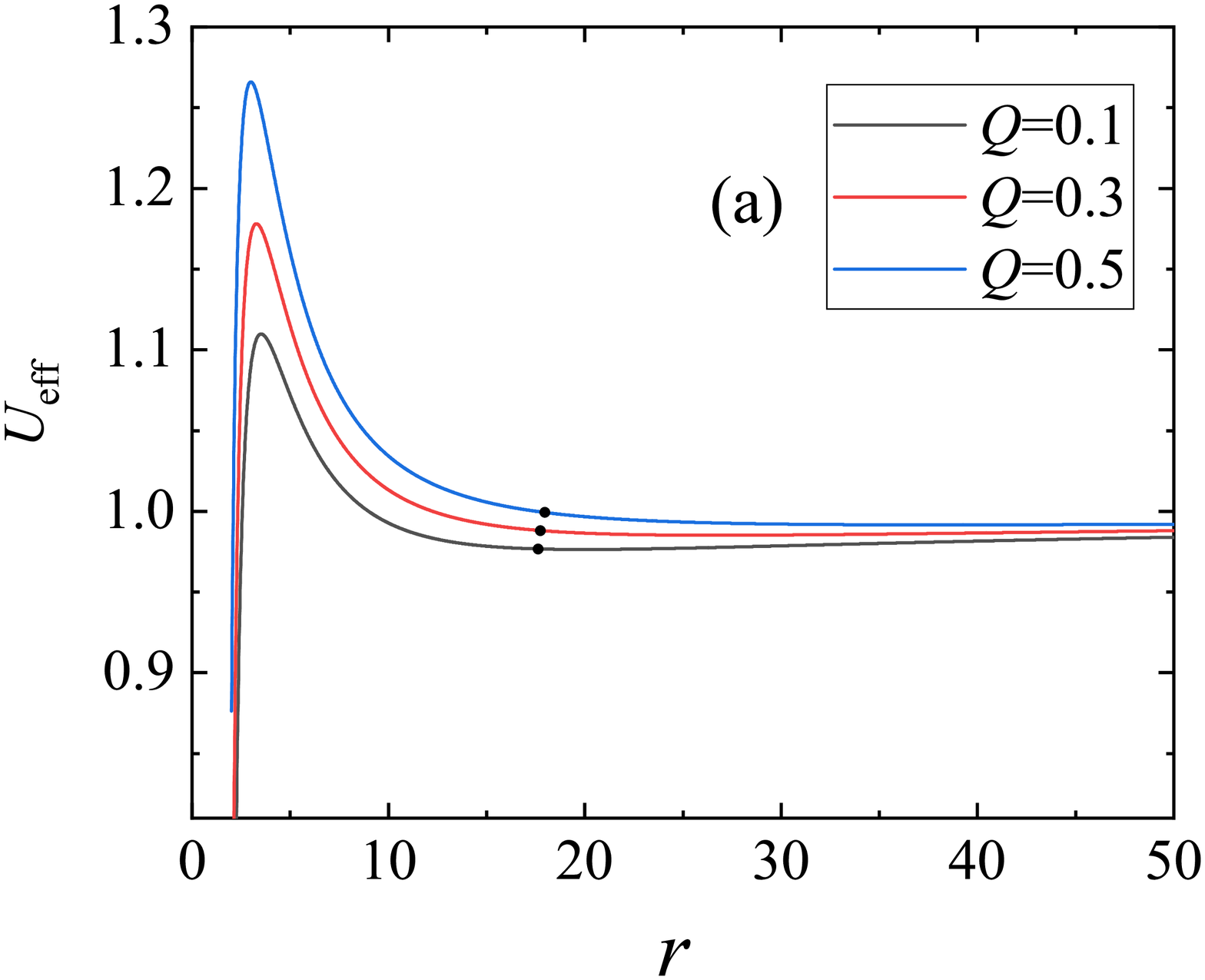}
        \includegraphics[width=18pc]{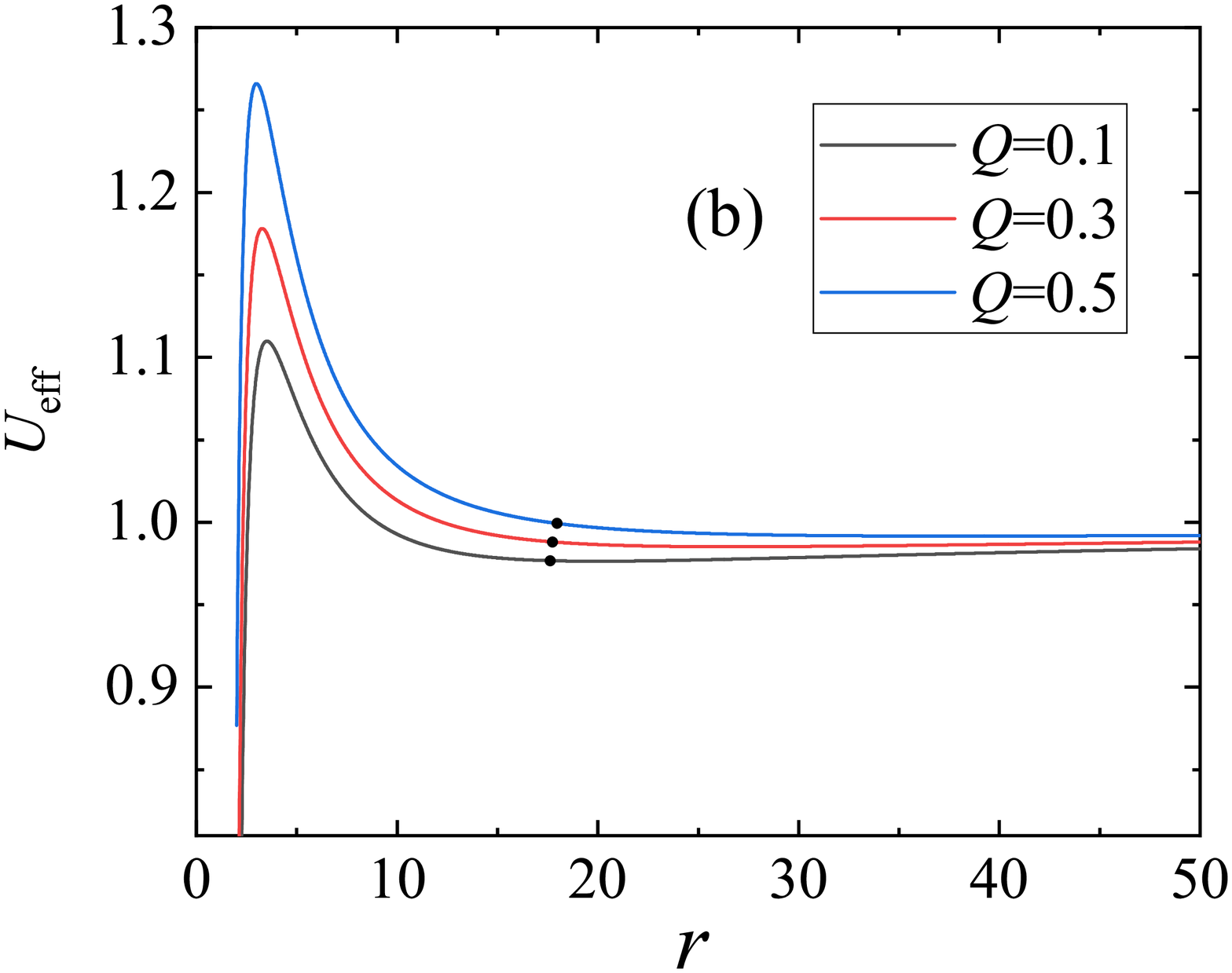}
\caption{Effective potentials of charged particles  at the
equatorial plane for several values of the black hole charges $Q$.
The other parameters are $L=4.6$, $Q^*=10^{-4}$ and
$\beta=10^{-3}$. (a): The  potentials relate to the regular black
hole. (b): The  potentials relate to the RN black hole.
            }}
\end{figure*}

\begin{figure*}
    \centering{
\includegraphics[width=12pc]{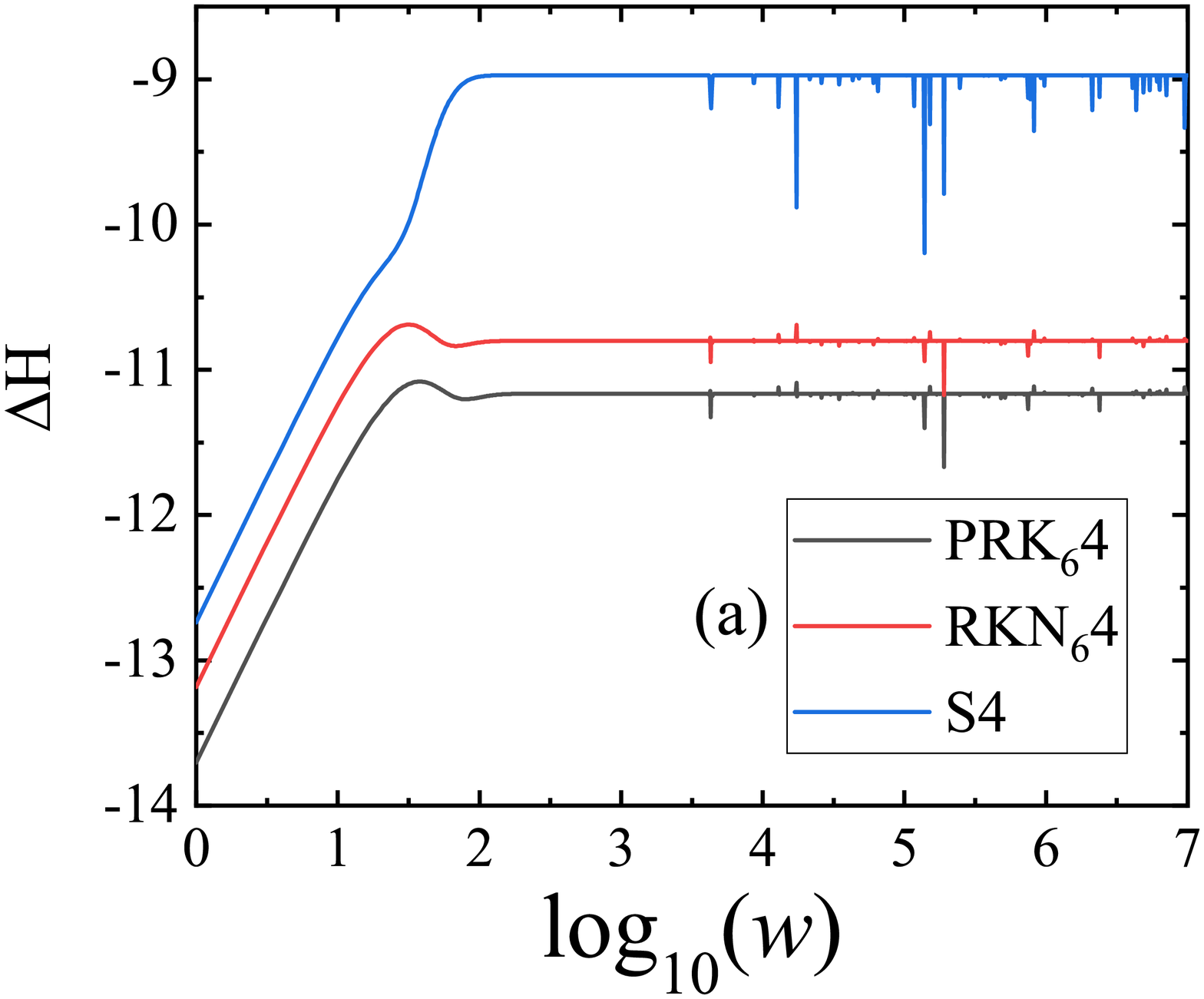}
\includegraphics[width=12pc]{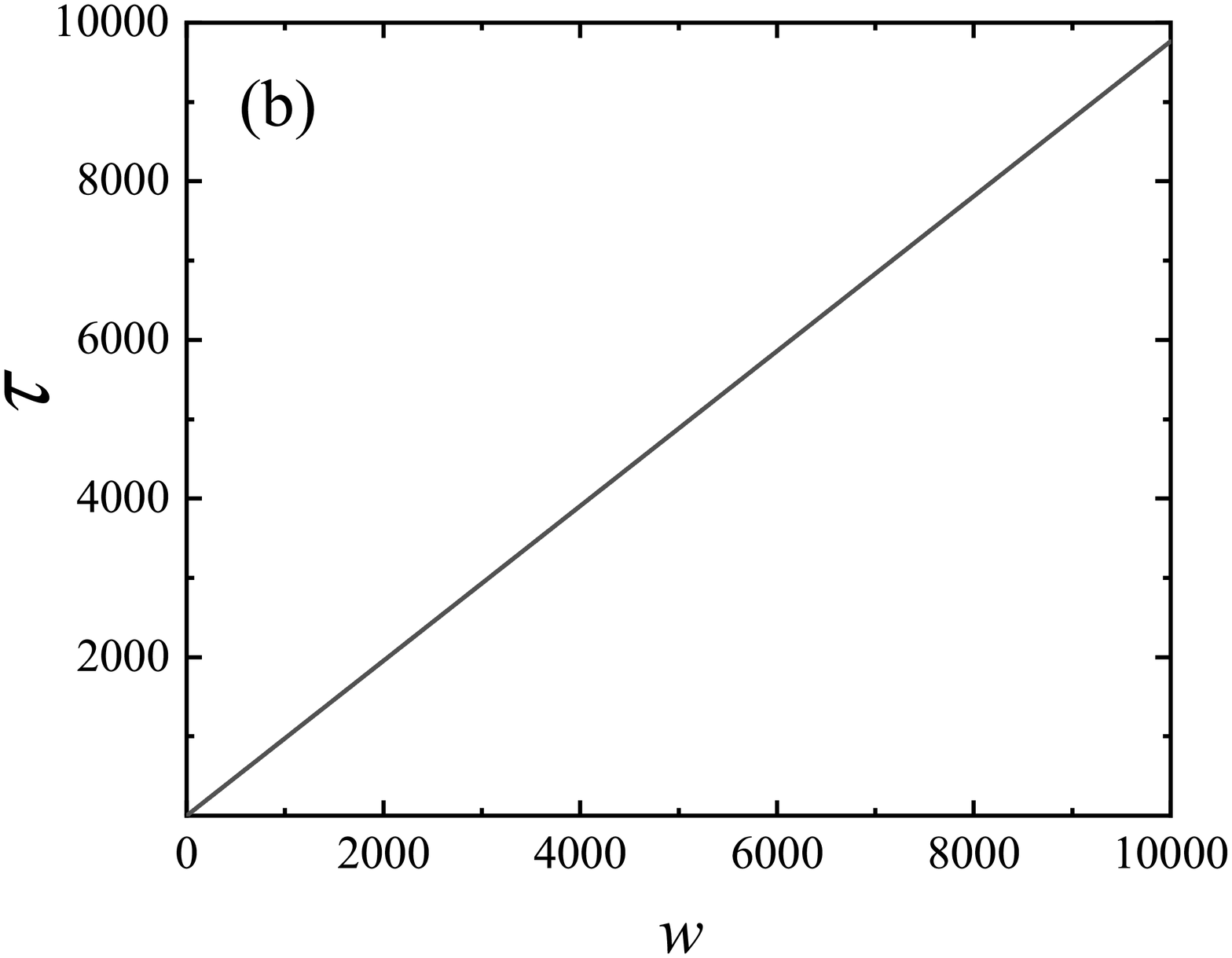}
\includegraphics[width=12pc]{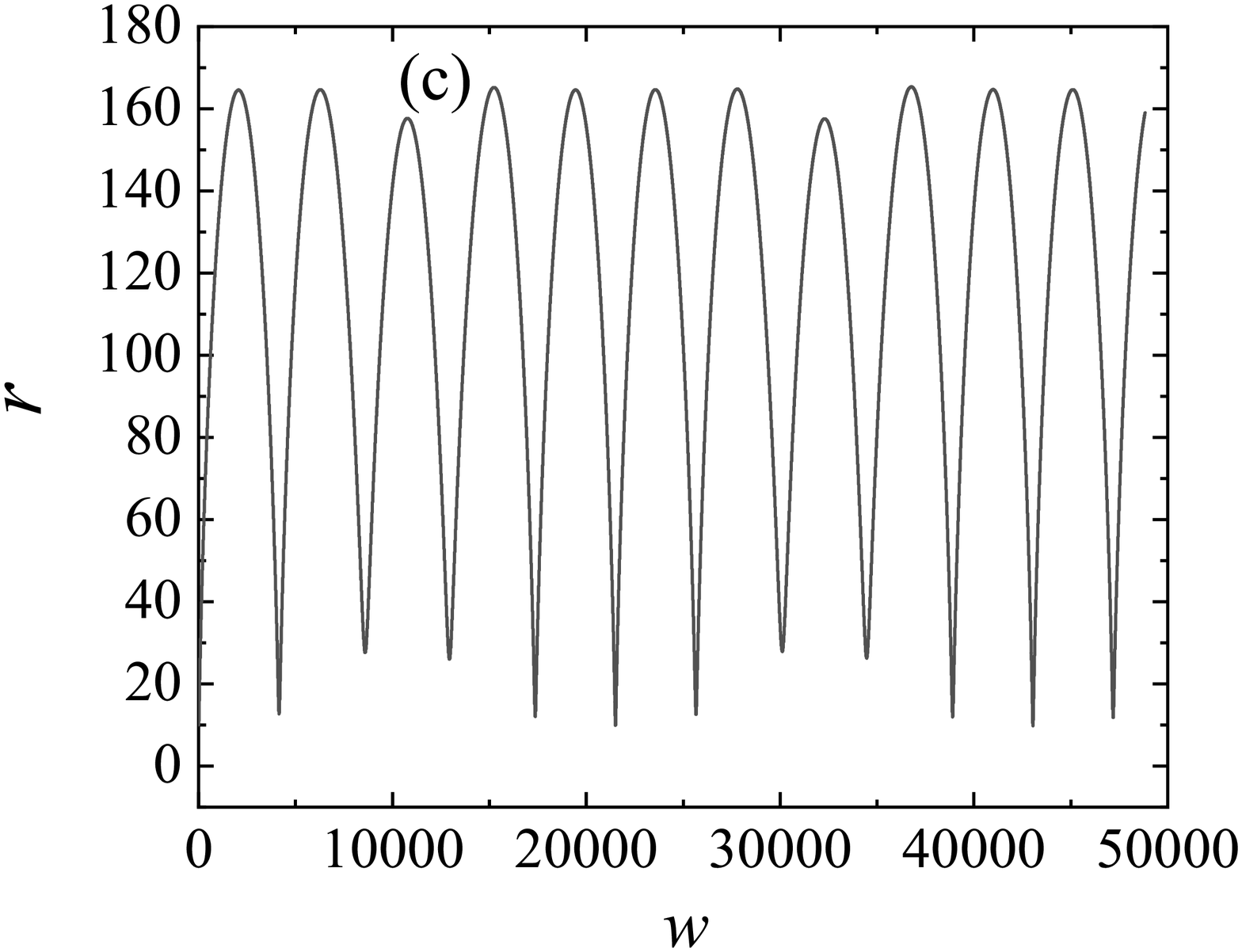}
\caption{(a): Errors of the Hamiltonian $\Delta H=\mathbb{H}$ (18)
corresponding to the magnetized regular black hole for the three
fourth-order explicit symplectic integrators $PRK_64$, $RKN_64$
and S4 with the time step $h=1$. The parameters are $E=0.995$,
$L=4.6$, $Q=0.1$, $Q^*=0.06$ and $\beta=8\times10^{-4}$; the
initial conditions are $\theta =\pi/2$, $ p_r=0$  and  $r=15$.
(b): Relation between the proper time $\tau $ and the new time
$w$.  (c) Variation of $r$ with the time $w$.
        }
       }
\end{figure*}

\begin{figure*}
    \centering{
        \includegraphics[width=12pc]{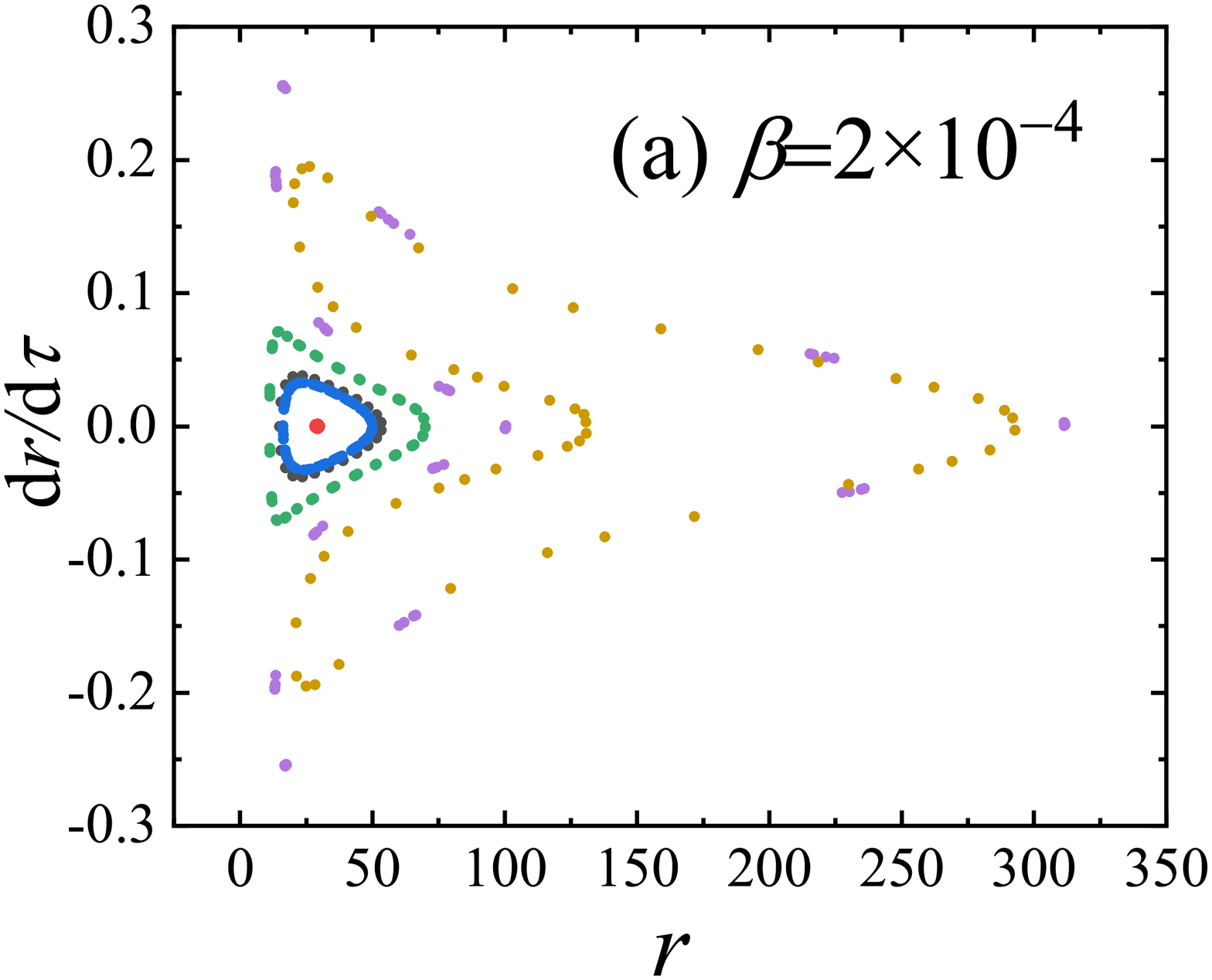}
        \includegraphics[width=12pc]{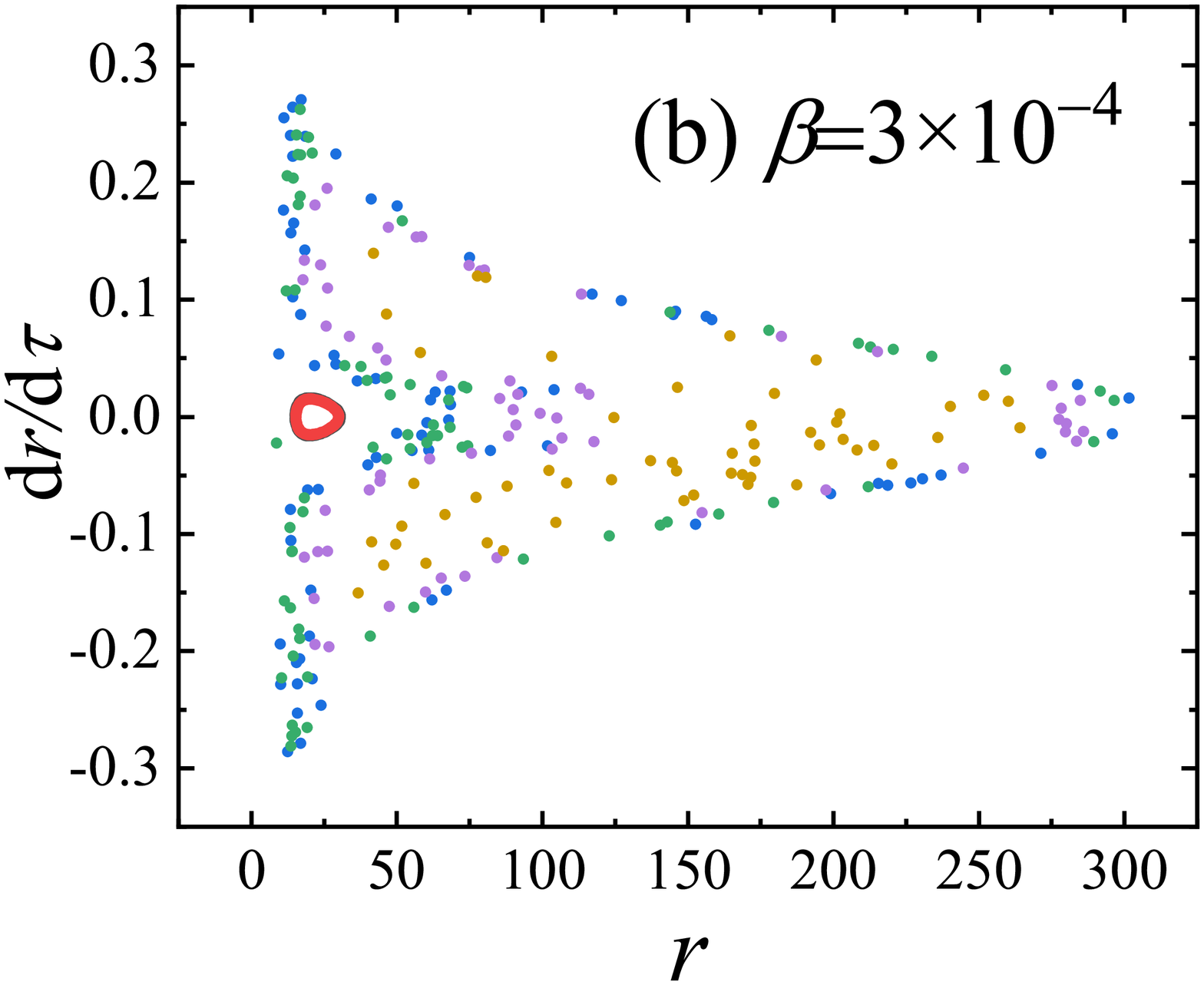}
        \includegraphics[width=12pc]{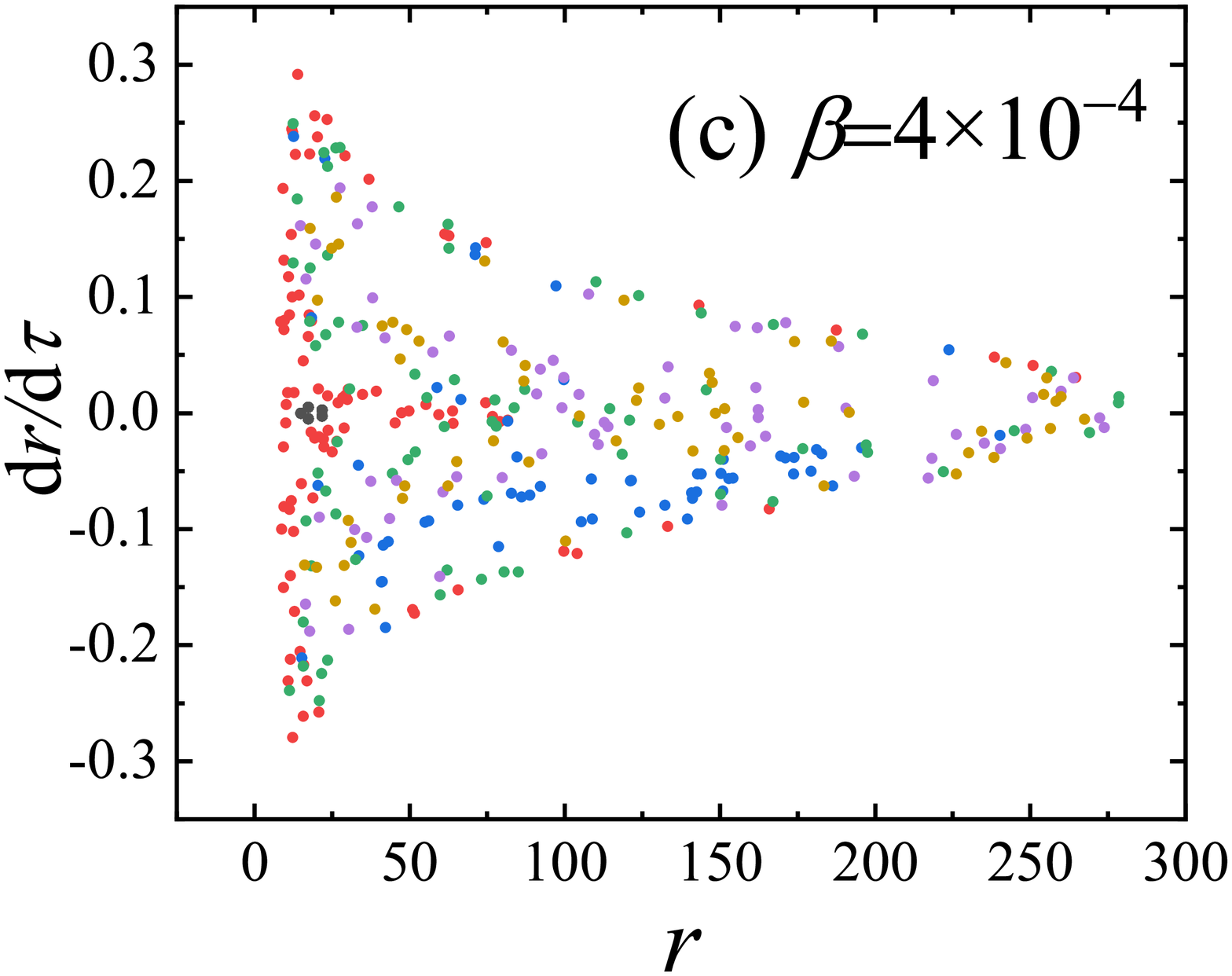}
        \includegraphics[width=12pc]{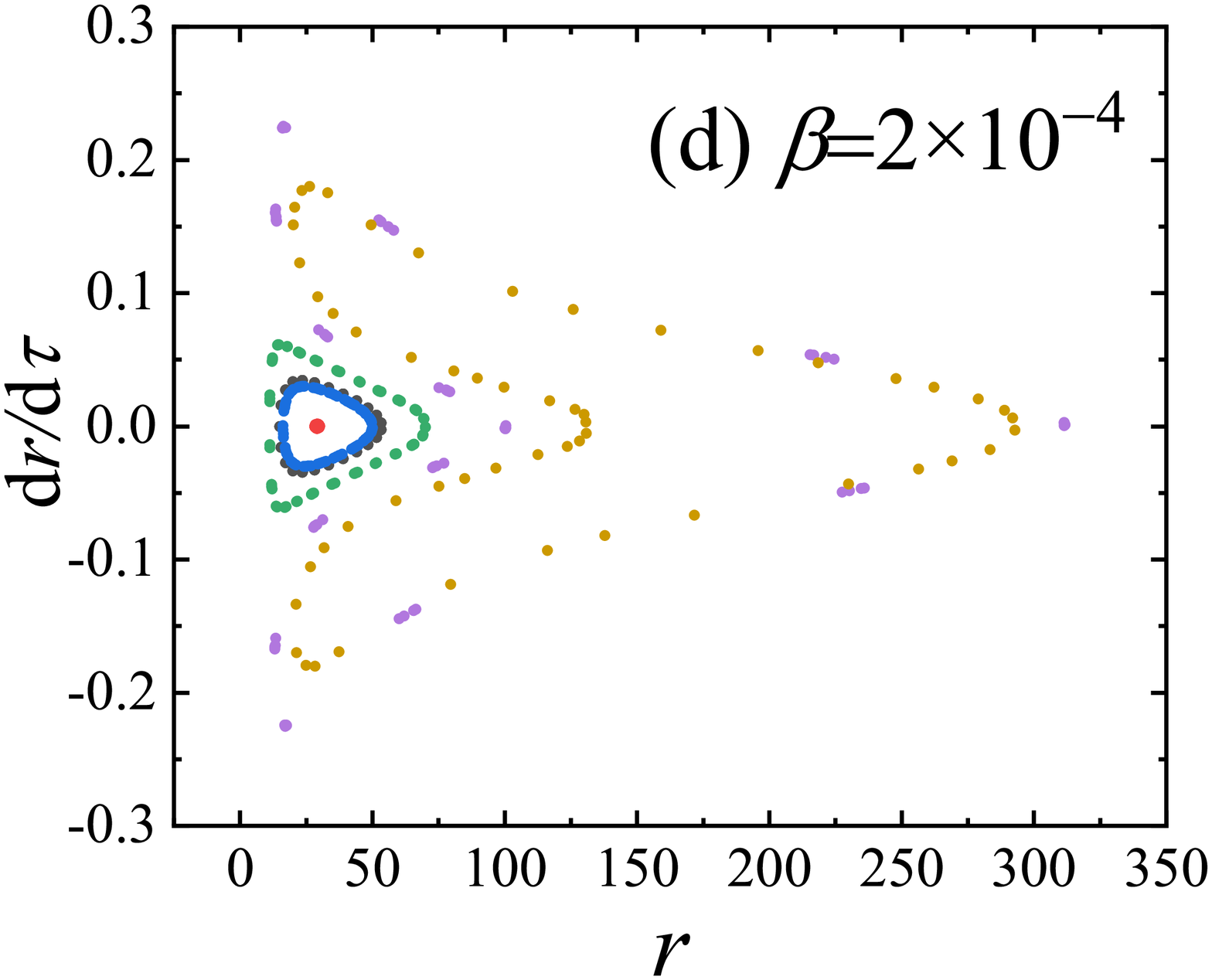}
        \includegraphics[width=12pc]{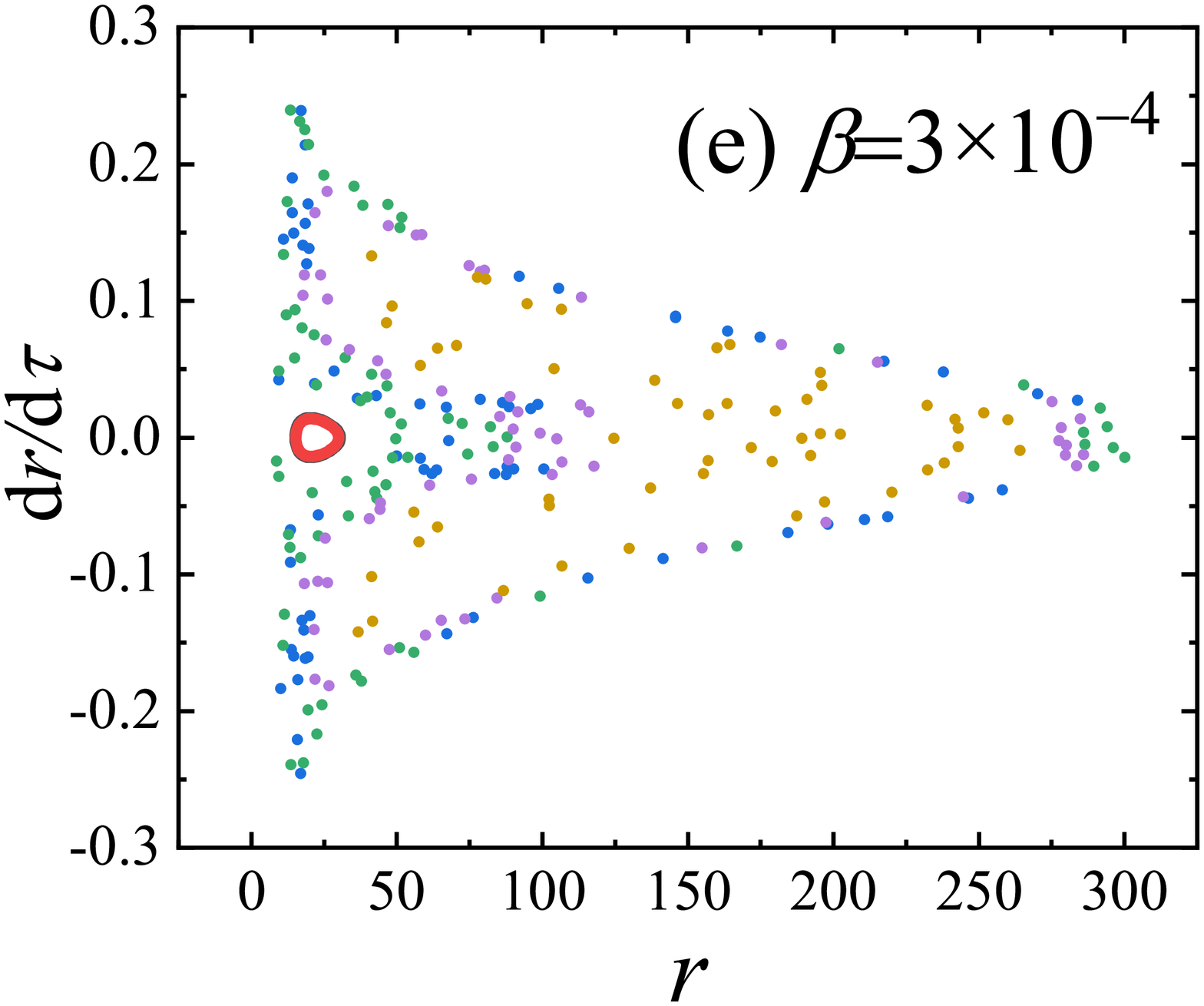}
        \includegraphics[width=12pc]{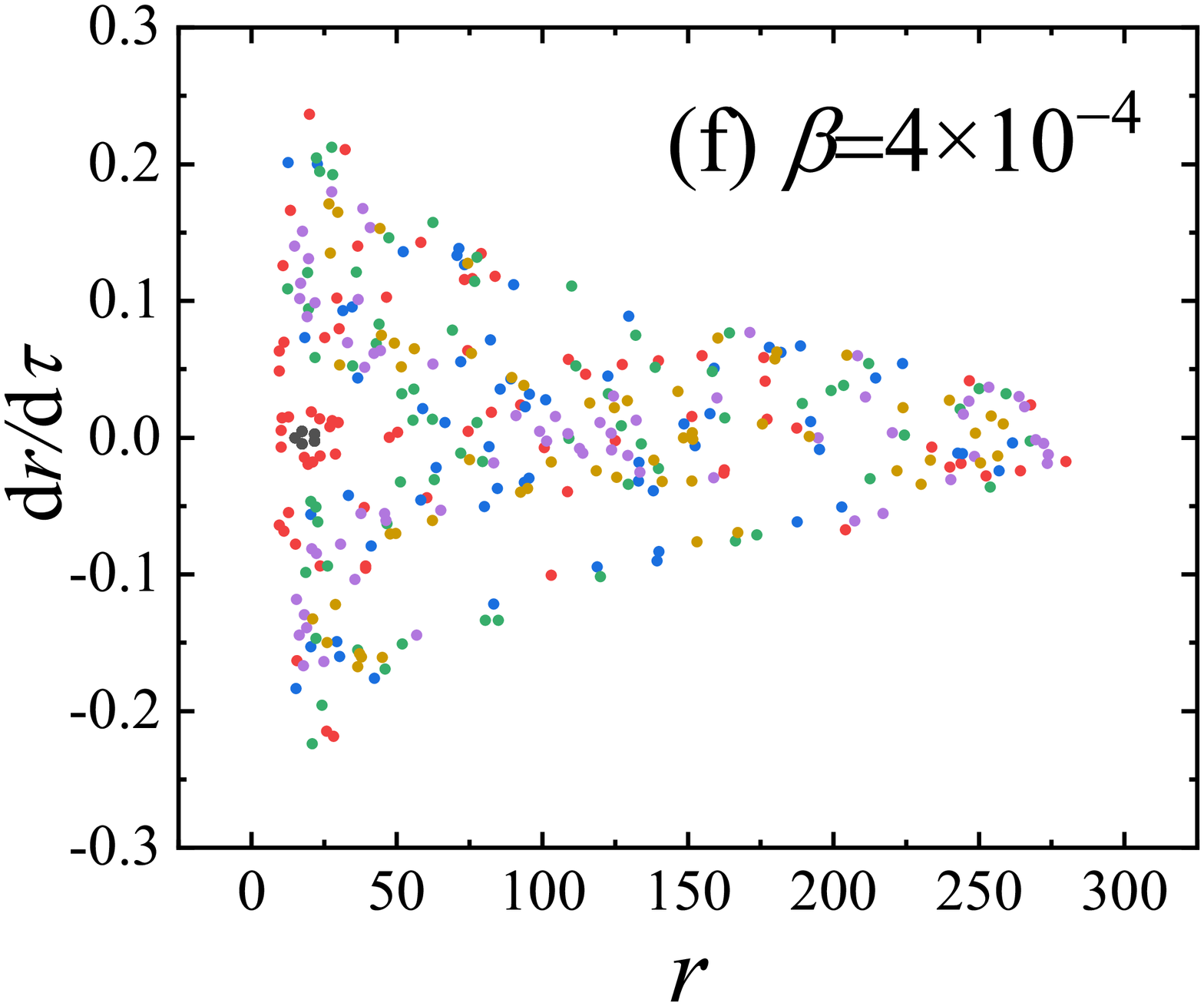}
\caption{Poincar\'{e} sections at the plane $\theta=\pi/2$ with
$p_{\theta}>0$. Three values are given to the magnetic field
parameter $\beta$, and the other parameters are $E=0.9975$, $L=4.6
$, $Q=0.3$ and $Q^*=0.001 $. (a)-(c): Corresponding to the regular
black hole with the metric function $f(r)$. (d)-(f): Corresponding
to the RN black hole with the metric function $f^{\star}(r)$.
            }}
\end{figure*}

\begin{figure*}
    \centering{
        \includegraphics[width=18pc]{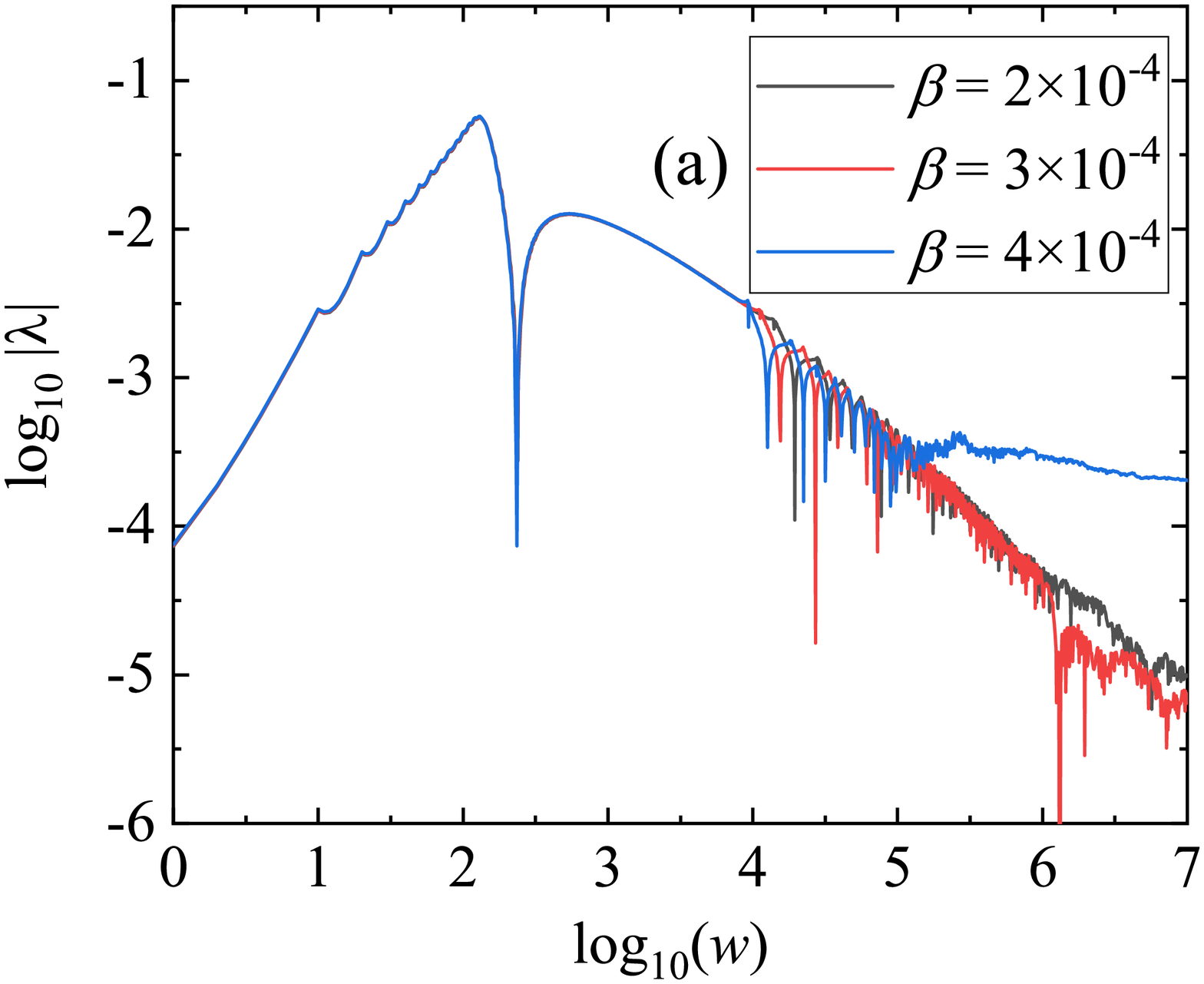}
        \includegraphics[width=18pc]{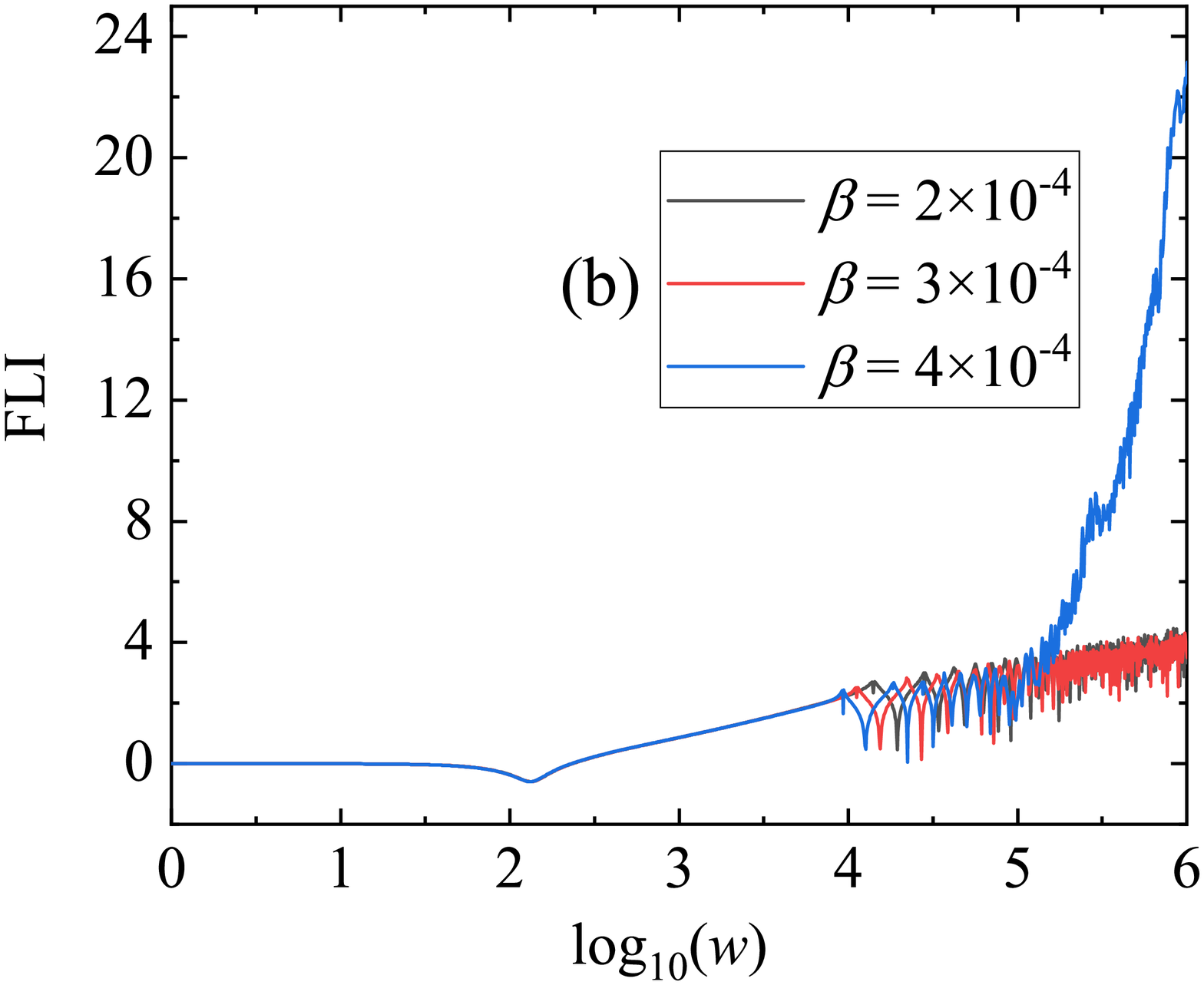}
        \includegraphics[width=18pc]{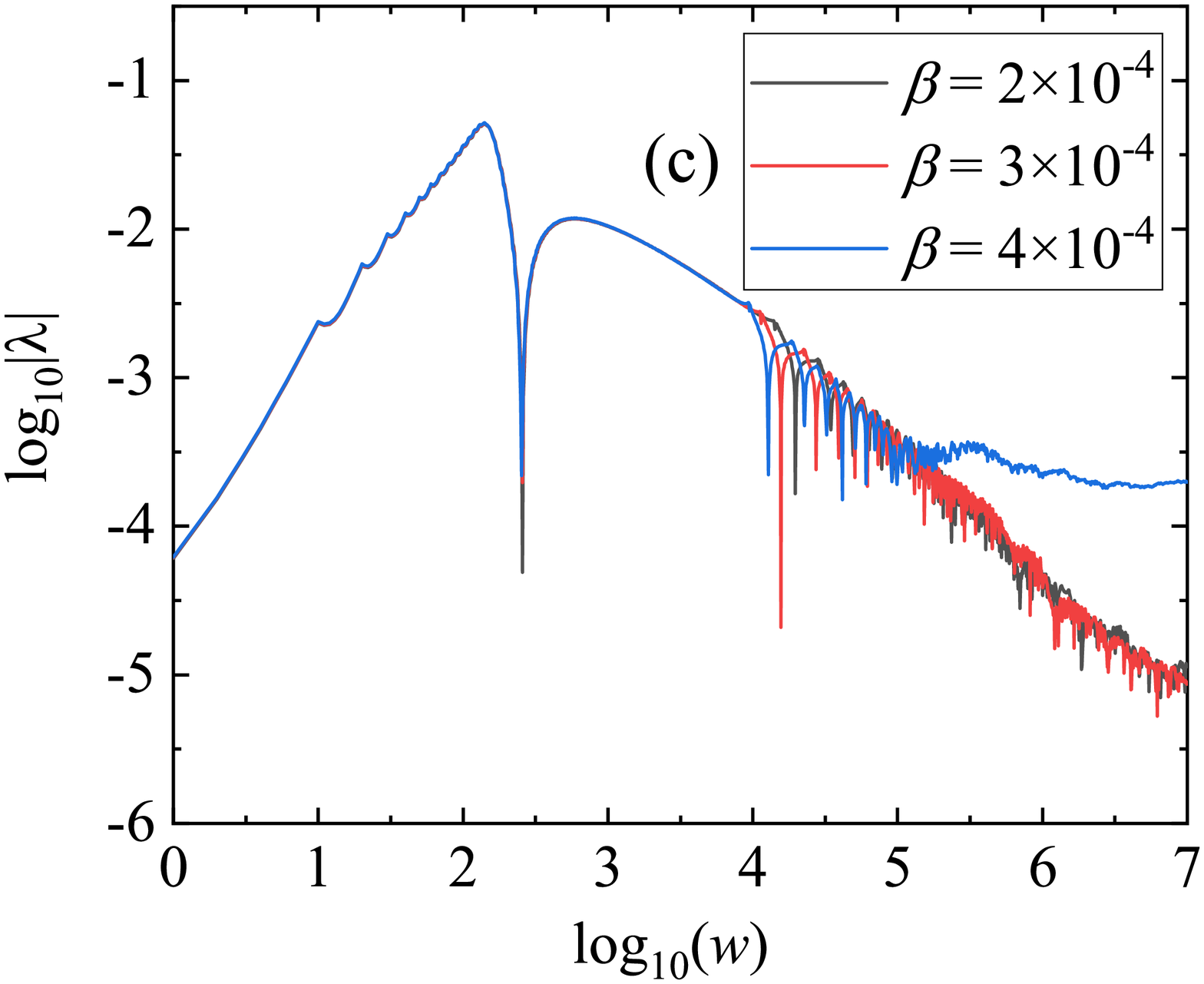}
        \includegraphics[width=18pc]{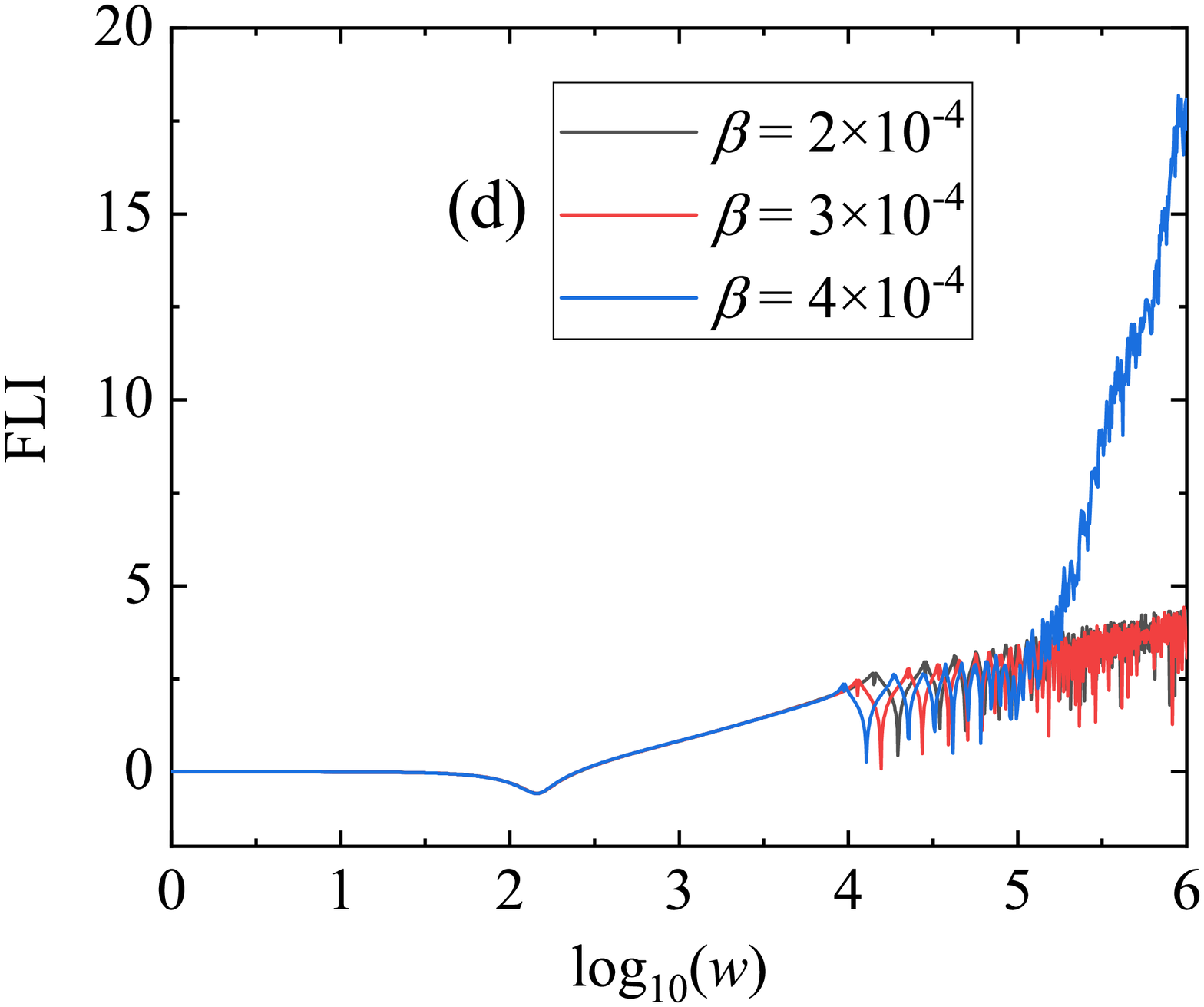}
\caption{(a) The largest Lyapunov exponents $\lambda$ of the
orbits with the initial separation $r=30$ in the regular black
hole system of Figs. 4 (a)-(c). (b) Same as (a), but the fast
Lyapunov indicators (FLIs) are used instead of $\lambda$. (c) Same
as (a), but the regular black hole gives place to the RN black
hole. (d) Same as (c), but the FLIs are used.
            }}
\end{figure*}

\begin{figure*}
    \centering{
        \includegraphics[width=18pc]{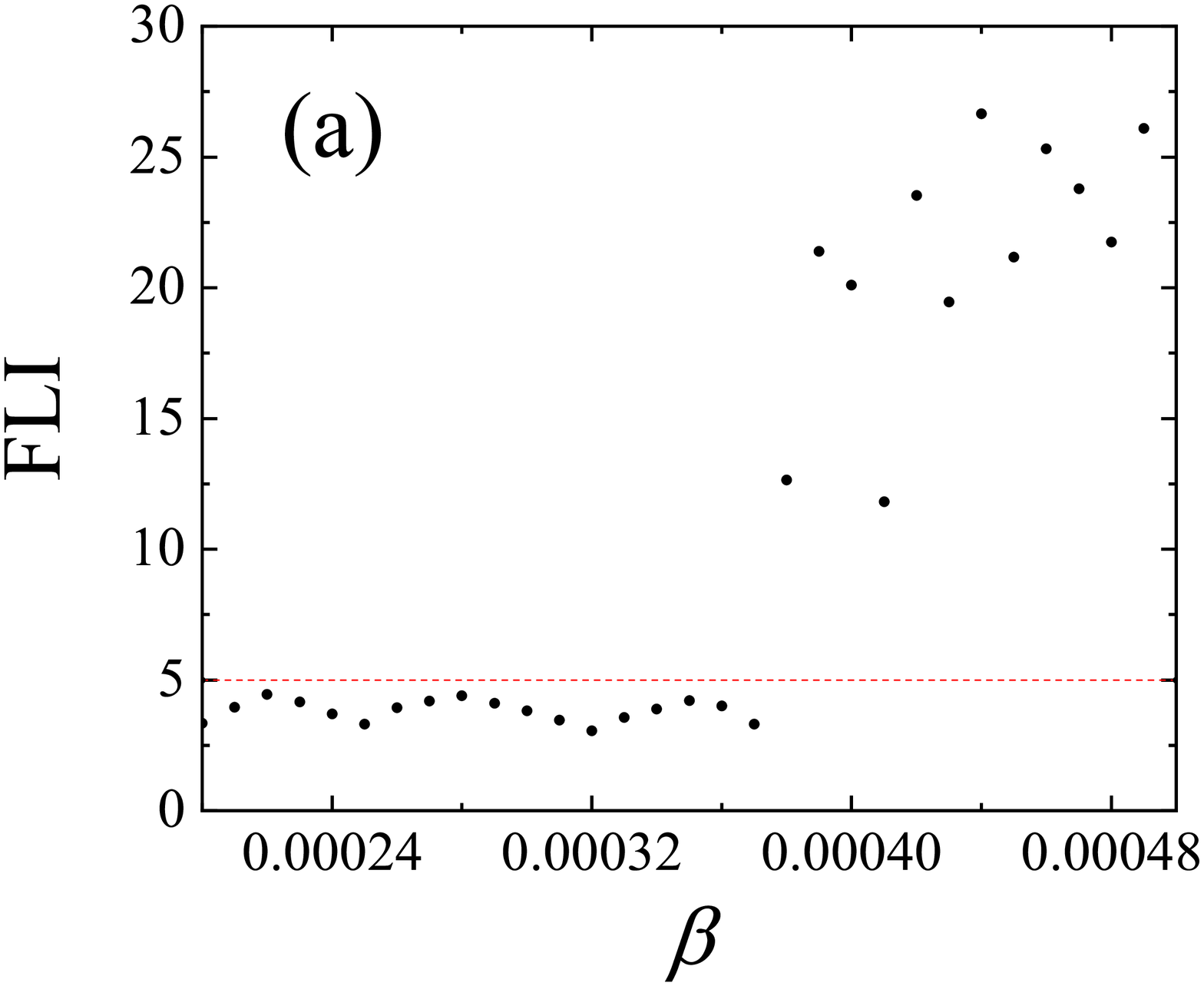}
        \includegraphics[width=18pc]{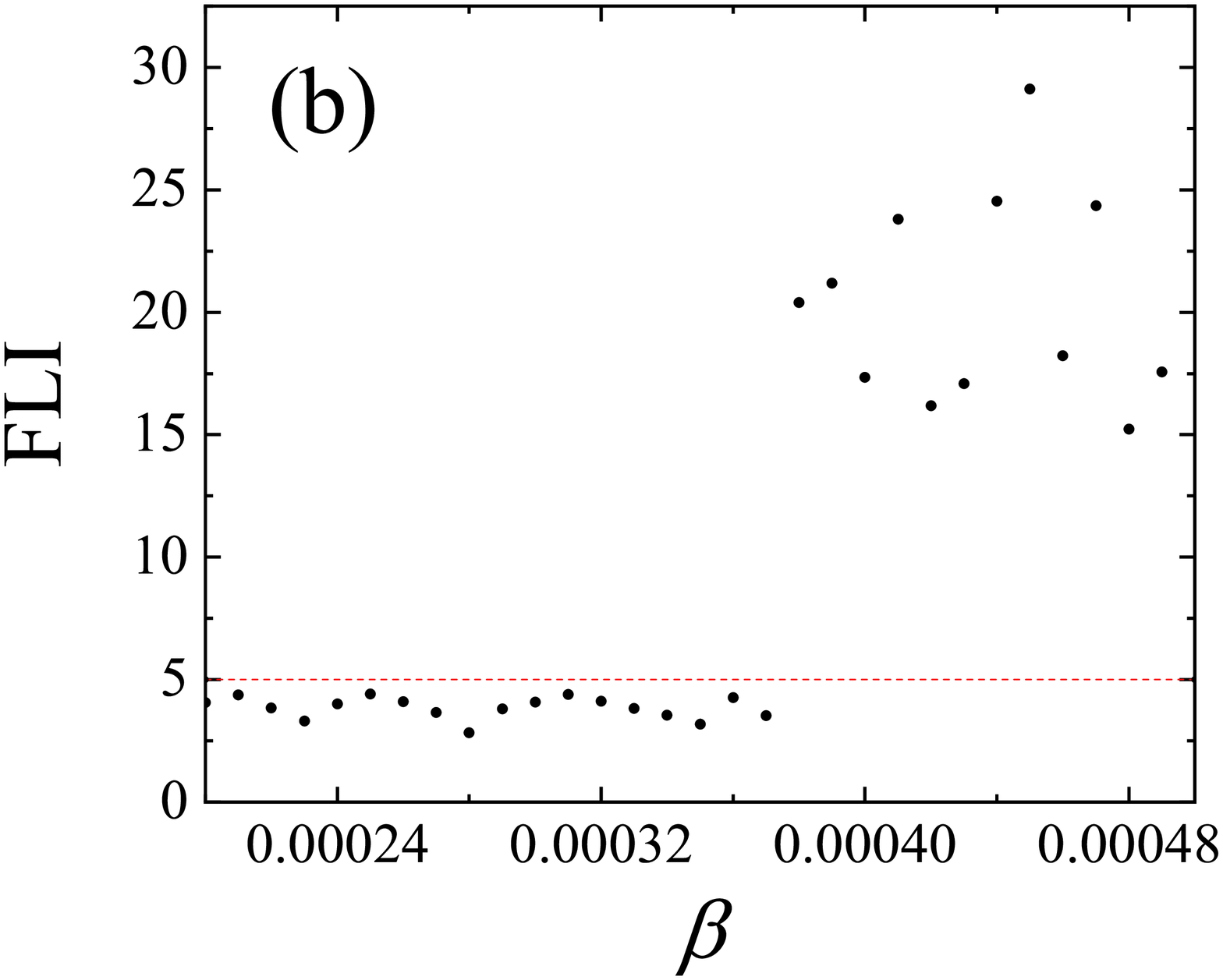}
\caption{Relation between the FLI and the magnetic field parameter
$\beta$. The tested orbit is that in Fig. 5. Each of the FLIs is
obtained after the integration time $w=10^6$. (a): Corresponding
to the regular black hole with the metric function $f(r)$. (b):
Corresponding to the RN black hole with the metric function
$f^{\star}(r)$.
            }}
\end{figure*}

\begin{figure*}
    \centering{
        \includegraphics[width=12pc]{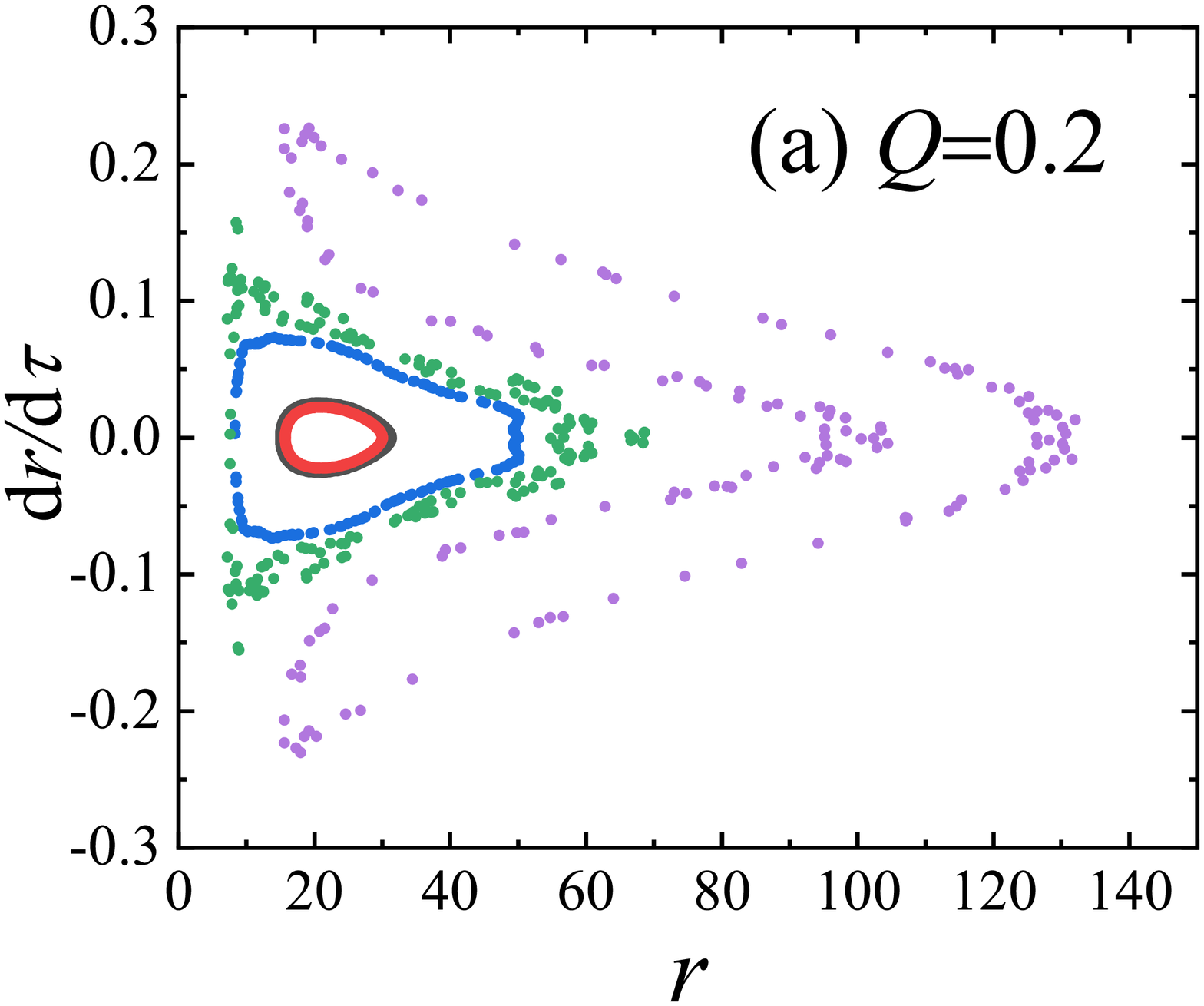}
        \includegraphics[width=12pc]{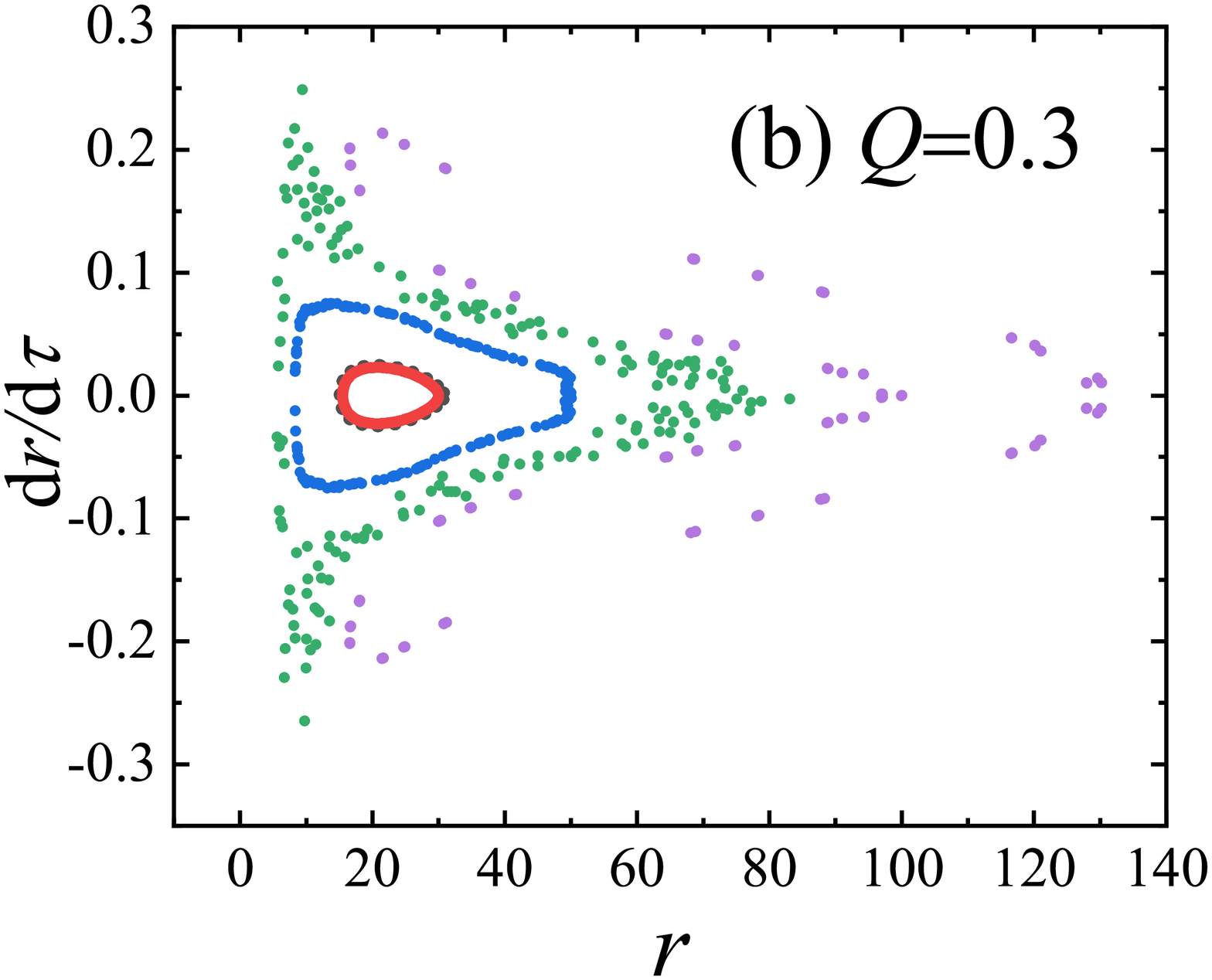}
        \includegraphics[width=12pc]{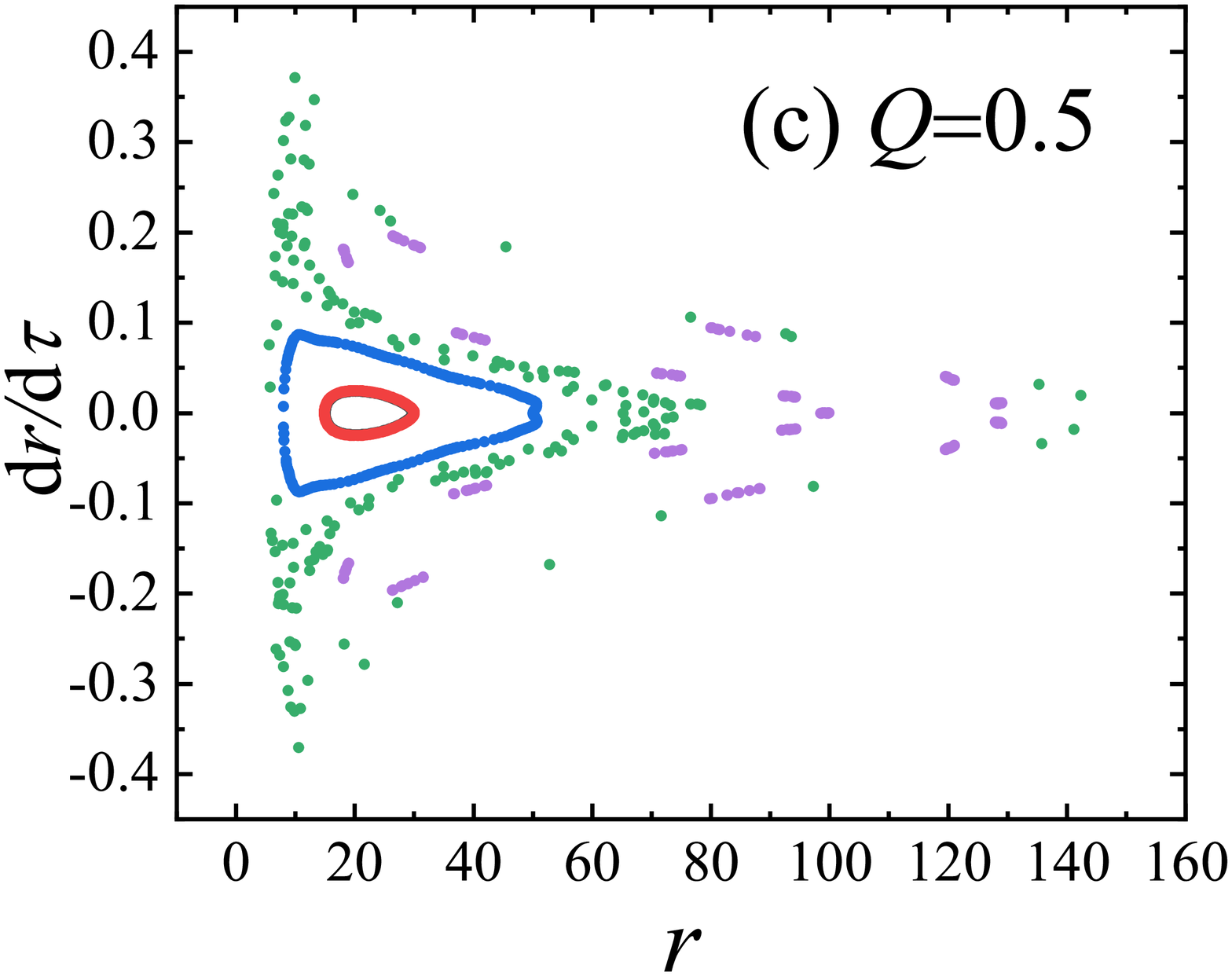}
        \includegraphics[width=12pc]{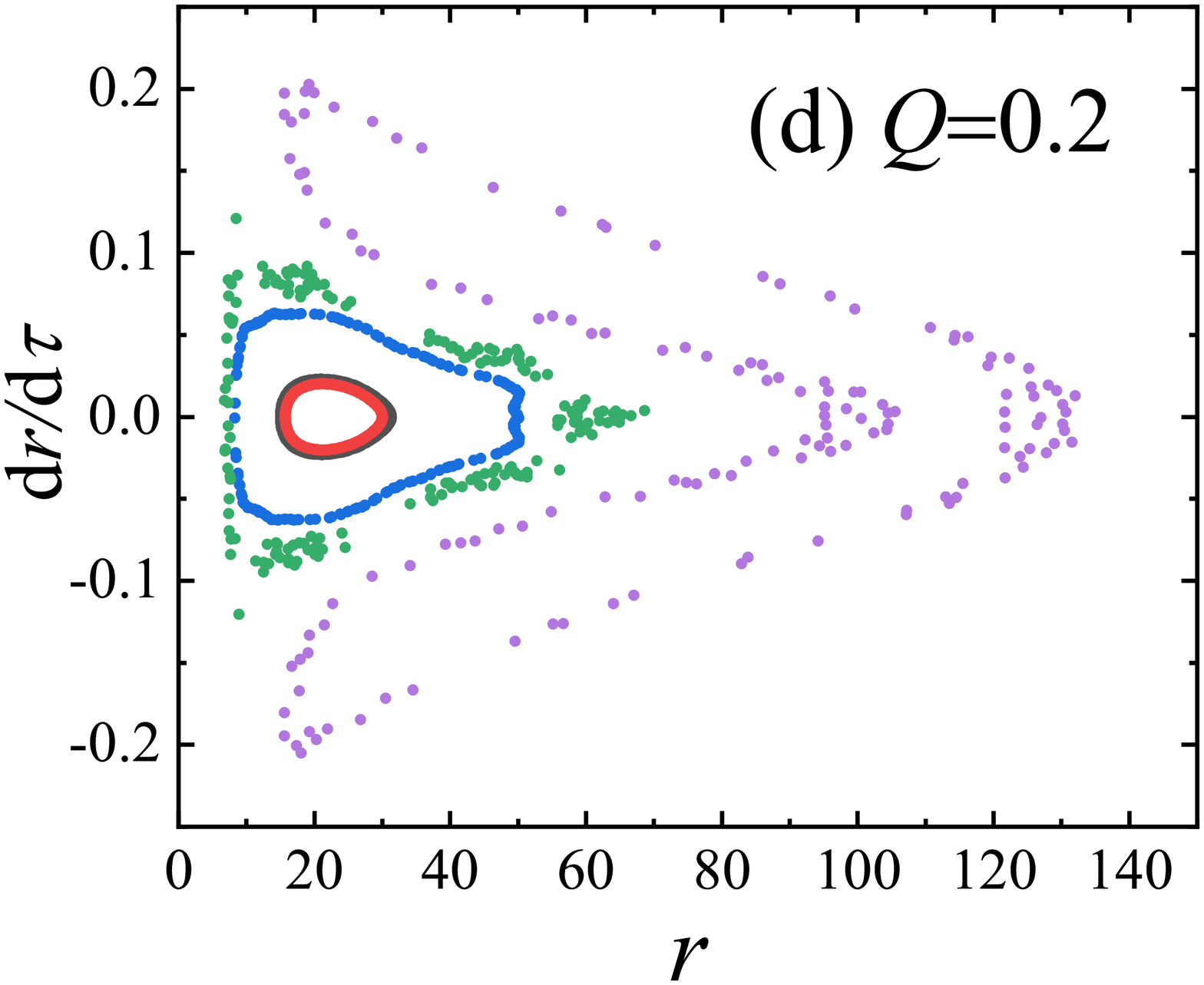}
        \includegraphics[width=12pc]{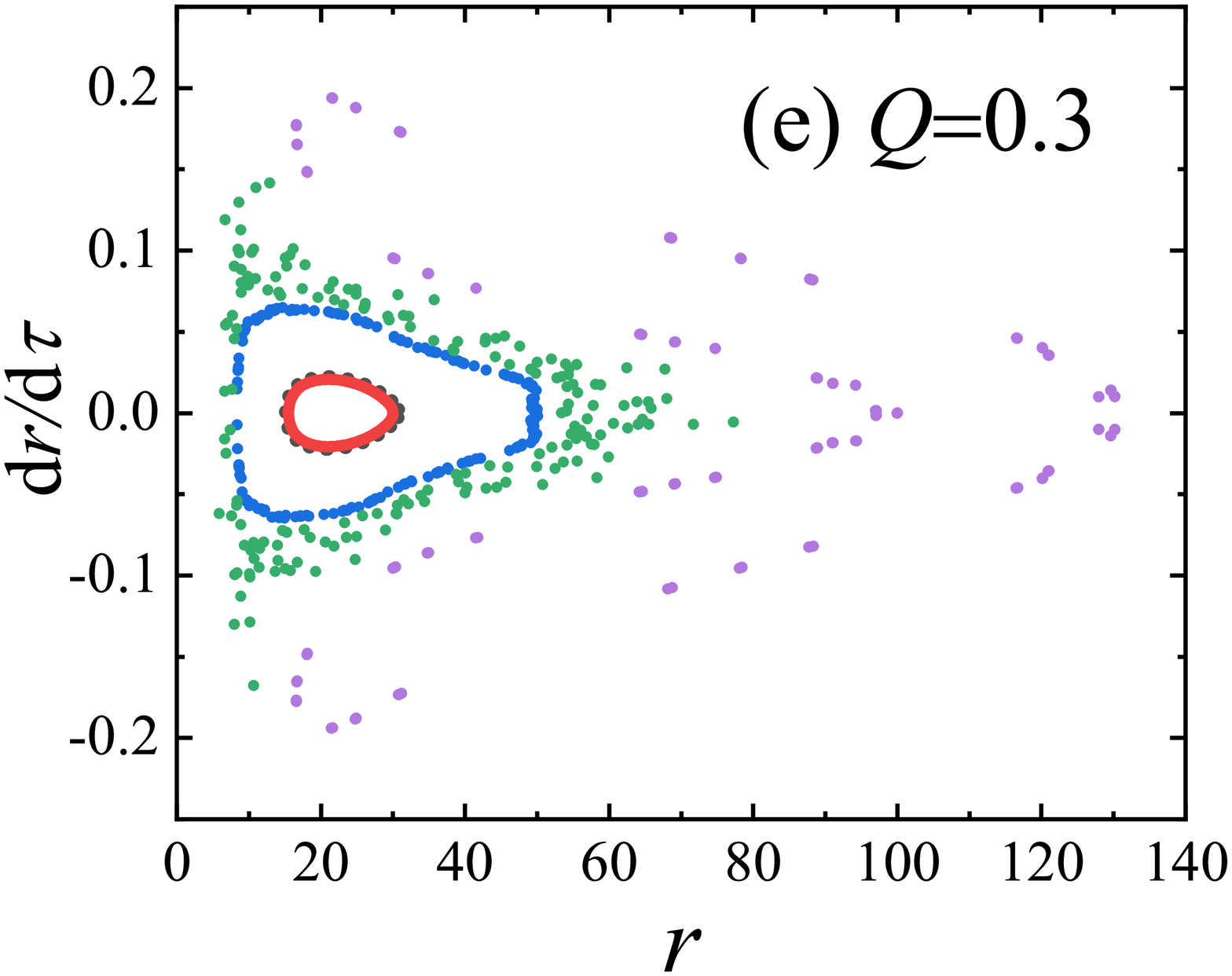}
        \includegraphics[width=12pc]{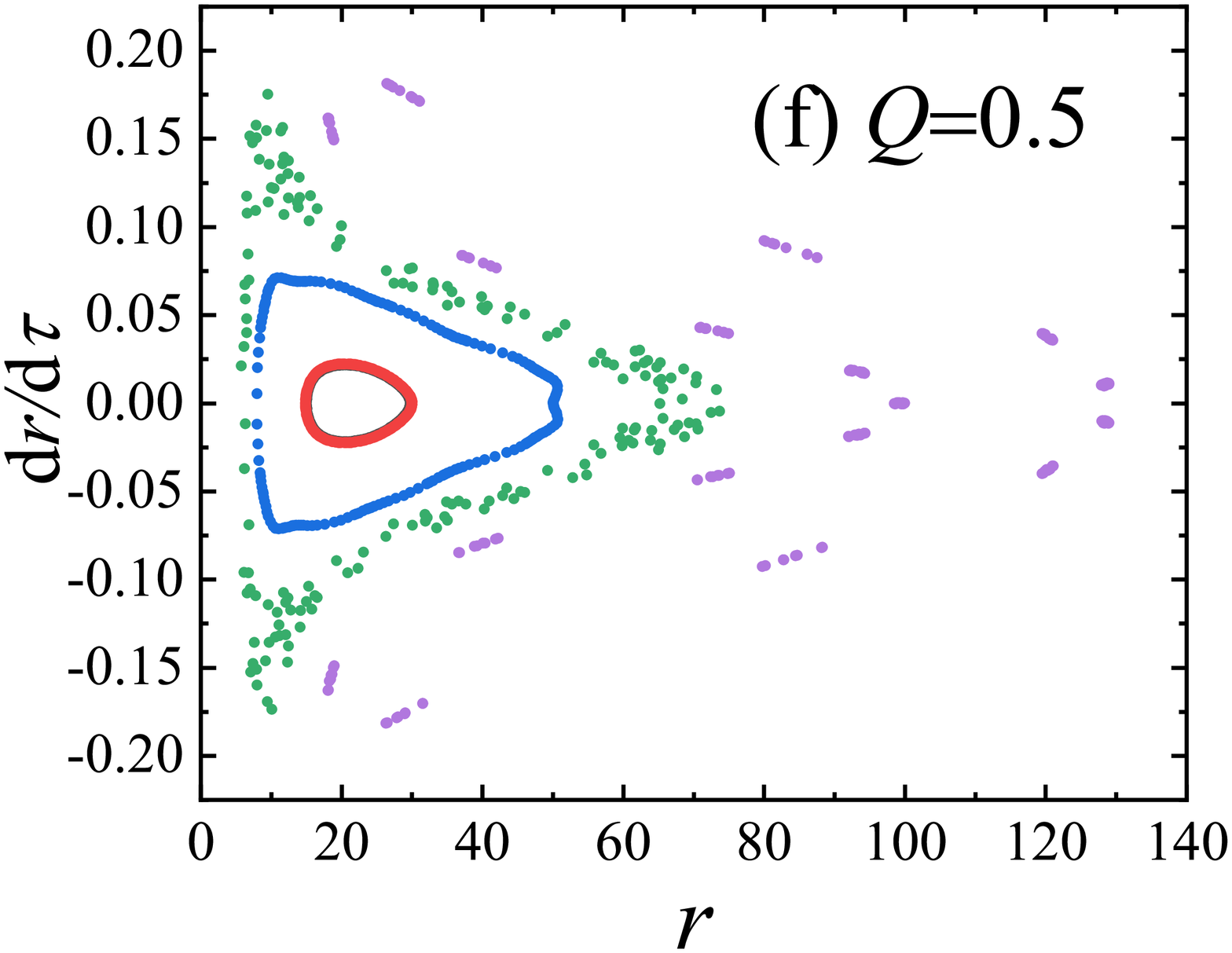}
\caption{Poincar\'{e} sections. The parameters are $E=0.995$,
$L=4$, $\beta=6\times 10^{-4}$ and $Q^*=0.001$, and $Q$ is given
three values. (a)-(c): Corresponding to the regular black hole
with the metric function $f(r)$. (d)-(f): Corresponding to the RN
black hole with the metric function $f^{\star}(r)$.
            }}
\end{figure*}

\begin{figure*}
    \centering{
        \includegraphics[width=18pc]{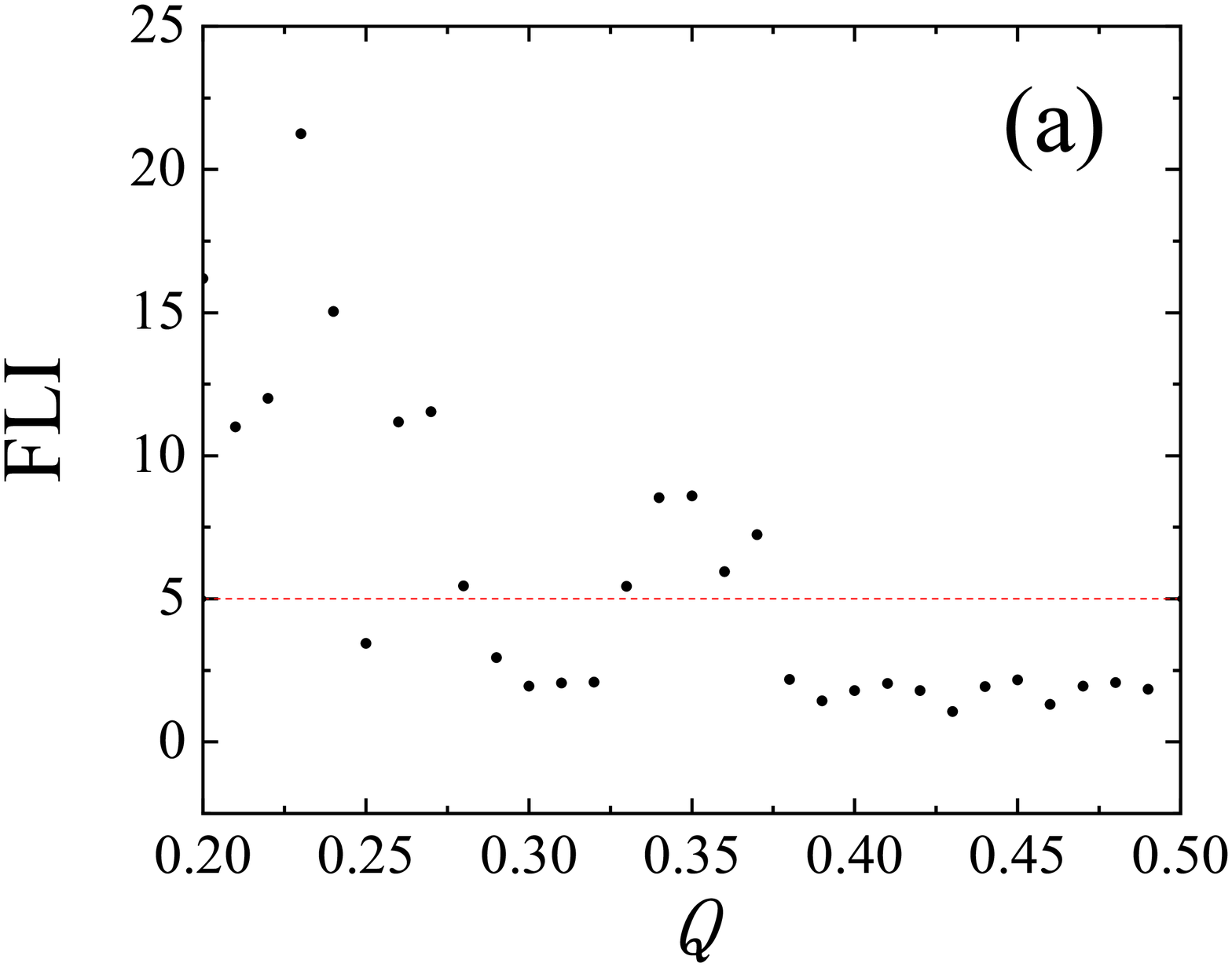}
        \includegraphics[width=18pc]{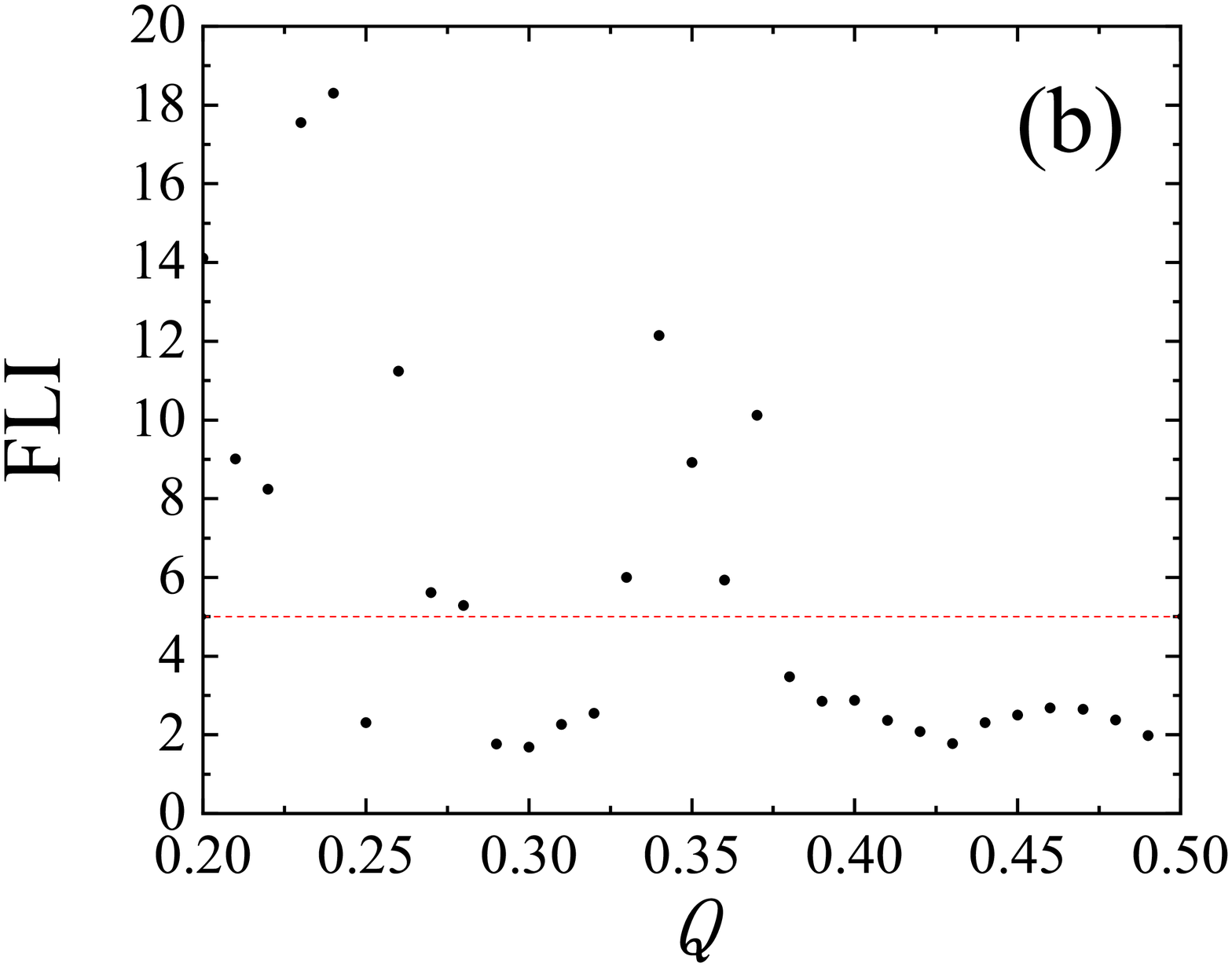}
\caption{Dependence of FLI on the black hole charge $Q$. The other
parameters are consistent with those of Fig. 7, and the initial
separation is $r=100$. The dependence of FLI on the regular black
hole charge $Q$ in panel (a) is the same as that of FLI on the RN
one in panel (b).
            }}
\end{figure*}

\begin{figure*}
    \centering{
\includegraphics[width=12pc]{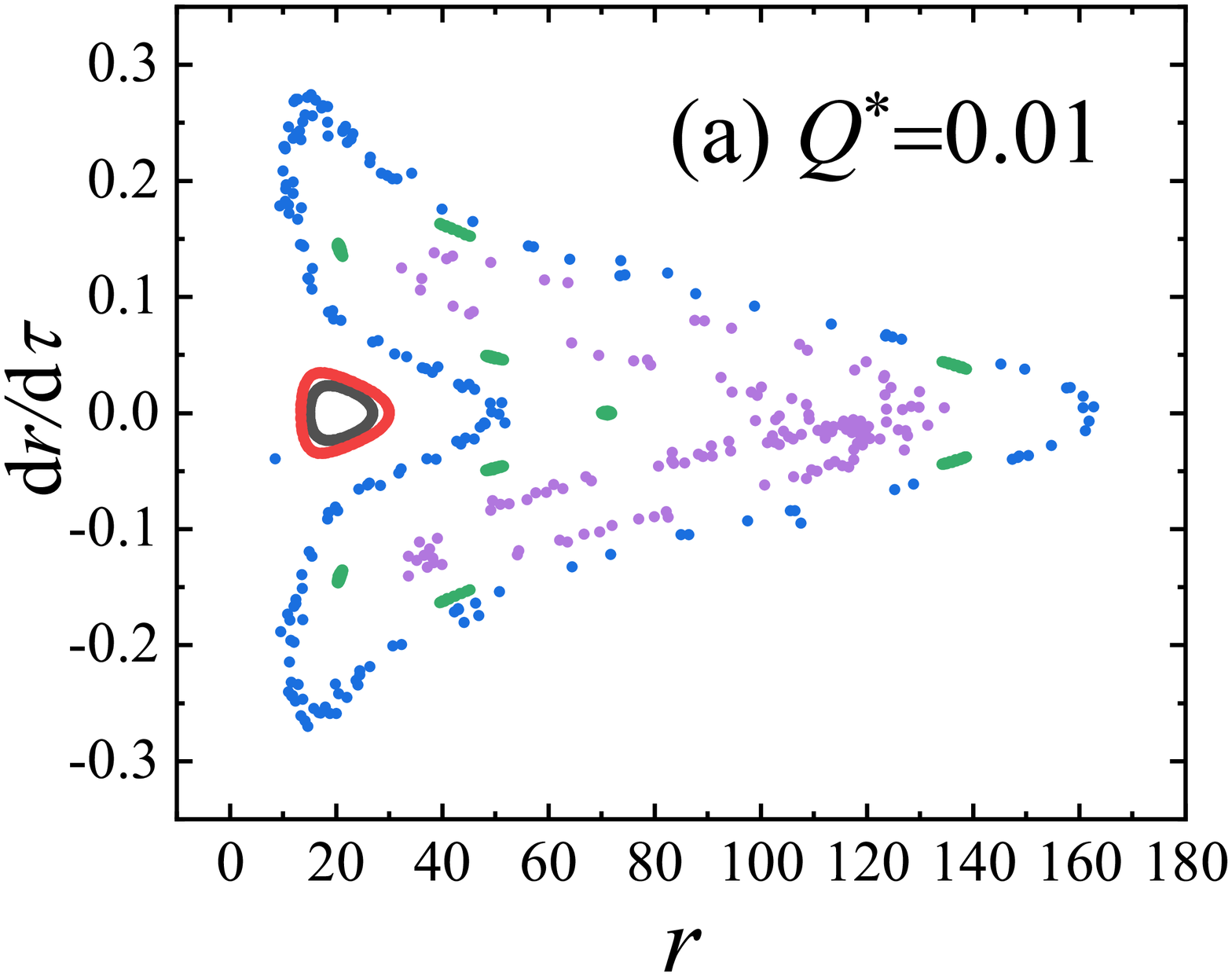}
\includegraphics[width=12pc]{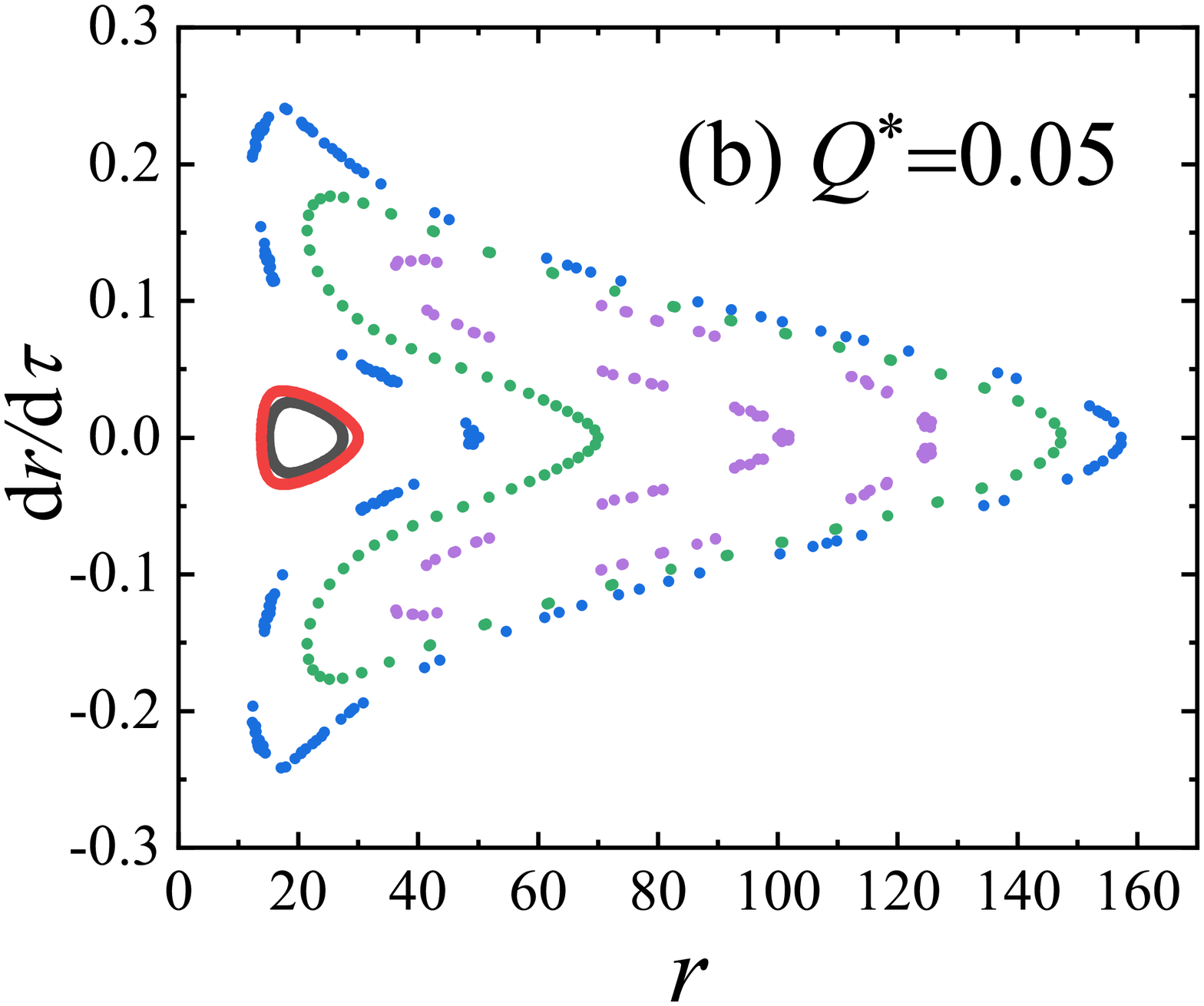}
\includegraphics[width=12pc]{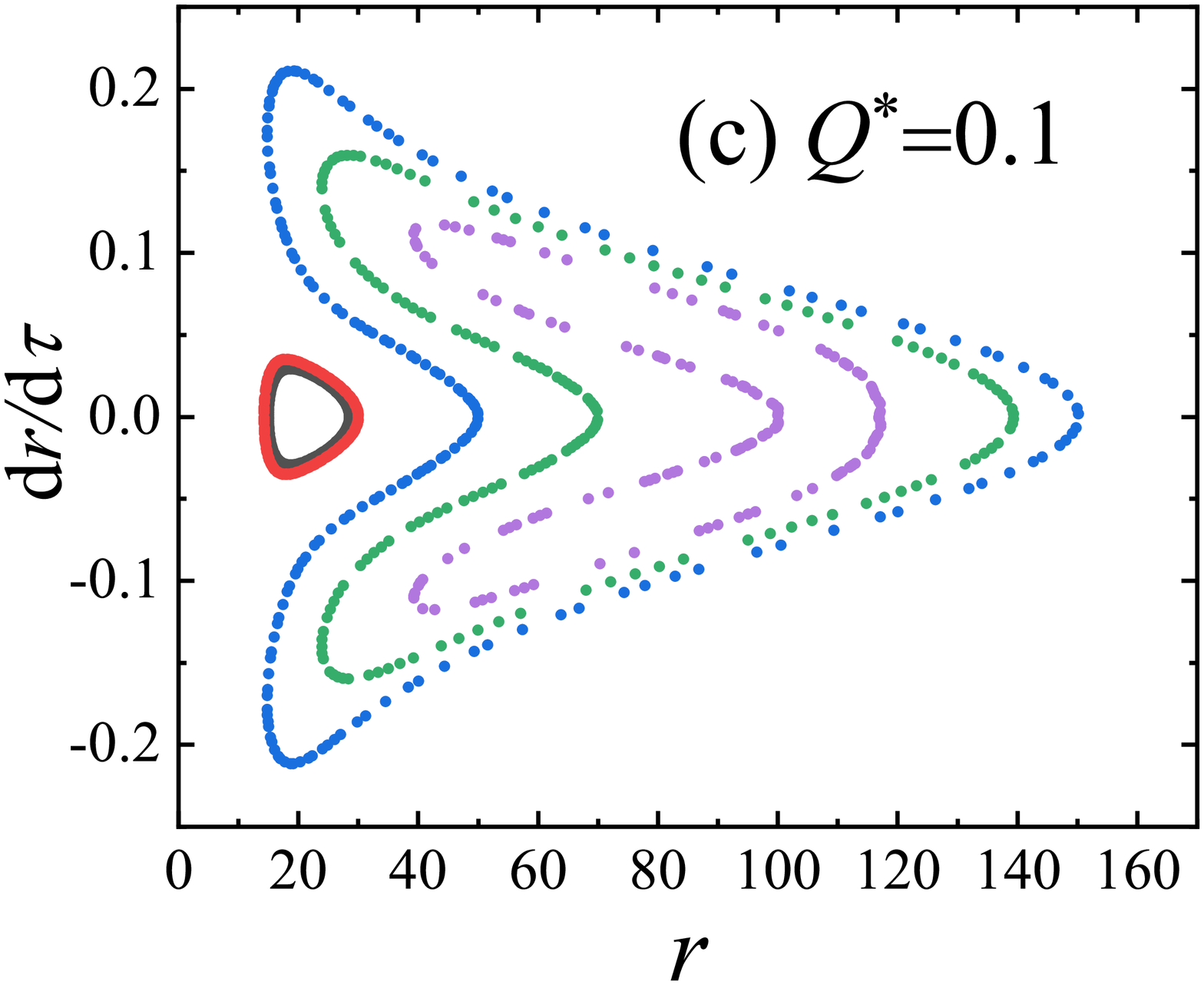}
\includegraphics[width=12pc]{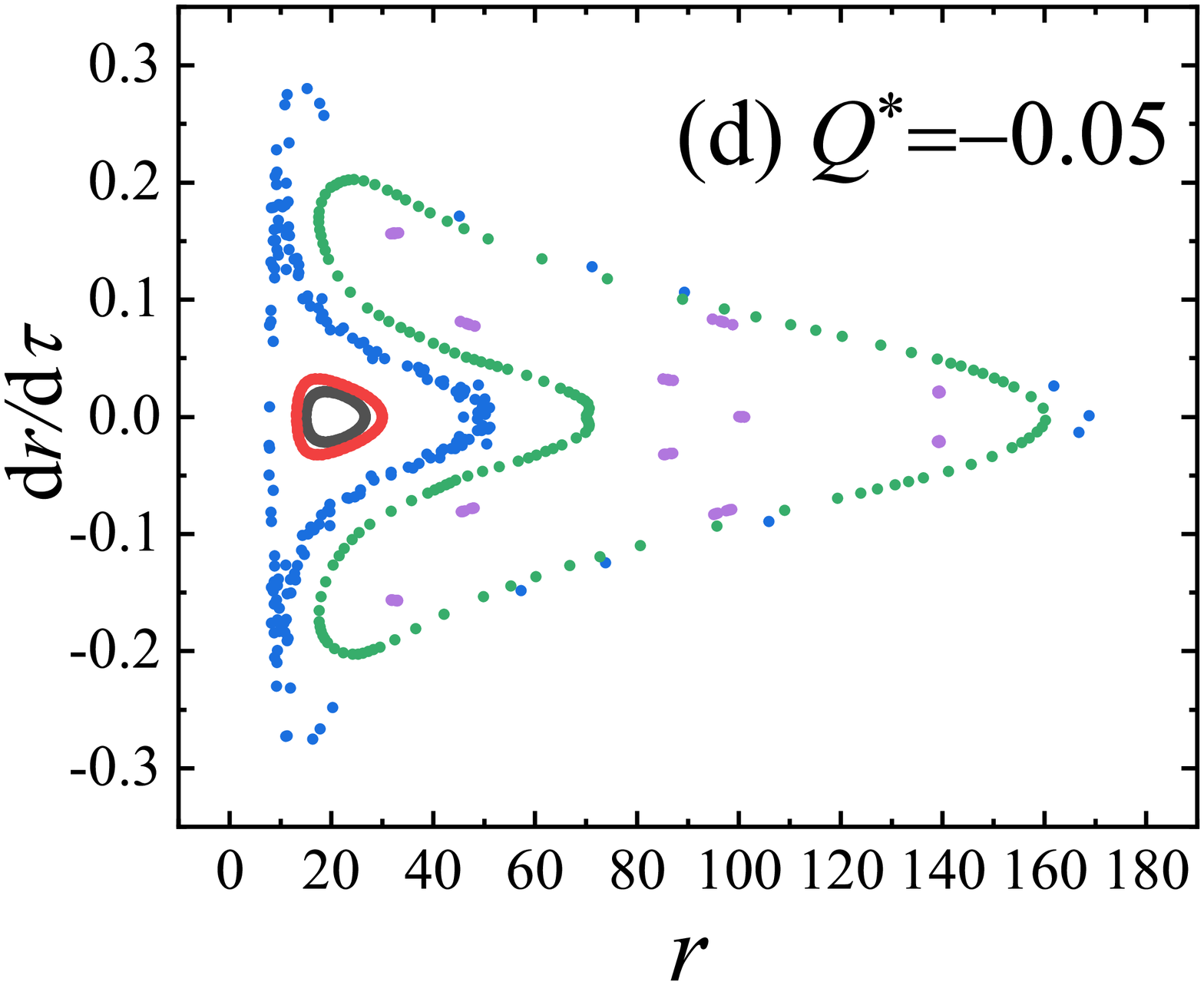}
\includegraphics[width=12pc]{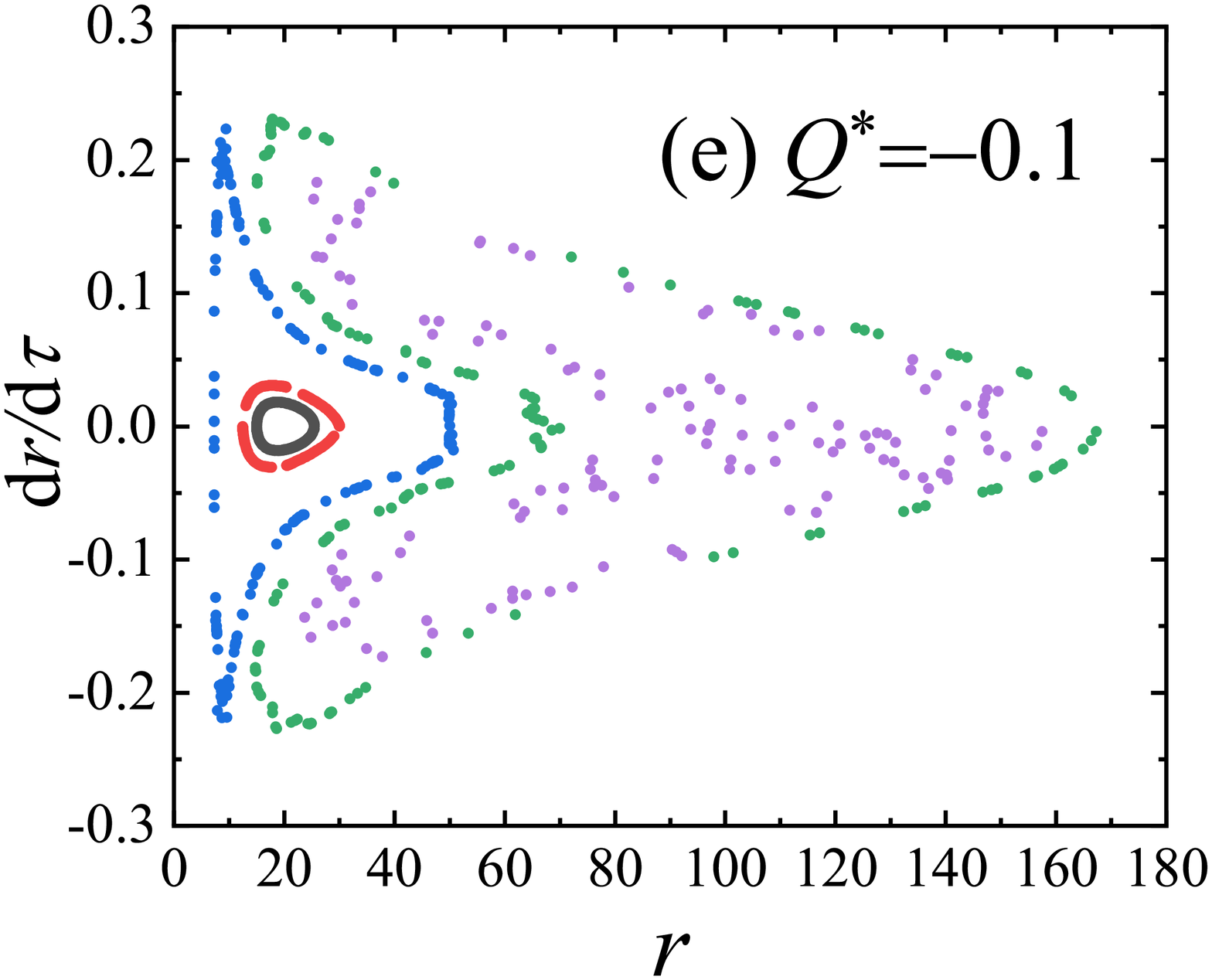}
\includegraphics[width=12pc]{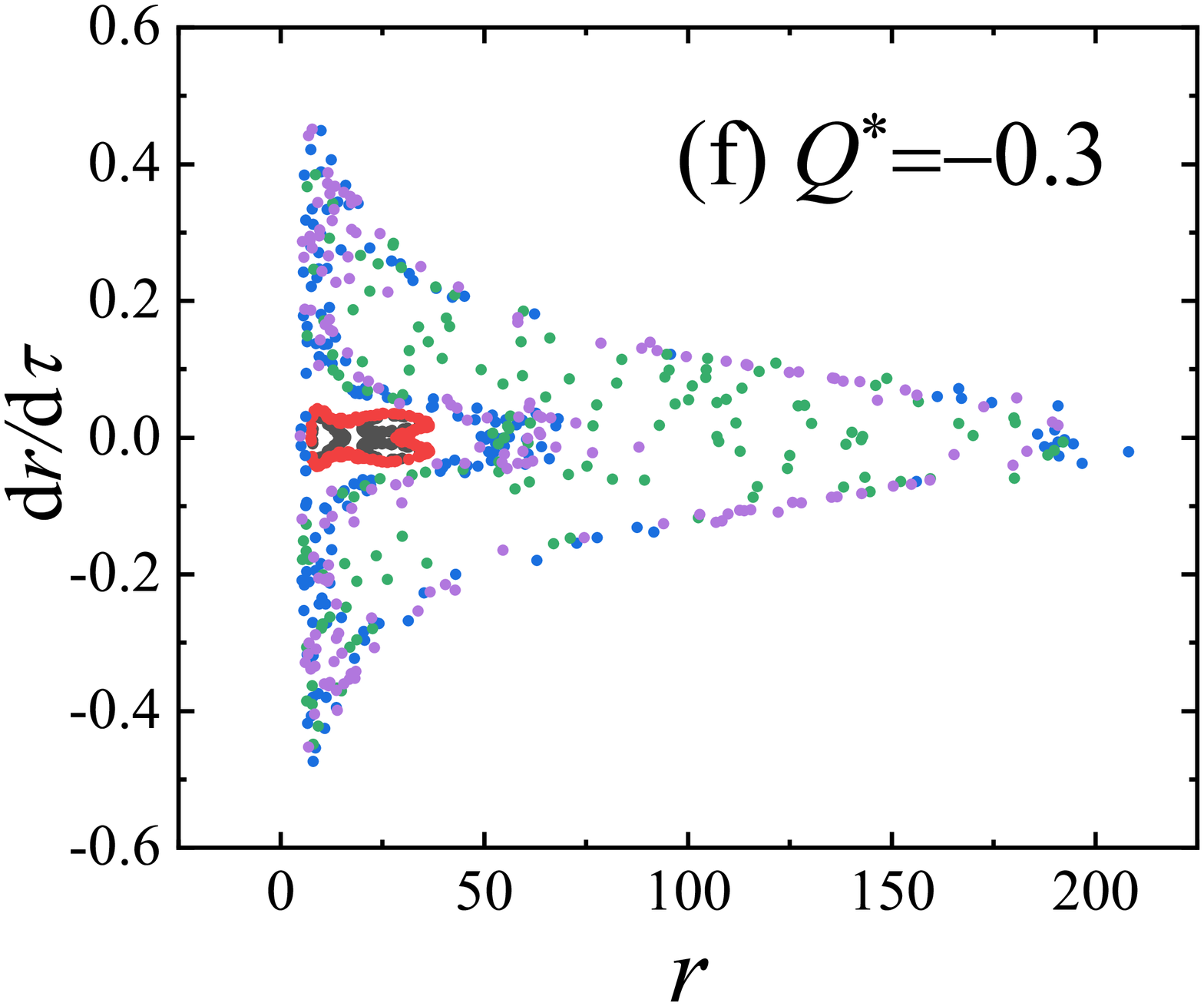}
\caption{Poincar\'{e} sections for three values of  $Q^*$ in the
magnetized regular black hole systems. The other parameters are
$E=0.995$, $L=4.6$, $ Q=0.3$ and $\beta=6.5\times10^{-4}$. The
strength of chaos decreases with an increase of $Q^*>0$ in panels
(a)-(c), but increases with a decrease of $Q^*<0$ in panels
(d)-(f).
            }}
\end{figure*}

\begin{figure*}
    \centering{
\includegraphics[width=12pc]{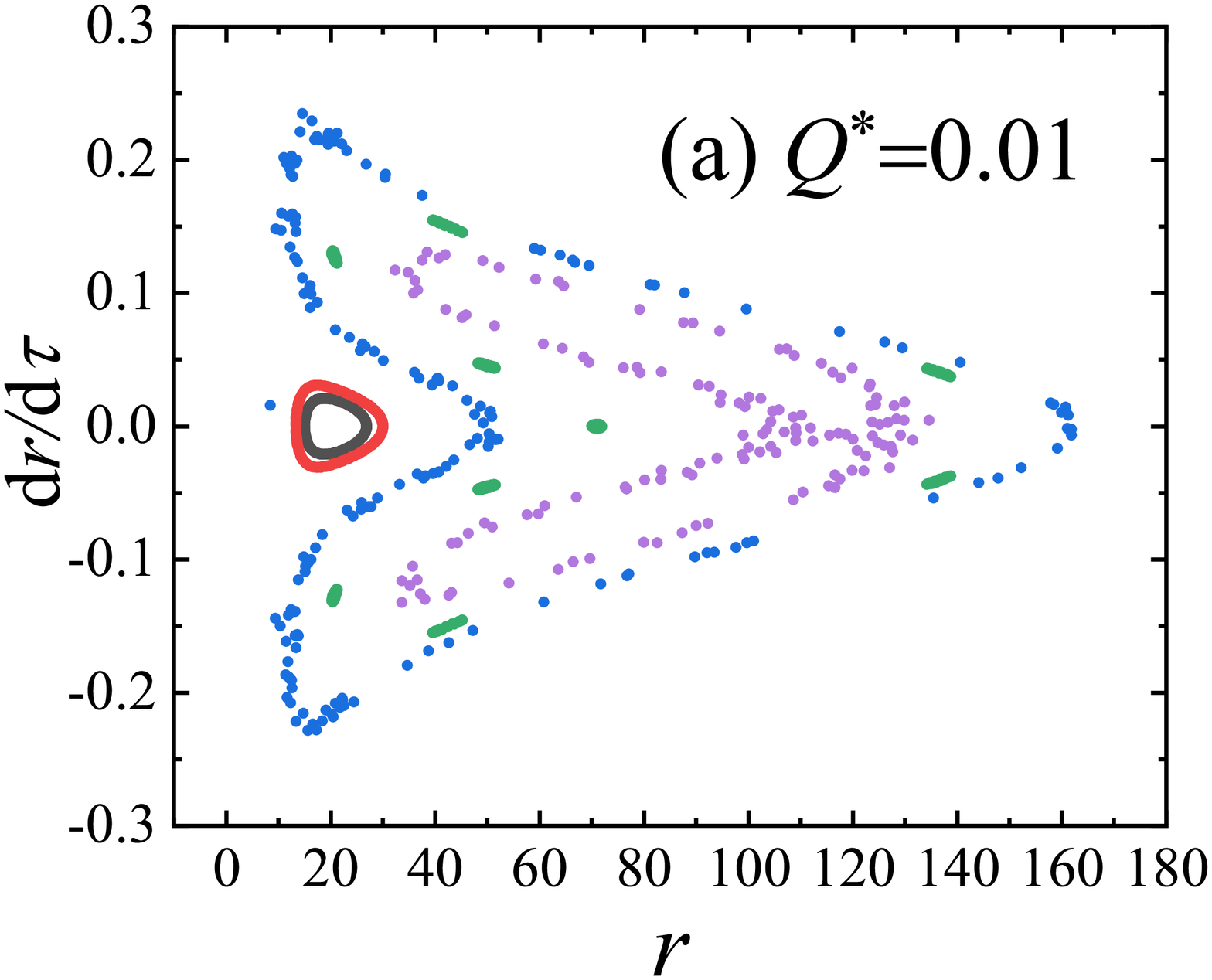}
\includegraphics[width=12pc]{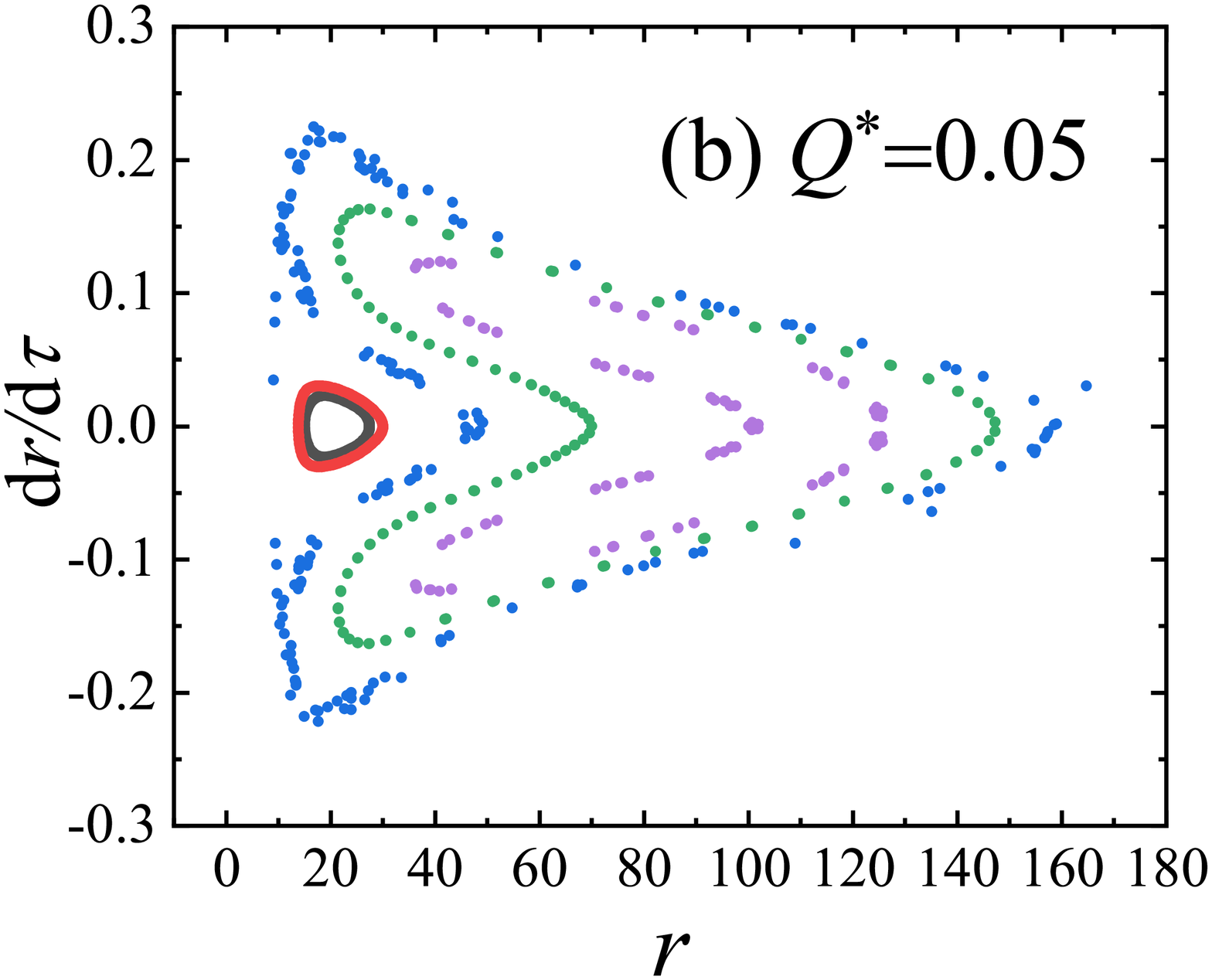}
\includegraphics[width=12pc]{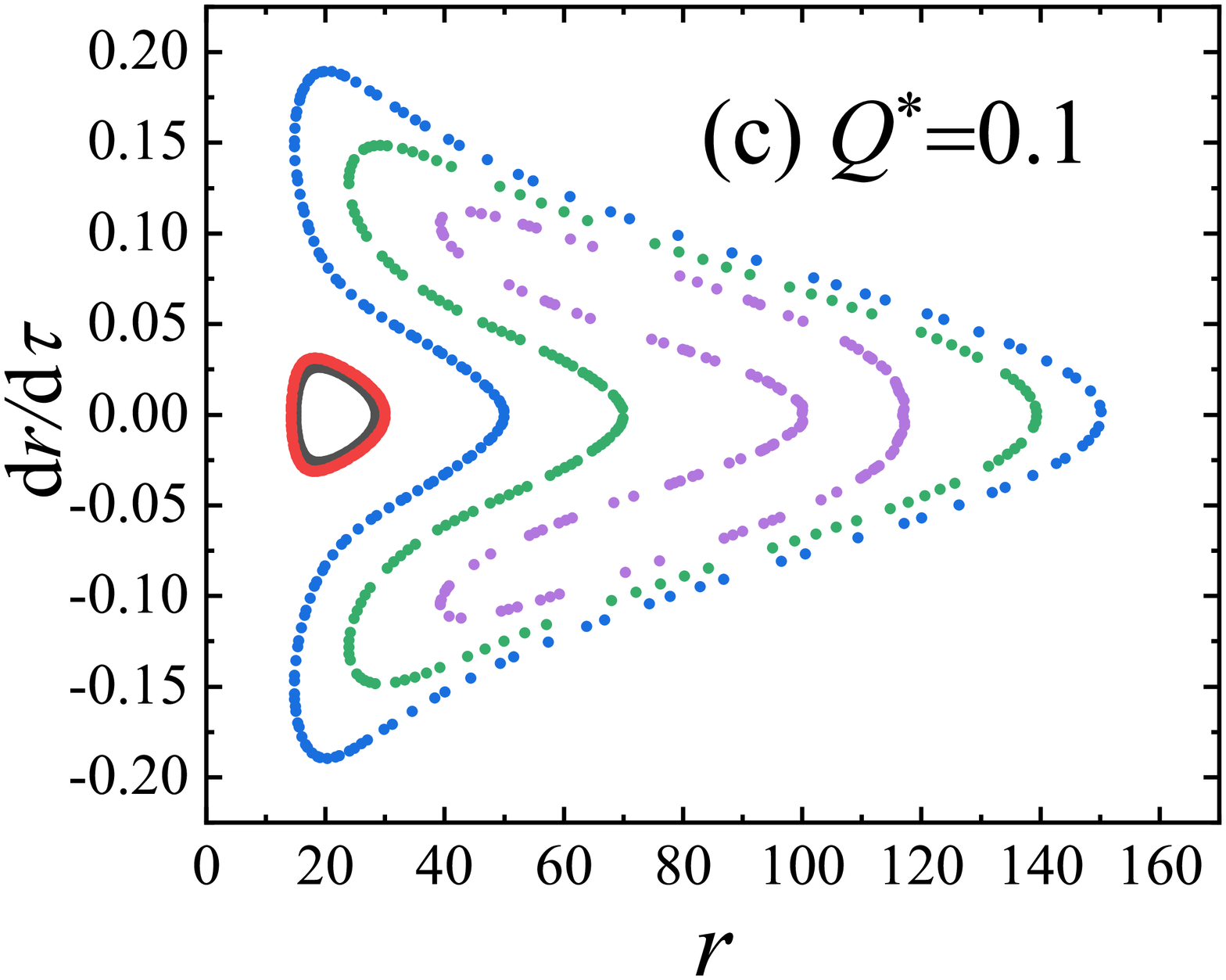}
\includegraphics[width=12pc]{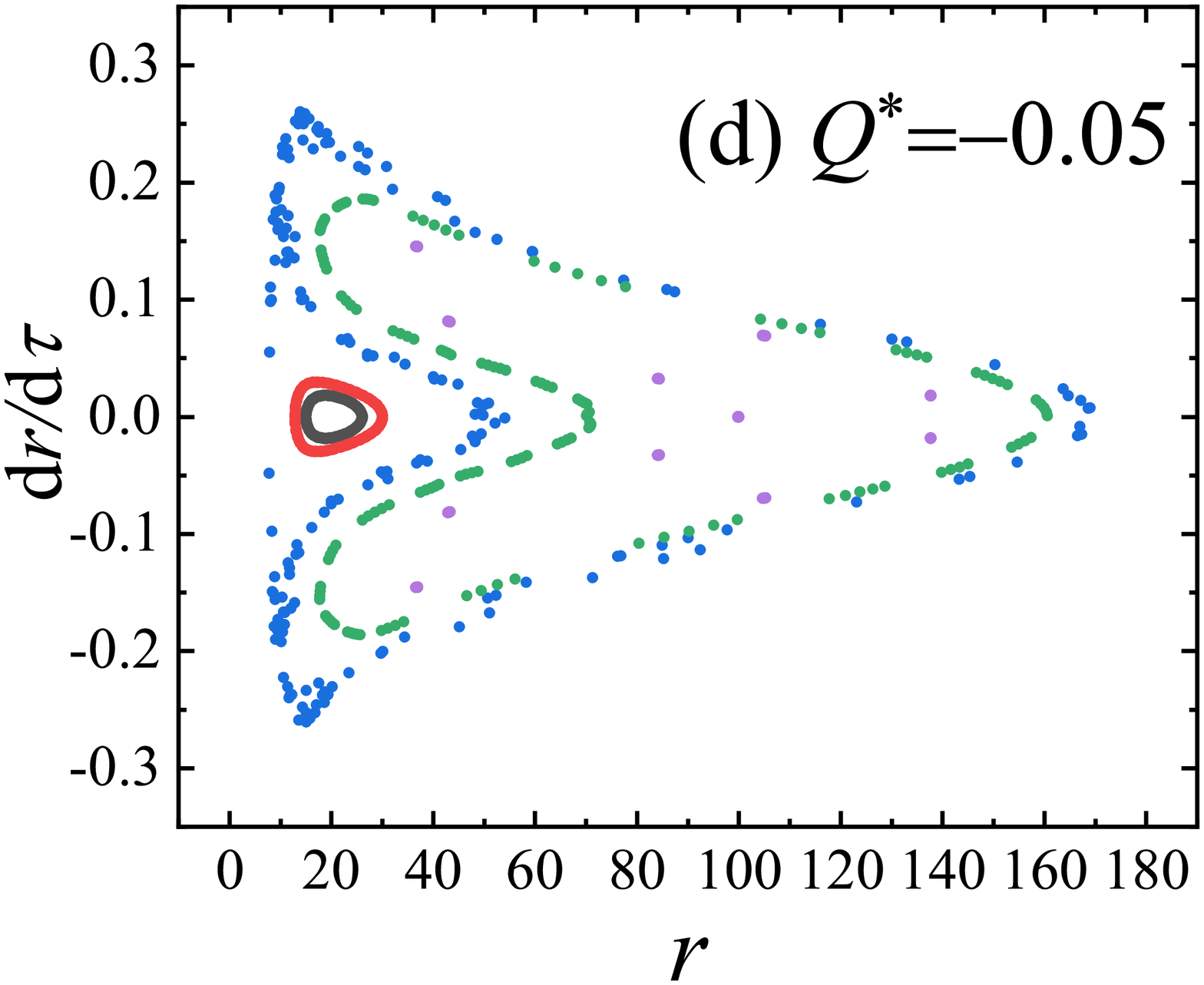}
\includegraphics[width=12pc]{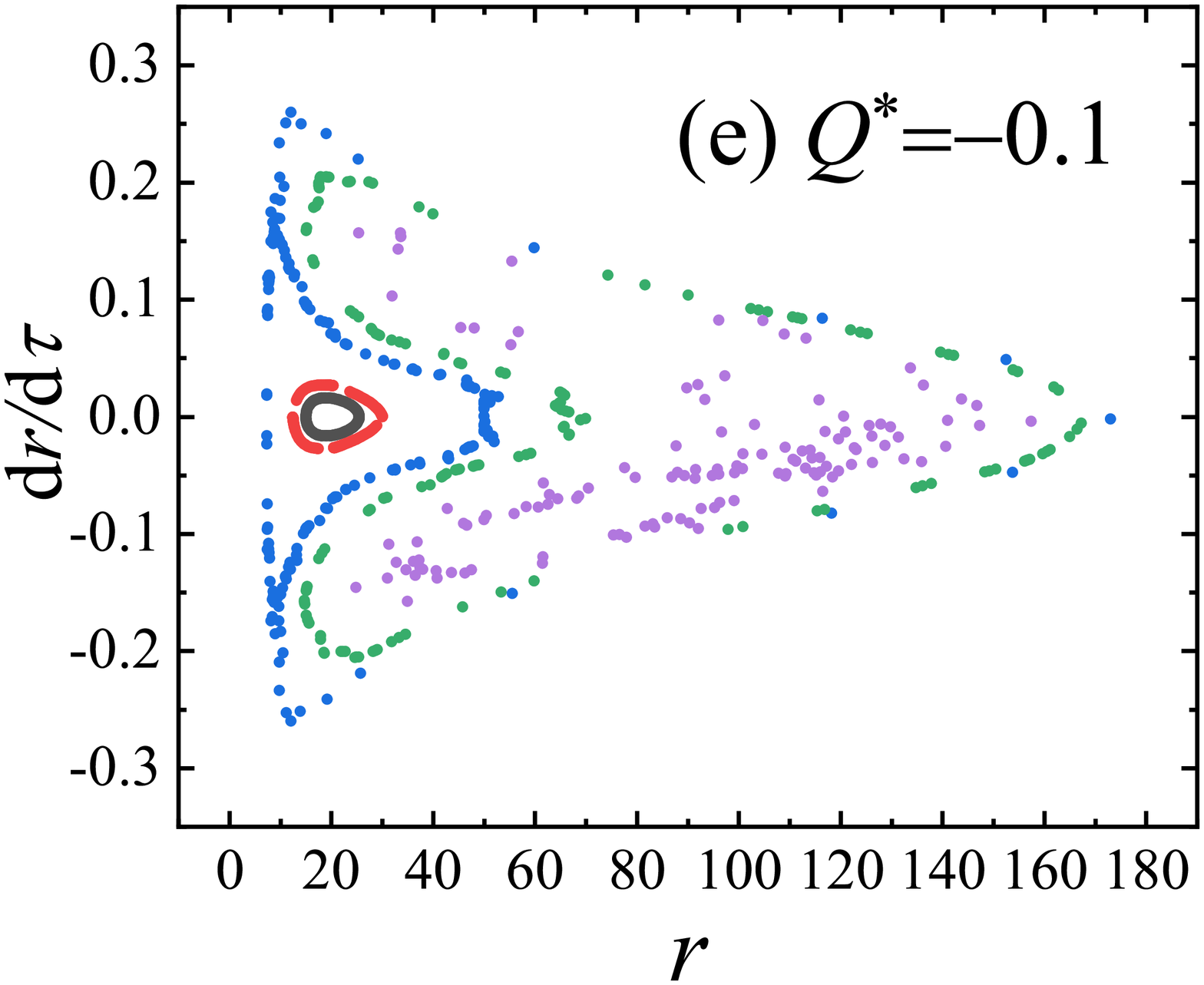}
\includegraphics[width=12pc]{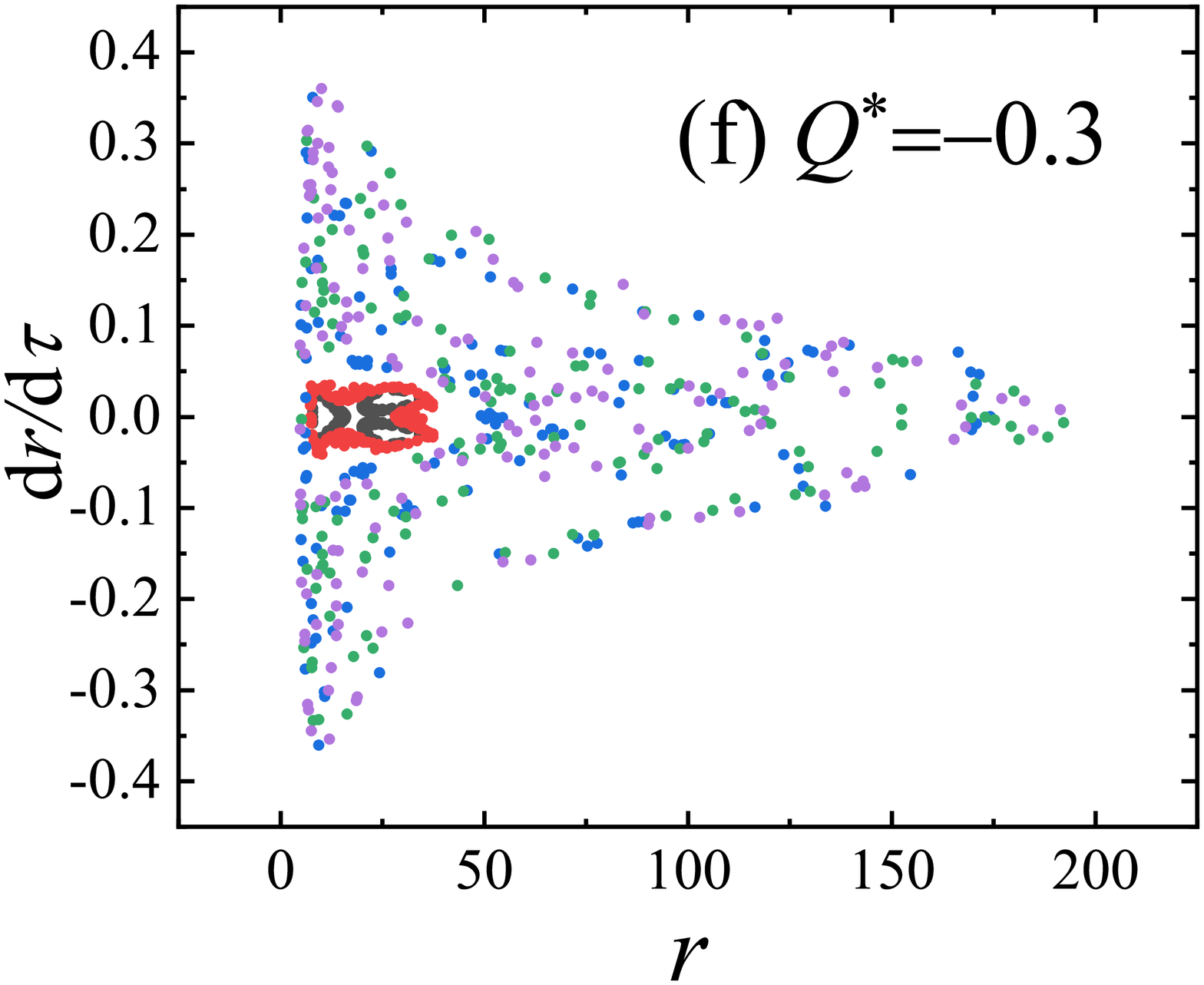}
\caption{Same as Fig. 9, but the magnetized RN black hole systems
are considered. The phase space structures for the RN black hole
are in agreement with those for the regular black hole when the
two black hole systems take the same parameters.
            }}
\end{figure*}

\begin{figure*}
    \centering{
        \includegraphics[width=18pc]{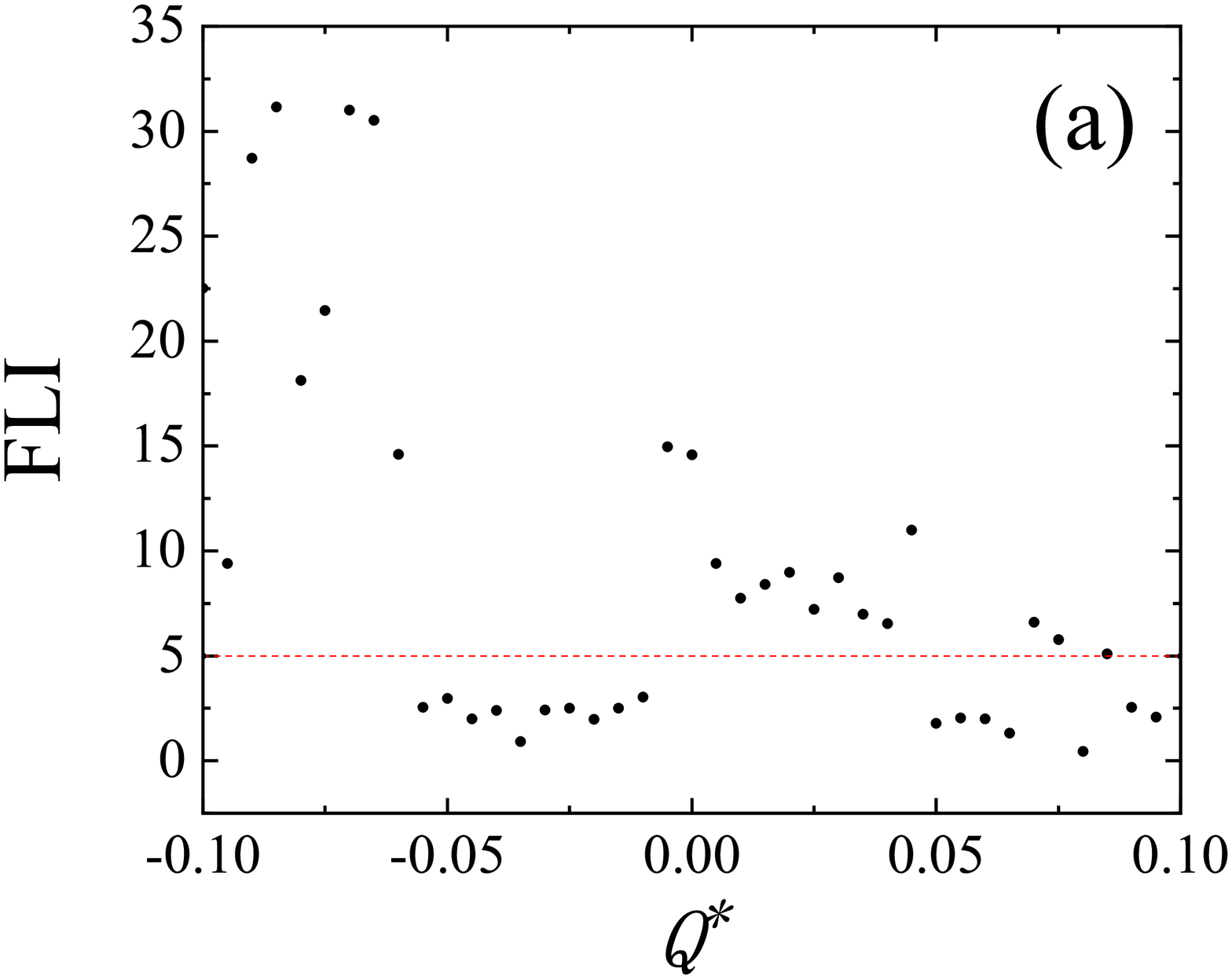}
        \includegraphics[width=18pc]{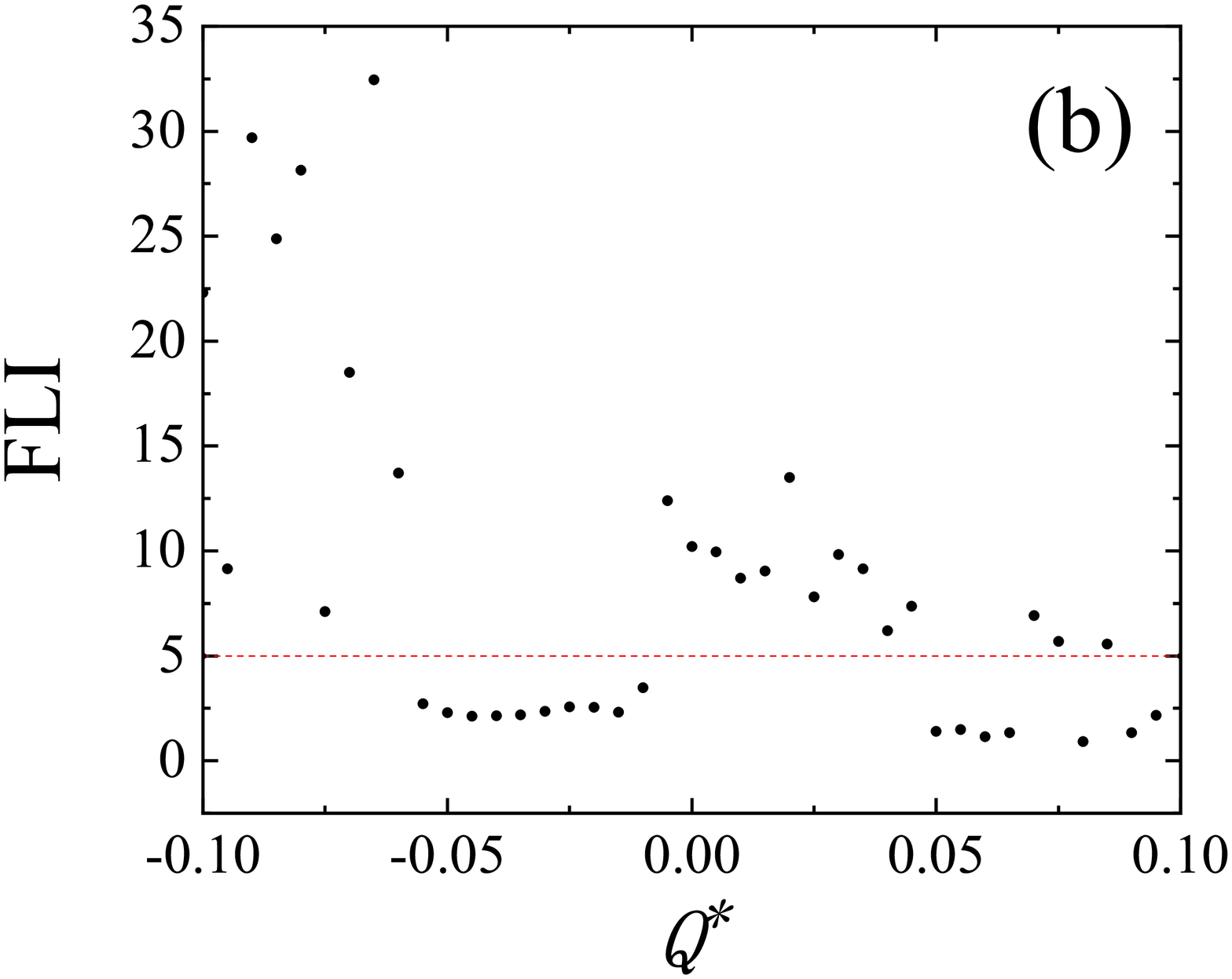}
\caption{Effects of  $Q^*$ on chaos in terms of the FLIs. The
other parameters are those of Fig. 9, and the initial separation
is $r=100$. (a): Corresponding to the regular black hole with the
metric function $f(r)$. (b): Corresponding to the RN black hole
with the metric function $f^{\star}(r)$. The results on the
regular and chaotic behaviors between the two black hole systems
are the same. They are also consistent with those in Figs. 9 and
10.
            }}
\end{figure*}

\begin{figure*}
    \centering{
        \includegraphics[width=18pc]{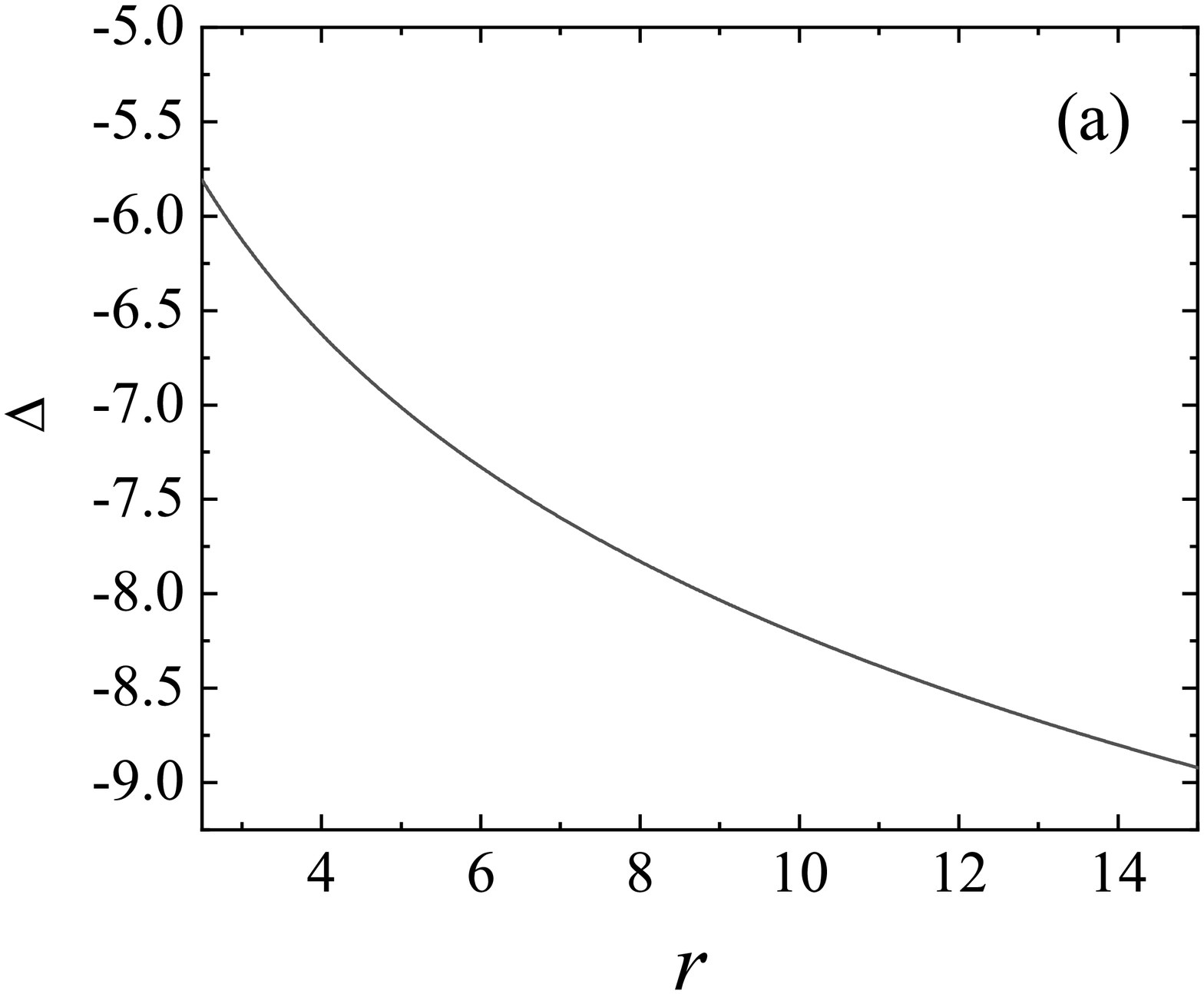}
        \includegraphics[width=18pc]{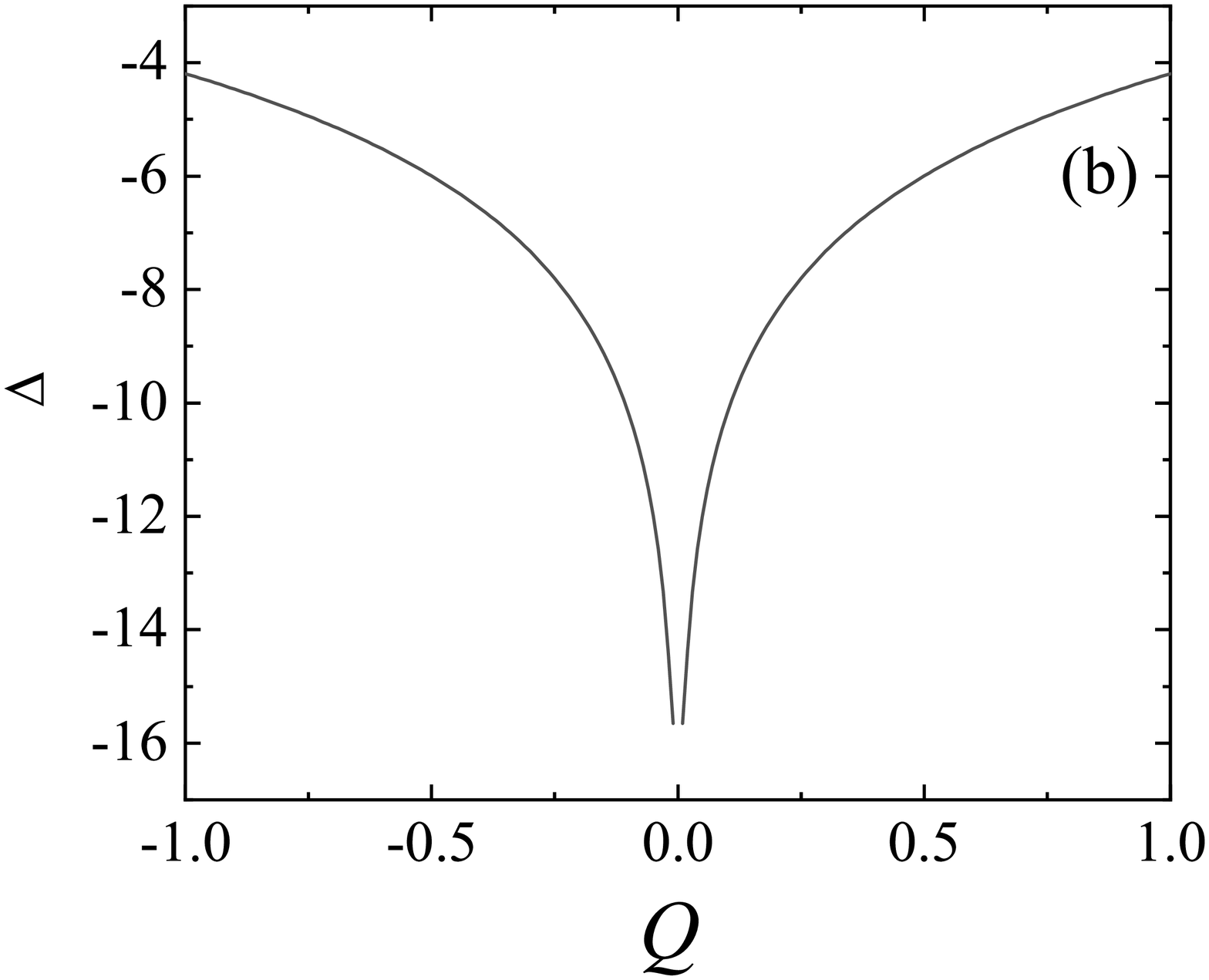}
        \includegraphics[width=18pc]{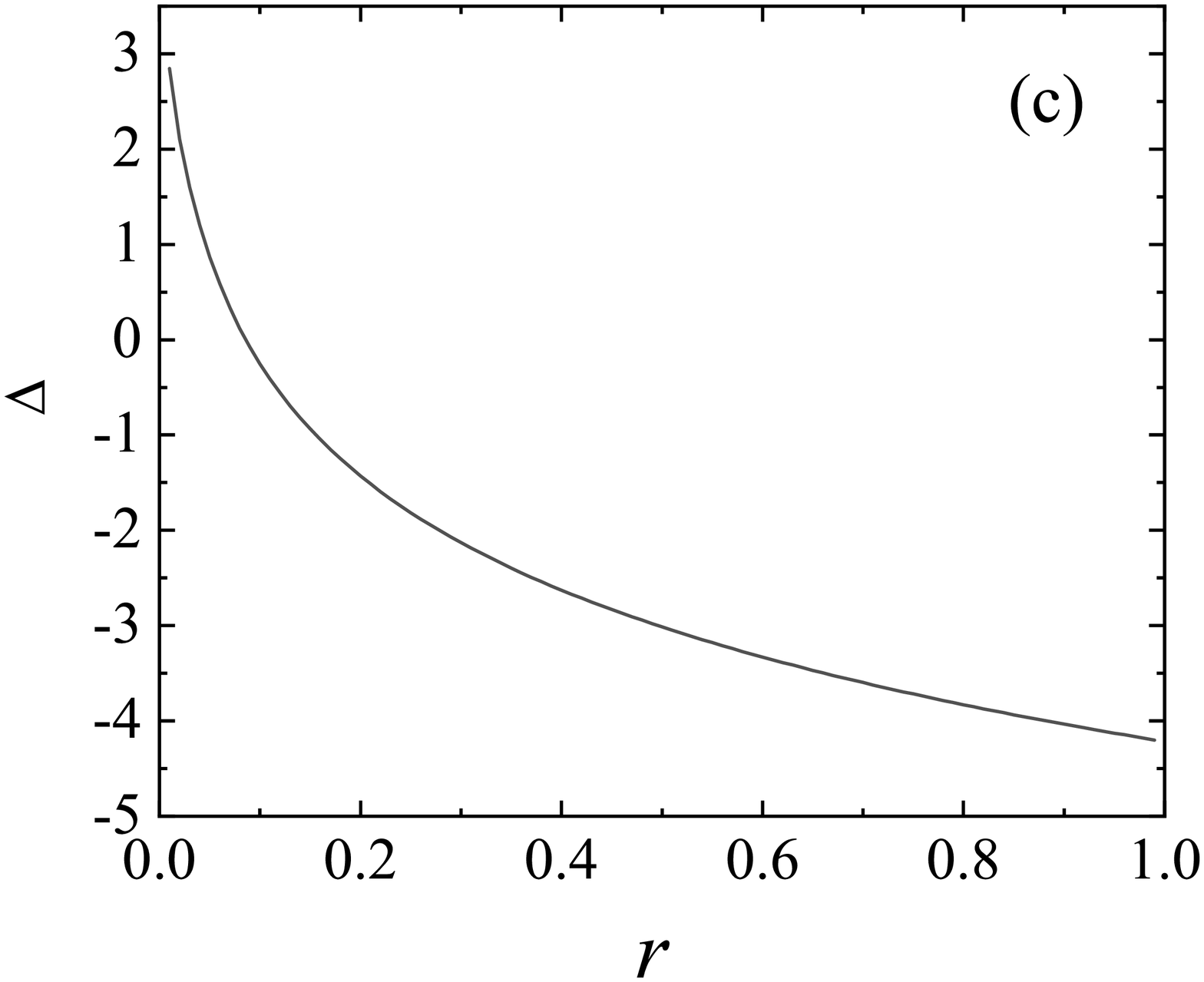}
        \includegraphics[width=18pc]{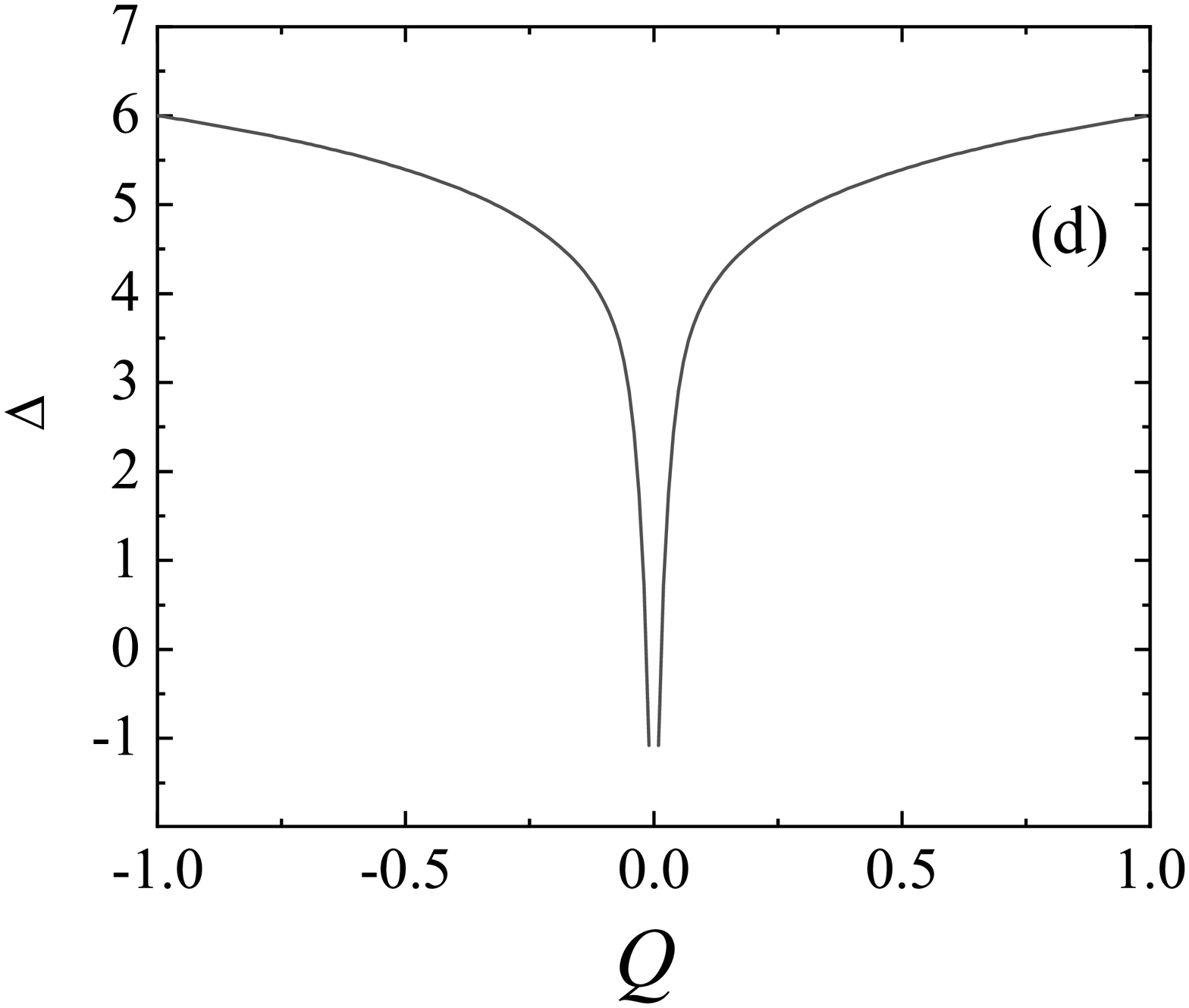}
\caption{The difference between the regular black hole metric
function $f(r)$ and the RN black hole metric function
$f^{\star}(r)$, $|f(r)-f^{\star}(r)|=10^{\Delta}$. (a): Dependence
of $\Delta$ on $r$, where $|Q|=0.3$. When $r$ is given in the
range $2.5\leq r\leq 6$. (b): Dependence of $\Delta$ on $Q$, where
$r=6$.  (c): Dependence of $\Delta$ on $r$, where $|Q|=0.3$ and
$r$ is given in the range $0< r<1$. (d): Dependence of $\Delta$ on
$Q$, where $r=0.001$.
            }}
\end{figure*}

\end{document}